\documentclass[aps,pra,superscriptaddress,10pt,tightenlines,twocolumn,nofootinbib]{revtex4}
\usepackage{amsmath,amsthm,amssymb,verbatim}
\usepackage{graphicx}
\usepackage{lscape}
\usepackage{url}
\usepackage{hyperref}
\usepackage{xifthen} 
\hypersetup{colorlinks=true,urlcolor=blue}

\newcommand{\tr}[1]{{\ensuremath{\hbox{tr}\, #1}}}
\newcommand{\ket}[1]{{\ensuremath{\left| #1 \right\rangle}}}
\newcommand{\bra}[1]{{\ensuremath{\left\langle #1 \right|}}}
\newcommand{\braket}[2]{{\ensuremath{\left\langle #1 \middle| #2
      \right\rangle}}}

\newcommand{\veec}[4]{%
\begin{array}{c}{\!\!#1\!\!}\\{\!\!#2\!\!}\\{\!\!#3\!\!}\\{\!\!#4\!\!}\end{array}}
\newcommand{\ticket}[1]{\boxed{\hbox{#1}}}

\begin{document}

\title{QBism: Quantum Theory as a Hero's Handbook}
\author{Christopher A.\ Fuchs}
\affiliation{Physics Department, University of Massachusetts Boston, Boston, MA 02125, USA}
\author{Blake C.\ Stacey}
\affiliation{Physics Department, University of Massachusetts Boston, Boston, MA 02125, USA}


\begin{abstract}
This article situates QBism among other well-known interpretations of quantum mechanics.  QBism has its roots in personalist Bayesian probability theory, is crucially dependent upon the tools of quantum information, and in latest developments has set out to investigate whether the physical world might be of a type sketched in the philosophies of pragmatism, pluralism, nonreductionism, and meliorism.  Beyond conceptual issues, the technical side of QBism is focused on the mathematically hard problem of finding a {\it good\/} representation of quantum mechanics purely in terms of probabilities, without amplitudes or Hilbert-space operators.  The best candidate representation involves an entity called a symmetric informationally complete quantum measurement, or SIC\@.  Contemplation of it gives a way to think of the Born rule as an {\it addition\/} to the {\it rules\/} of probability theory, applicable when an agent considers gambling on the consequences of her actions on the external world, duly taking into account a new universal capacity:  namely, Hilbert-space dimension.  The article ends by showing that the egocentric elements in QBism represent no impediment to pursuing quantum cosmology and even open up possibilities never dreamt of in the usual formulations.
\end{abstract}

\maketitle

\section{Introduction}

\begin{flushright}
\baselineskip=13pt
\parbox{2.8in}{\baselineskip=13pt\small
Chauncey Wright a nearly forgotten philosopher of real merit, taught me when young that I must not say {\it necessary\/} about the universe, that we don't know whether anything is necessary or not.  So I describe myself as a {\it bet\/}tabilitarian.  I believe that we can {\it bet\/} on the behavior of the universe in its \underline{contact} with us. [Our underlining; his italics.]}
\medskip\\
\small --- Oliver Wendell Holmes, Jr.
\end{flushright}

Quantum physics works.  But why?

The mathematical apparatus that we heft onto physics students is
astonishingly successful.  It guides our understanding of phenomena
from the submicroscopic clustering of quarks to the spectra of
quasars, and it underpins technological advances that affect our lives
more dramatically by the day.  But when we take a moment and
ask---whether in a dorm room or the pages of a philosophy
journal---what the theory is all about, we find ourselves thrashing
about in decades of accumulated murk.  How much of the mathematical
gadgetry is human convention and historical happenstance, and how much
of it truly indicates the character of the natural world that was here
before we were and would endure in our absence?  Can we take discourse
about observership and agency, about what it means to be an agent
whose actions have consequences, about the relation between truth and
what works in practice, and make honest mathematics of it?

According to the research program of QBism~\cite{RMP, Fuchs10,
  Mermin13, Fuchs10b, AJP, Mermin14, stacey-vonneumann, Baeyer16,
  sep-quantum-bayesian}, the modern development of the older Quantum
Bayesianism~\cite{Caves02, Fuchs02, Fuchs04, Caves07}, the answer is
wholeheartedly {\it Yes.}  On one hand, QBism is a way of investing
meaning in the abstract structure of quantum theory: It tells us that
quantum theory is fundamentally about agency, actions, consequences
and expectations.  On the other, QBism points out the virtue of
reconstructing quantum theory from deep, physical principles.  Of all
the ideas and theorems that look important, which are {\it the\/}
ideas, the captivating and compelling seed from which all the formulae
would grow given only careful thinking?

We can illustrate the trouble with quantum mechanics by comparing it with other areas of physics in which we have collectively honed our understanding to a high degree of sophistication.  Two examples that come to mind are the science of thermodynamics and the special theory of relativity.  An old joke has it that the three laws of thermodynamics are ``You can't win,'' ``You can't break even,'' and ``You can't get out of the game.''  To these, we ought to prepend the zeroth law, which we could state as, ``At least the scoring is fair.''  But consider the premise of the joke, which is really rather remarkable:  There {\it are\/} laws of thermodynamics---a concise list of deep physical principles that underlie and nourish the entire subject.  Likewise for special relativity:  Inertial observers Alice and Bob can come to agree on the laws of physics, but no experiment they could ever do can establish that one is ``really moving'' and the other ``really standing still''---not even measuring the speed of light.  We invest a little mathematics, and then close and careful consideration of these basic principles yields all the details of the formal apparatus, with its nasty square roots or intermingling partial derivatives.

This level of understanding brings many advantages.  Having the deep principles set out in explicit form points out how to {\it test\/} a theory in the most direct manner.  Moreover, it greatly aids us when we {\it teach\/} the theory.  We do not have to slog through all the confusions that bedeviled the physicists who first developed the subject, to say nothing of the extra confusions created by the fact that ``historical'' pedagogy is almost inevitably a caricature.  In addition, a principled understanding helps us {\it apply\/} a theory.  As we work our way into a detailed calculation, we can cross-check against the basic postulates.  Does our calculation imply that signals travel faster than light?  Does our seventeenth equation imply that entropy is flowing the wrong way?  We must have made an error!  And, when we found our theory upon its deep principles, we have a guide for {\it extending\/} our theory, because we know what properties must obtain in order for a new, more general theory to reduce to our old one in a special case.

To our great distress, we must admit that in the matter of quantum mechanics, the physics profession lacks this level of understanding.

Instead, we have a mathematical apparatus and day-to-day experience on how to apply it successfully.  But the deep principles remain elusive.  What we have in their place is almost a century of ``interpretations'' cooked up for the theory---campfire stories told to give meaning to the mathematics and say what it is ``all about.''  And what a host of tales have there been!
\begin{flushleft}
\begin{itemize}
\item Copenhagen Interpretations -- Bohr, Heisenberg, Pauli, von Weizs\"acker, Peierls, Wheeler (they're actually all different!)
\item Nonlocal Hidden Variables -- de Broglie, Bohm, Bub (early on), Maudlin, Goldstein, Valentini, Norsen, Hardy
\item Stochastic Mechanics -- Nelson, Smolin (Lee)
\item Modal Interpretations -- Kochen, Dieks, van Fraassen, Bub (later on), Healey (for a time),
  Spekkens \& Sipe (briefly)
\item Quantum Logics -- Birkhoff \& von Neumann, Mackey, Jauch, Piron, Finkelstein, Putnam
\item Consciousness-Induced Collapse Theories -- Wigner, von Neumann
\item Objective Collapse Models -- Ghirardi-Rimini-Weber, Pearl, Penrose, Frigg, Gisin, Tumulka, Albert
\item Consistent Histories -- Griffiths, Omn\`es
\item Transactional Interpretation -- Cramer, Kastner
\item Relational Interpretations -- Rovelli, Spekkens (TBA)
\item Ensemble Interpretation -- Einstein (?), Ballentine
\item Informational Interpretations -- Zeilinger, Brukner, Bub (presently)
\item Superdeterminism -- Bell (as a joke), 't Hooft (in full seriousness)
\item Many-Worlds Interpretations -- Everett, Albert \& Loewer, Barbour,
  Coleman, Deutsch (early version), Deutsch (later version), DeWitt, Gell-Mann \& Hartle, Geroch,
  Graham, Greaves, Lockwood, Papineau, Saunders,
  Smolin (John), Tegmark, Vaidman, Wallace, Zurek \ldots
\end{itemize}
\end{flushleft}
This list is far from exhaustive, and as already hinted abundantly, two authors listed under the same bullet may well disagree on significant issues.\footnote{Over the years, the many-worlders in particular have done a remarkably poor job of agreeing with one another about what specifically there are supposed to be many of.}

It is often said that ``all interpretations of quantum mechanics make the same predictions, so we cannot tell them apart by experiment.''  This is false for at least two reasons, a small one and a big one.  The small point is that some ideas classified among the ``interpretations'' are deliberately fashioned to depart from quantum theory in some way.  Objective Collapse models are the chief examples of this.  But more importantly, it is not clear that all of the interpretations {\it can be made to yield predictions at all,} when they are thought upon with adequate stringency.

We should also mention the attitude that there is no {\it point\/} to ``interpreting'' quantum theory, because quantum theory will turn out to be wrong anyway.  Perhaps the reason why we have yet to fit quantum physics together with gravity is that the quantum side is fundamentally defective, in some fashion that has only manifested when we put the full weight of general relativity upon it.  (Feynman even speculated that gravity itself could be the result of quantum mechanics breaking down, in which case ``quantizing gravity'' would be meaningless~\cite{Feynman03}.)  Fair enough!  But how, then, do we modify quantum theory in such a way that it still works in all the many places it has worked so far?

So the field of quantum foundations is not unfounded; it is absolutely vital to physics as a whole.  But what constitutes ``progress'' in quantum foundations?  How would one know progress if one saw it?  Through the years, it seems the most popular strategy has been to remove the observer from the theory just as quickly as possible, and with surgical precision.  In practice this has generally meant to keep the {\it mathematical structure\/} of quantum theory as it stands (complex Hilbert spaces, operators, tensor products, etc.), but, by hook or crook, find a way to tell a story about the mathematical symbols that involves no observers at all.

In short, the strategy has been to reify or objectify all the
mathematical symbols of the theory and then explore whatever comes of
the move.  All the various interpretations that result see quantum states
as physical entities, like a blob of $\psi$-flavored gelatin, sliding
about in accord with its own dynamical laws.
Three examples suffice to give a feel:  In the
de~Broglie--Bohm ``pilot wave'' version of quantum theory, there are
no fundamental measurements, only ``particles'' flying around in a
$3N$-dimensional configuration space, pushed around by a wave function
regarded as a real physical field in that space.  In ``spontaneous
collapse'' versions, systems are endowed with quantum states that
generally evolve unitarily, but from time-to-time collapse without any
need for measurement.  In Everettian or ``many-worlds'' quantum
mechanics, it is only the world as a whole---they call it a
multiverse---that is really endowed with an intrinsic quantum state,
and that quantum state evolves deterministically, with only an {\it
  illusion from the inside\/} of probabilistic ``branching.''

The trouble with all these interpretations as quick fixes for quantum strangeness is that they look to be just that, {\it really quick fixes}.  They look to be interpretive strategies hardly compelled by the particular details of the quantum formalism, giving only more or less arbitrary appendages to it.  This already explains in part why we have been able to exhibit three such different strategies, but it is worse:  Each of these strategies gives rise to its own set of tough-to-swallow ideas.  Pilot-wave theories, for instance, give instantaneous action at a distance, but not actions that can be harnessed to send detectable signals.  If so, then what a delicately balanced high-wire act nature presents us with.  And how much appeal does the idea of waves pushing particles about really have when it turns out that, in order to be consistent, ``position measurements'' don't really measure particle positions~\cite{Gisin15}?  Or take the Everettians.  Their world purports to have no observers, but then it has no probabilities either.  What are we then to do with the Born rule for calculating quantum probabilities?  Throw it away and say it never mattered?  It is true that quite an effort has been made by the Everettians to rederive the rule by one means or another.  But these attempts may have re-imported at least as much vagueness as they claim to eliminate \cite{Caves05, Kent14, Kastner14, Adlam14, Jansson16}. To many in the outside world, it looks like the success of these derivations depends upon where they are assessed---for instance, whether in Oxford \cite{Saunders05, Wallace09} or Cambridge \cite{Price08, Kent09}.

QBists hold that the way forward is to own up to the following lesson.  Before there were people using quantum {\it theory\/} as a branch of physics, before they were {\it calculating\/} neutron-capture cross-sections for uranium and working on all the other practical problems the theory suggests, there were no quantum states.  The world may be full of stuff and things of all kinds, but among all the stuff and all the things, there is no unique, observer-independent, {\it quantum-state kind of stuff}.

The immediate payoff of this strategy is that it eliminates the conundrums arising in the various objectified-state interpretations.  A paraphrase of a quote by James Hartle makes the point decisively~\cite{Hartle68}:
\begin{quote}
\noindent A quantum-mechanical state being a summary of the observers' information about an individual physical system changes both by
dynamical laws, and whenever the observer acquires new information about the system through the process of measurement.  The existence
of two laws for the evolution of the state vector becomes problematical only if it is believed that the state vector is an objective property of the system.   If, however, the state of a system is defined as a list of [experimental] propositions together with their [probabilities of occurrence], it is not surprising that after a measurement the state must be changed to be in accord with [any] new information.  The ``reduction of the wave packet'' does take place in the consciousness of the observer, not because of any unique physical process which takes place there, but only because the state is a construct of the observer and not an objective property of the physical system.
\end{quote}

The objective quantum state is the latter-day equivalent of the
luminiferous {\ae}ther.  But recognizing this is only the first step
of an adventure.  Luckily the days for this expedition are ripe, thanks in large part to the development of the field of quantum information theory in the last 25 years---that is, the multidisciplinary field that has brought about quantum cryptography, quantum teleportation, and will one day bring about full-blown quantum computation.  Terminology can say it all:  A practitioner in this field, whether she has ever thought an ounce about quantum foundations, is just as likely to say ``quantum information'' as ``quantum state'' when talking of any $|\psi\rangle$.  ``What does the quantum teleportation protocol do?''  A now completely standard answer is: ``It transfers {\it quantum information\/} from Alice's site to Bob's.''  What we have here is a change of mindset~\cite{Fuchs10}.

What the facts and figures, protocols and theorems of quantum information pound home is the idea that quantum states look, act, and feel like information in the technical sense of the word---the sense provided by probability theory and Shannon's information theory.  There is no more beautiful demonstration of this than Robert Spekkens's ``toy model'' for mimicking various features of quantum mechanics \cite{Spekkens07}.  In that model, the ``toys'' are each equipped with four possible mechanical configurations; but the players, the manipulators of the toys, are consistently impeded from having more than one bit of information about each toy's actual configuration. (Or a total of two bits for each two toys, three bits for each three toys, and so on.)  The only things the players can know are their own states of uncertainty about the configurations.  The wonderful thing is that these states of uncertainty exhibit many of the characteristics of quantum information: from the no-cloning theorem to analogues of quantum teleportation, quantum key distribution, entanglement monogamy, and even interference in a Mach--Zehnder interferometer.  More than two dozen quantum phenomena are reproduced {\it qualitatively}, and all the while one can always pinpoint the underlying cause of this:  The phenomena arise in the uncertainties, never in the mechanical configurations.  It is the states of uncertainty that mimic the formal apparatus of quantum theory, not the toys' so-called {\it ontic states\/} (states of reality).

What considerations like this tell the $\psi$-ontologists\footnote{This beautiful word, not to be confused with a practitioner of Scientology, was coined by Christopher Granade \cite{LordVoldemort}.}---i.e., those who to attempt to remove the observer too quickly from quantum mechanics by giving quantum states an unfounded ontic status---was well put by Spekkens:
\begin{quote}
[A] proponent of the ontic view might argue that the phenomena in question are not mysterious if one abandons certain preconceived notions about physical reality.  The challenge we offer to such a person is to present a few simple physical principles by the light of which all of these phenomena become conceptually intuitive (and not merely mathematical consequences of the formalism) within a framework wherein the quantum state is an ontic state. Our impression is that this challenge cannot be met.  By contrast, a single information-theoretic principle, which imposes a constraint on the amount of knowledge one can have about any system, is sufficient to derive all of these phenomena in the context of a simple toy theory \ldots
\end{quote}
The point is, far from being an appendage cheaply tacked on to the theory, the idea of quantum states as information has a simple unifying power that goes some way toward explaining why the theory has the very mathematical structure it does.\footnote{We say ``goes some way toward'' because, though the toy model makes about as compelling a case as we have ever seen that quantum states are states of information (an extremely valuable step forward), it gravely departs from quantum theory in other aspects. For instance, by its nature, it can give no Bell inequality violations or analogues of the Kochen--Specker noncolorability theorems.  Later sections of this paper will indicate that the cause of the deficit is that the toy model differs crucially from quantum theory in its answer to the question {\it Information about what?}}  By contrast, who could take the many-worlds idea and derive any of the structure of quantum theory out of it?  This would be a bit like trying to regrow a lizard from the tip of its chopped-off tail:  The Everettian conception never purported to be more than a reaction to the formalism in the first place.

But there are still deep puzzles left outstanding.  Above all, there are the old questions of {\it Whose information?}\ and {\it Information about what?}---these certainly must be addressed before any resolution of the quantum mysteries can be declared a success.  It must also be settled whether quantum theory is obligated to give a criterion for what counts as an observer.  Finally, because no one wants to give up on physics, we must tackle head-on the most crucial question of all:  If quantum states are not part of the stuff of the world, then what is?  What sort of stuff does quantum mechanics say the world {\it is\/} made of?

An understanding of the quantum, like all things worth having, will not come easily.  But this much is sure:  The glaringly obvious (that a large part of quantum theory, the central part in fact, is about information) should not be abandoned rashly.  To do so is to lose grip of the theory as it is applied in practice, with no better grasp of reality in return.  If on the other hand, one holds fast to the central point about information, initially frightening though it may be, one may still be able to reconstruct a picture of reality from the unfocused edge of one's vision.  Often the best stories come from there anyway.

\begin{figure}[h]
\begin{center}
\includegraphics[height=4.0in]{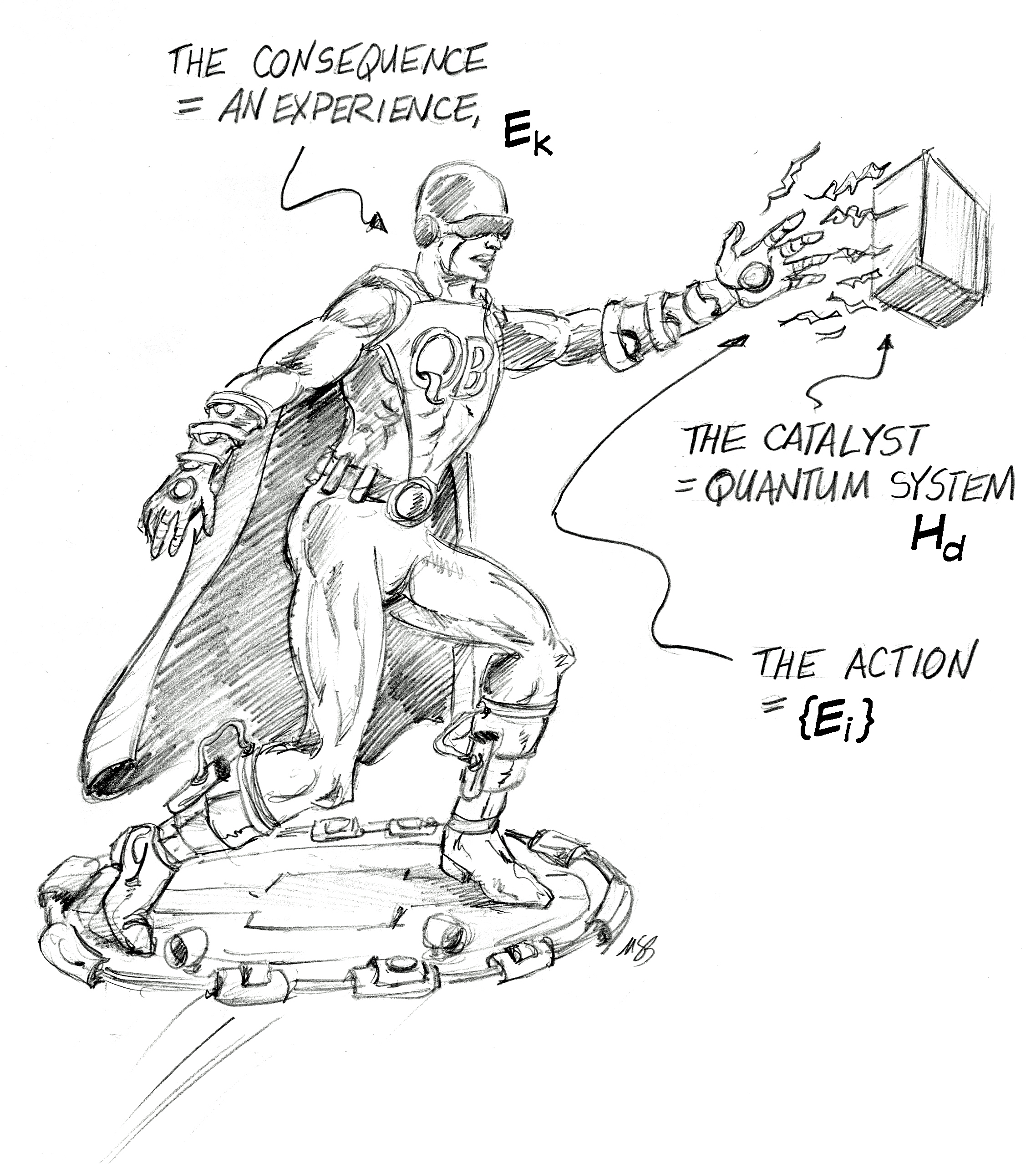}
\end{center}
\caption{``May not the {\it creatia\/} of a quantum observer's actions likewise be such additions to the universe as to enhance its total value?  And on this view, is not the QBist quantum observer---the agent---a kind of superhero for the universe as a whole, making {\it extra\/} things happen wherever, whenever he is called to duty?'' {\footnotesize (Drawing courtesy of Mark Staff Brandl.)}}
\end{figure}

So, what is the overarching story of QBism, and what does the QB stand for anyway? The Q clearly stands for Quantum, but the B\@? Initially, we had taken the B to stand for Bayesian, the interpretation of probability from which our efforts grew.  This is reflected in much of the literature on the subject---see, for instance, the title of the {\sl Stanford Encyclopedia of Philosophy\/} article on QBism \cite{sep-quantum-bayesian}. However, QBists eventually became dissatisfied with this meaning for the B, as there are just too many varieties of Bayesianism \cite{Good83}---QBism only represents one very specific strain of it~\cite{stacey-vonneumann}.  So, for a time we would say jokingly that the B rather stands for Bruno de Finetti, our hero in probability theory.  But eventually and with good reason, we landed on the idea of B for Bettabilitarian! This rolling word, coined by U.S.\ Supreme Court Justice Oliver Wendell Holmes, Jr., seemed to capture QBism perfectly.  Author Louis Menand, in his magisterial book {\sl The Metaphysical Club:\ A Story of Ideas in America} \cite{Menand01}, put it like this:
\begin{quote}
`The loss of certainty' is a phrase many intellectual historians have used the characterize the period in which Holmes lived. But the phrase has it backward. It was not the loss of certainty that stimulated the late-nineteenth-century thinkers with whom Holmes associated; it was the discovery of uncertainty. Holmes was, in many respects, a materialist. He believed, as he put it, that ``the law of the grub \ldots\ is also the law for man.''  And concerning the hope of social betterment, he was something worse than a pessimist. ``I despise,'' he said, ``the upward and onward.'' But he was not entirely a determinist, because he did not think that the course of human events was fixed \ldots. Complete certainty was an illusion; of that he was certain. There were only greater and lesser degrees of certainty, and that was enough.  It was, in fact, better than enough; for although we always want to reduce the degree of uncertainty in our lives, we never want it to disappear entirely, since uncertainty is what puts the play in the joints.  Imprecision, the sportiveness, as it were, of the quantum, is what makes life interesting and change possible. Holmes liked to call himself a ``bettabilitarian'':  we cannot know what consequences the universe will attach to our choices, but we can bet on them, and we do it every day.
\end{quote}
And that is it---{\it that\/} is QBism!  To be a QBist is to use quantum theory to be a ``better bettabilitarian'' in this world in which we are all immersed and which we shape with our every action.

Let us unroll this big idea.

\vfill

\section{Exactly How Quantum States Fail To Exist}
\begin{flushright}
\baselineskip=13pt
\parbox{2.8in}{\baselineskip=13pt
\small An experimental physicist usually says that an ``experimentally
determined'' probability has an ``error,'' and writes $P(H) = N_H/N
\pm 1/2\sqrt{N}$.  There is an implication in such an expression that there is a ``true'' or ``correct'' probability which could be computed if we knew enough, and that the observation may be in ``error'' due to a fluctuation. There is, however, no way to make such thinking logically consistent. It is probably better to realize that the probability concept is in a sense subjective, that it is always based on uncertain knowledge, and that its quantitative evaluation is subject to change as we obtain more information.}
\medskip\\
\small --- Richard P. Feynman\\
\small {\sl The Feynman Lectures on Physics}
\end{flushright}

Every area of human endeavor has its bold extremes.  Ones that say, ``If this is going to be done right, we must go this far.  Nothing less will do.''  In probability theory, the bold extreme is the personalist Bayesian account of probability \cite{Bernardo94}.  It says that probability theory is of the character of formal logic---a set of criteria for testing consistency.  In the case of formal logic, the consistency is between truth values of propositions.  However logic itself does not have the power to {\it set\/} the truth values it manipulates.  It can only say if various truth values are consistent or inconsistent; the actual values come from another source.  Whenever logic reveals a set of truth values to be inconsistent, one must dip back into the source to find a way to alleviate the discord.  But precisely in which way to alleviate it, logic gives no guidance.  ``Is the truth value for this one isolated proposition correct?''  Logic itself is powerless to say.

The key idea of personalist Bayesian probability theory is that it too is a calculus of consistency (or ``coherence'' as the practitioners call it), but this time for one's decision-making degrees of belief.  Probability theory can only say if various degrees of belief are consistent or inconsistent with each other. The actual beliefs come from another source, and there is nowhere to pin their responsibility but on the agent who holds them.  Dennis Lindley put it nicely in his book {\sl Understanding Uncertainty\/} \cite{Lindley06}:
\begin{quote}
The Bayesian, subjectivist, or coherent, para\-digm is egocentric. It is a tale of one person contemplating the world and not wishing to be stupid (technically, incoherent). He realizes that to do this his statements of uncertainty must be probabilistic.
\end{quote}
A probability {\it assignment\/} is a tool an agent uses to make gambles and decisions---it is a tool she uses for navigating life and responding to her environment.  Probability {\it theory\/} as a whole, on the other hand, is not about a single isolated belief, but about a whole mesh of them.  When a belief in the mesh is found to be incoherent with the others, the theory flags the inconsistency.  However, it gives no guidance for how to mend any incoherences it finds.  To alleviate the discord, one can only dip back into the source of the assignments---specifically, the agent who attempted to sum up all her history, experience, and expectations with those assignments in the first place.  This is the reason for the terminology that a probability is a ``degree of belief'' rather than a ``degree of truth'' or ``degree of facticity.''

To give an example of how mere internal consistency can yield the mathematical rules of probability is to quantify the consequences of acting upon beliefs in terms of costs and benefits.  An agent, whom we can call Alice, does business with a bookie, whose goal is to profit by exposing inconsistencies in Alice's mesh of beliefs.  Alice's goal is to avoid gambling in a way that forces her into a sure loss.  The bookie buys and sells lottery tickets of the form
\begin{equation}
\ticket{Worth \$1 if the event $E$ occurs.}
\end{equation}
Based on her own expectations about the event $E$, Alice assigns a number $p(E)$, which is the price in dollars at which she is willing to buy or to sell a lottery ticket of this form.  The normative rule that Alice should avoid a sure loss implies bounds on $p(E)$: If $p(E) < 0$, then Alice would be willing to pay money to have the bookie take the ticket off her hands, whereas if $p(E) > 1$, she would be willing to pay more for a ticket than it could ever be worth.  Furthermore, consider two events $E$ and $F$, which Alice believes to be mutually exclusive. Alice must strive for consistency in her pricing for the following three tickets.  First,
\begin{equation}
\ticket{Worth \$1 if $E$.}
\end{equation}
Second,
\begin{equation}
\ticket{Worth \$1 if $F$.}
\end{equation}
And finally,
\begin{equation}
\ticket{Worth \$1 if $E$ or $F$.}
\end{equation}
The value of the third ticket should be the sum total value of the first two.  If Alice sets her prices such that $p(E \hbox{ or } F) > p(E) + p(F)$, then the bookie can have a set of transactions that lead her into a sure loss.  In the jargon, Alice is vulnerable to a ``Dutch book'':  The bookie sells Alice the third ticket and buys the first two, leaving Alice in debt.  Whichever event happens, Alice cannot recoup the loss.  Likewise, the bookie can lead Alice into a sure loss if Alice chooses $p(E \hbox{ or } F) < p(E) + p(F)$.  In this way, striving to satisfy the normative rule of avoiding sure loss, Alice builds up the theory of probabilities:  Her personal probability for an event $E$ is simply her fair price for a gamble upon $E$.

Suppose now that Alice feels compelled to making assignments $p(E)$, $p(F)$, and $p(E \hbox{ or } F)$ such that $p(E \hbox{ or } F) \ne p(E) + p(F)$.  She detects that something is amiss. What to do to fix the problem?  For this, probability theory says nothing:  It's now up to Alice to make an adjustment somewhere, somehow, or else she will be vulnerable to a sure loss.

Where personalist Bayesianism breaks away the most from other developments of probability theory is that it says there are no {\it external\/} criteria for declaring an isolated probability assignment right or wrong.  The only basis for a judgment of adequacy comes from the {\it inside}, from the greater mesh of beliefs Alice may have the time or energy to access when appraising coherence.

It was not an arbitrary choice of words to title this section {\it
  Exactly How Quantum States Fail To Exist}, but a hint of what we must
elaborate to develop a perfected vaccine against the fever of quantum
interpretations.  This is because the phrase has a precursor in a
slogan that Bruno de Finetti, the founder of personalist Bayesianism, used
to vaccinate probability theory itself.  In the preface to his seminal
book \cite{DeFinetti90}, de Finetti writes, centered in the page and
in all capital letters,
\begin{center}
PROBABILITY$\,$ DOES$\,$ NOT$\,$ EXIST.
\end{center}
It is a powerful statement, constructed to put a finger on the single most-significant cause of the conceptual problems in pre-Bayesian probability theory.  A probability is not a solid object, like a rock or a tree that the agent might bump into, but a feeling, an estimate inside herself.

Previous to  Bayesianism, probability was often thought to be a physical property\footnote{Witness Richard von Mises, who even went so far as to write, ``Probability calculus is part of {\it theoretical physics\/} in the same way as classical mechanics or optics, it is an entirely self-contained theory of certain phenomena \ldots''\cite{Mises22}.}---something objective and having nothing to do with decision-making or agents at all.  But when thought so, it could be thought only inconsistently so.  And hell hath no fury like an inconsistency scorned.
The trouble is always the same in all its varied and complicated forms:  If probability is to be a physical property, it had better be a rather ghostly one---one that can be told of in campfire stories, but never quite prodded out of the shadows.  Here's a sample dialogue:
\begin{quote}
\begin{description}

\item[Pre-Bayesian:]  \ Ridiculous, probabilities are without doubt objective.  They can be seen in the relative frequencies they cause.
\item[Bayesian:]  So if $p=0.75$ for some event, after 1000 trials we'll see exactly 750 such events?
\item[Pre-Bayesian:]  \ You might, but most likely you won't see that exactly.  You're just likely to see something close to it.
\item[Bayesian:]  \ ``Likely''?  ``Close''?  How do you define or quantify these things without making reference to your degrees of belief for what will happen?
\item[Pre-Bayesian:]  \ Well, in any case, in the infinite limit the correct frequency will definitely occur.
\item[Bayesian:]  \ How would I know?  Are you saying that in one billion trials I could not possibly see an ``incorrect'' frequency?  In one trillion?
\item[Pre-Bayesian:]  \ OK, you can in principle see an {\it incorrect\/} frequency, but it'd be ever less {\it likely}!
\item[Bayesian:]  \ Tell me once again, what does ``likely'' mean?
\end{description}
\end{quote}
This is a cartoon of course, but it captures the essence and the futility of every such debate.  It is better to admit at the outset that probability is a degree of belief, and deal with the world on its own terms as it coughs up its objects and events.  What do we gain for our theoretical conceptions by saying that along with each actual event there is a ghostly spirit (its ``objective probability,'' its ``propensity,'' its ``objective chance'') gently nudging it to happen just as it did?  Objects and events are enough by themselves.

Similarly for quantum mechanics.  Here too, if ghostly spirits are imagined behind the actual events produced in quantum measurements, one is left with conceptual troubles to no end.  The defining feature of QBism~\cite{Caves02,Fuchs02,Fuchs04,Caves07,RMP,Fuchs10,Mermin13,Fuchs10b,AJP,Mermin14,stacey-vonneumann,Baeyer16} is that it says along the lines of de Finetti, ``If this is going to be done right, we must go this far.''  Specifically, there can be no such thing as a right and true quantum state, if such is thought of as defined by criteria {\it external\/} to the agent making the assignment:  Quantum states must instead be like personalist, Bayesian probabilities.

The direct connection between the two foundational issues is this.  Quantum states, through the Born rule, can be used to calculate probabilities.  Conversely, if one assigns probabilities for the outcomes of a well-selected set of measurements, then this is mathematically equivalent to making the quantum-state assignment itself.  The two kinds of assignments determine each other uniquely.  Just think of a spin-$\frac{1}{2}$ system.  If one has elicited one's degrees of belief for the outcomes of a $\sigma_x$ measurement, and similarly one's degrees of belief for the outcomes of $\sigma_y$ and $\sigma_z$ measurements, then this is the same as specifying a quantum state itself:  For if one knows the quantum state's projections onto three independent axes, then that uniquely determines a Bloch vector, and hence a quantum state.  Something similar is true of all quantum systems of all sizes and dimensionality.  There is no mathematical fact embedded in a quantum state $\rho$ that is not embedded in an appropriately chosen set of probabilities.\footnote{See Section \ref{SeekingSICs} where this statement is made precise in all dimensions.}  Thus generally, if probabilities are personal in the Bayesian sense, then so too must be quantum states.

What this buys interpretatively, beside airtight consistency with the best understanding of probability theory, is that it gives each quantum state a home.  Indeed, a home localized in space and time---namely, the physical site of the agent who assigns it!  By this method, one expels once and for all the fear that quantum mechanics leads to ``spooky action at a distance,'' and expels as well any hint of a problem with ``Wigner's friend''~\cite{Wigner71}.  It does this because it removes the very last trace of confusion over whether quantum states might still be objective, agent-independent, physical properties.

The innovation here is that, for most of the history of efforts to take an informational point of view about quantum states, the supporters of the idea have tried to have it both ways:  that on the one hand quantum states are not real physical properties, yet on the other there is a right and true quantum state independent of the agent after all. For instance, one hears things like, ``The {\it right\/} quantum state is the one the agent should adopt if he had all the information.''  The tension in these two desires leaves their holders open to attack on both flanks and general confusion all around.

Take first instantaneous action at a distance---the horror of this idea is often one of the strongest motivations for those seeking to take an informational stance on quantum states.  But, now an opponent can say:
\begin{quote}
\noindent If there is a {\it right quantum state}, then why not be done with all this squabbling and call the state a physical fact to begin with?  It is surely external to the agent if the agent can be wrong about it.  But, once you admit that (and you should admit it), you're sunk: For, now what recourse do you have to declare no action at a distance when a delocalized quantum state changes instantaneously?

 Here I am with a physical system right in front of me, and though {\it my\/} probabilities for the outcomes of measurements {\it I\/} can do on it might have been adequate a moment ago, there is an objectively better way to gamble {\it now\/} because of something that happened far in the distance?  (Far in the distance and just now.)  How could that not be the signature of action at a distance?  You can try to defend yourself by saying ``quantum mechanics is all about relations''\footnote{A typical example is of a woman traveling far from home when her husband divorces her.  Instantaneously she becomes unmarried---marriage is a relational property, not something localized at each partner.  It seems to be popular to give this example and say, ``Quan\-tum mechanics might be like that.''  The conversation usually stops without elaboration, but let's carry it a little further:  Suppose the woman, Carol, is right in front of Alice.  Alice has a set of probabilities for what might happen should she buy Carol a celebratory bottle of tequila and congratulate Carol on losing the deadbeat.  Would the far-off divorce mean that there is instantaneously a different set of probabilities that Alice could use for weighing the consequences of those actions?  Not at all.  Alice would have no account to change her probabilities (not due to the divorce, anyway) until Alice became aware of Carol's changed relation, however long it might take that news to get to Alice.} or some other feel-good phrase, but I'm talking about measurements right here, in front of me, with outcomes I can see right now.  Ones entering my awareness---not outcomes in the mind of God who can see everything and all relations.  It is that which I am gambling upon with the help of the quantum formalism.  An objectively better quantum state would mean that my gambles and actions, though they would have been adequate a moment ago, are now simply wrong in the eyes of the world---they could have been better.  How could the quantum system in front of me generate outcomes instantiating that declaration without being privy to what the eyes of the world already see?  That's action at a distance, I say, or at least a holism that amounts to the same thing---there's nothing else it could be.
\end{quote}

Without the protection of truly personal quantum-state assignments, action at a distance is there as doggedly as it ever was.  And things only get worse with ``Wigner's friend'' if one insists there be a {\it right\/} quantum state.  As it turns out, the method of mending this conundrum displays one of the most crucial ingredients of QBism.  Let us put it in plain sight.

``Wigner's friend'' is the story of two agents, Wigner and his friend, and one quantum system---the only deviation we make from a more common presentation\footnote{For instance, see Ref.\ \cite{Albert94}.} is that we put the story in informational terms.  It starts off with the friend and Wigner having a conversation:  Suppose they both agree that some quantum state $|\psi\rangle$ captures their mutual beliefs about the quantum system.\footnote{Being Bayesians, of course, they don't have to agree at this stage---for recall $|\psi\rangle$ is not a physical fact for them, only a catalogue of {\it beliefs}.  But suppose they do agree.} Furthermore suppose they agree that at a specified time the friend will make a measurement on the system of some observable (outcomes $i=1,\ldots,d$).  Finally, they both note that if the friend gets outcome $i$, he will (and should) update his beliefs about the system to some new quantum state $|i\rangle$.  There the conversation ends and the action begins:  Wigner walks away and turns his back to his friend and the supposed measurement.  Time passes to some point beyond when the measurement should have taken place.

\begin{figure}
\begin{center}
\includegraphics[height=2.6in]{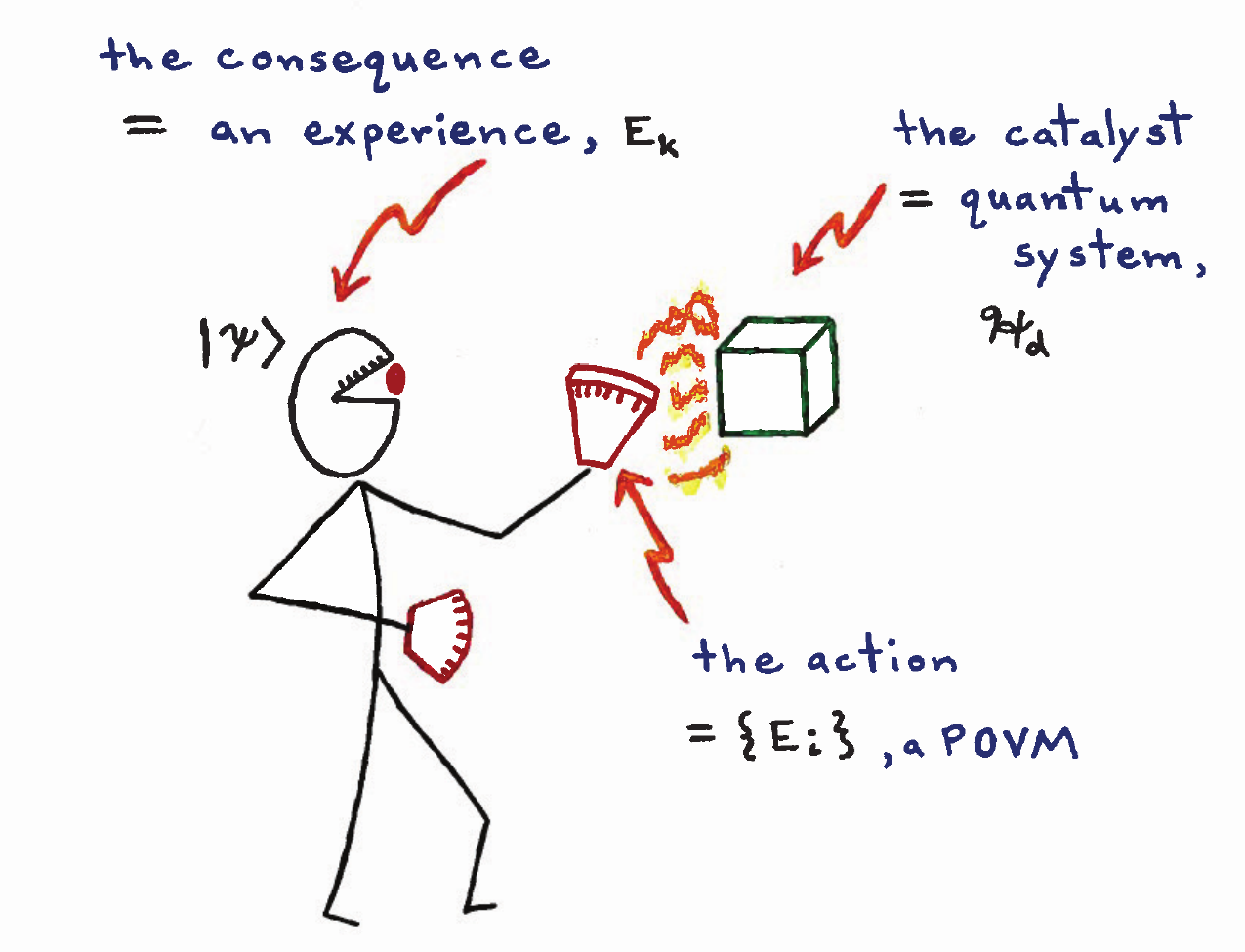}
\bigskip\caption{ In contemplating a quantum measurement, one makes a conceptual split in the world:  one part is treated as an agent, and the other as a kind of reagent or catalyst (one that brings about change in the agent itself).  The latter is a quantum system of some finite dimension $d$.  A quantum measurement consists first in the agent taking an {\it action\/} on the quantum system.  The action is represented formally by a set of operators $\{E_i\}$---a positive-operator-valued measure. The action generally leads to an incompletely predictable {\it consequence\/} $E_i$ for the agent.  The quantum state $|\psi\rangle$ makes no appearance but in the agent's head; for it captures his degrees of belief concerning the consequences of his actions, and, in contrast to the quantum system itself, has no existence in the external world.  Measurement devices are depicted as prosthetic hands to make it clear that they should be considered an integral part of the agent.  The sparks between the measurement-device hand and the quantum system represent the idea that the consequence of each quantum measurement is a unique creation within the previously existing universe.  Two points are decisive in distinguishing this picture of quantum measurement from a kind of solipsism:  1) The conceptual split of agent and external quantum system: If it were not needed, it would not have been made. 2) Once the agent chooses an action $\{E_i\}$ to take, the particular consequence $E_k$ of it is beyond his control---that is, the actual outcome is not a product of his whim and fancy.}
\end{center}
\end{figure}

What now is the ``correct'' quantum state each agent should have assigned to the quantum system?  We have already concurred that the friend will and
should assign some $|i\rangle$.  But what of Wigner?  If he were to
consistently dip into his mesh of beliefs, he would very likely treat
his friend as a quantum system like any other:  one with some initial
quantum state $\rho$ capturing his (Wigner's) beliefs of {\it it\/}
(the friend), along with a linear evolution operator\footnote{We
  suppose for the sake of introducing less technicality that $U$ is a
  unitary operation, rather than the more general completely positive
  trace-preserving linear maps of quantum information
  theory~\cite{Nielsen00}.  This, however, is not essential to the
  argument.} $U$ to adjust those beliefs with the flow of
time.\footnote{For an explanation of the status of unitary operations
  from the QBist perspective, as personal judgments directly analogous
  to quantum states themselves, see Section~\ref{SeekingSICs} and
  Refs.~\cite{Fuchs02,RMP,Leifer06}.}  Suppose this quantum state
includes Wigner's beliefs about everything he assesses to be
interacting with his friend---in old parlance, suppose Wigner treats
his friend as an isolated system. From this perspective, before any further interaction between himself and the friend or the other system, the quantum state Wigner would assign for the two together would be $U\big(\rho\otimes|\psi\rangle\langle\psi|\big)U^\dagger$ --- most generally an entangled quantum state.  The state of the system itself for Wigner would be gotten from this larger state by a partial trace operation; in any case, it will not be an $|i\rangle$.

Does this make Wigner's new state assignment incorrect?  After all, ``if he had all the information'' (i.e., all the facts of the world) wouldn't that include knowing the friend's measurement outcome? Since the friend should assign some $|i\rangle$, shouldn't Wigner himself (if he had all the information)?  Or is it the friend who is incorrect?  For if the friend had ``all the information,'' wouldn't he say that he is neglecting that Wigner could put the system and himself into the quantum computational equivalent of an iron lung and forcefully reverse the so-called measurement?  I.e., Wigner, if he were sufficiently sophisticated, should be able to force
\begin{equation}
U\big(\rho\otimes|\psi\rangle\langle\psi|\big)U^\dagger\;\;\longrightarrow\;\;\rho\otimes|\psi\rangle\langle\psi|\;.
\label{RatifiedLatified}
\end{equation}
And so the back and forth goes.  Who has the {\it right\/} state of information?  The conundrums simply get too heavy if one tries to hold to an agent-independent notion of correctness for otherwise personalistic quantum states.  A QBist dispels these and similar difficulties by being conscientiously forthright.  {\it Whose information?\/}  ``Mine!''  {\it Information about what?\/}  ``The consequences (for {\it me\/}) of {\it my\/} actions upon the physical system!''  It's all ``I-I-me-me mine,'' as the Beatles sang.

The answer to the first question surely comes as no surprise by now, but why on earth the answer for the second?   Why something so egocentric, anthro\-po\-centric, psychology-laden, and positivistic (we've heard any number of expletives) as {\it the consequences (for me) of my actions upon the system}?  Why not simply say something neutral like ``the outcomes of measurements''?  Or, why not fall in line with Wolfgang Pauli and say \cite{Pauli94}:
\begin{quote}
The objectivity of physics is \ldots\ fully ensured in quantum mechanics in the following sense.  Although in principle, according to the theory, it is in general only the statistics of series of experiments that is determined by laws, the observer is unable, even in the unpredictable single case, to influence the result of his observation---as for example the response of a counter at a particular instant of time.  {\it Further, personal qualities of the observer do not come into the theory in any way---the observation can be made by objective registering apparatus, the results of which are objectively available for anyone's inspection.}  [Our emphasis.]
\end{quote}
To the uninitiated, our answer for {\it Information about what?}\ surely appears to be a cowardly, unnecessary retreat from realism.  But it is the opposite.  The answer we give is the very injunction that keeps the potentially conflicting statements of Wigner and his friend in check,  at the same time as giving each agent a hook to the external world in spite of QBism's egocentric quantum states.  Pauli's statement certainly wouldn't have done that.  Results objectively available for anyone's inspection?  This is the whole issue with ``Wigner's friend'' in the first place.  If both agents could just ``look'' at the counter simultaneously with negligible effect in principle, we would not be having this discussion.

You see, for the QBist, the real world, the one both agents are embedded in---with its objects and events---is taken for granted.  What is not taken for granted is each agent's access to the parts of it he has not touched.  Wigner holds two thoughts in his head: 1) that his friend interacted with a quantum system, eliciting some consequence of the interaction for himself, and 2) after the specified time, for any of Wigner's own further interactions with his friend or system or both, he ought to gamble upon their consequences according to $U\big(\rho\otimes|\psi\rangle\langle\psi|\big)U^\dagger$.  One statement refers to the friend's potential experiences, and one refers to Wigner's own.  So long as it is explicit that $U\big(\rho\otimes|\psi\rangle\langle\psi|\big)U^\dagger$ refers to the latter---i.e., how Wigner should gamble upon the things that might happen to him---making no statement whatsoever about the former, there is no conflict.  The world is filled with all the same things it was before quantum theory came along, like each of our experiences, that rock and that tree, and all the other things under the sun; it is just that quantum theory provides a calculus for gambling on each agent's own experiences---it doesn't give anything else than that.  It certainly doesn't give one agent the ability to conceptually pierce the other agent's personal experience.  It is true that with enough effort Wigner could enact Eq.~(\ref{RatifiedLatified}), causing him to predict that his friend will have amnesia to any future questions on his old measurement results.  But we always knew Wigner could do that---a mallet to the head would have been good enough.

The key point is that quantum theory, from this light, takes nothing away from the usual world of common experience we already know.  It only {\it adds}.\footnote{This point will be much elaborated on in Section \ref{HSpaceDim}.}  At the very least it gives each agent an extra tool with which to navigate the world.  More than that, the tool is here for a reason.  QBism says that when an agent reaches out and touches a quantum system---when he performs a {\it quantum measurement}---this process gives rise to birth in a nearly literal sense.  With the action of the agent upon the system, the no-go theorems of Bell and Kochen--Specker assert that something new comes into the world that wasn't there previously:  It is the ``outcome,'' the unpredictable consequence for the very agent who took the action.  John Archibald Wheeler said it this way, and we follow suit, ``Each elementary quantum phenomenon is an elementary act of `fact creation'.''  \cite{Wheeler82c}

With this much, QBism has a story to tell on both quantum {\it states\/} and quantum {\it measurements}, but what of quantum {\it theory\/} as a whole?  The answer is found in taking it as a {\it universal\/} single-user theory in much the same way that Bayesian probability theory itself is.  It is a user's manual that {\it any\/} agent can pick up and use to help make wiser decisions in this world of inherent uncertainty.\footnote{\label{Macca} Most of the time one sees Bayesian probabilities characterized (even by very prominent Bayesians like Edwin T. Jaynes \cite{Jaynes03}) as measures of ignorance or imperfect knowledge.  But that description carries with it a metaphysical commitment that is not at all necessary for the personalist Bayesian, where probability theory is an extension of logic.  Imperfect knowledge?  It sounds like something that, at least in imagination, could be perfected, making all probabilities zero or one---one uses probabilities only because one does not know the true, pre-existing state of affairs.  Language like this, the reader will notice, is never used in this paper.  All that matters for a personalist Bayesian is that there is {\it uncertainty\/} for whatever reason.  There might be uncertainty because there is ignorance of a true state of affairs, but there might be uncertainty because the world itself does not yet know what it will give---i.e., there is an objective indeterminism.  As will be argued in later sections, QBism finds its happiest spot in an unflinching combination of ``subjective probability'' with ``objective indeterminism.''}  To say it in a more poignant way:  In my case, it is a world in which $I$ am forced to be uncertain about the consequences of most of {\it my\/} actions; and in your case, it is a world in which {\it you\/} are forced to be uncertain about the consequences of most of {\it your\/} actions.  ``And what of God's case?  What is it for him?''  Trying to give {\it him\/} a quantum state was what caused this trouble in the first place!  In a quantum mechanics with the understanding that each instance of its use is strictly single-user---``My measurement outcomes happen right here, to me, and I am talking about my uncertainty of them.''---there is no room for most of the standard, year-after-year quantum mysteries.

\begin{figure*}
\begin{center}
\includegraphics[height=3in]{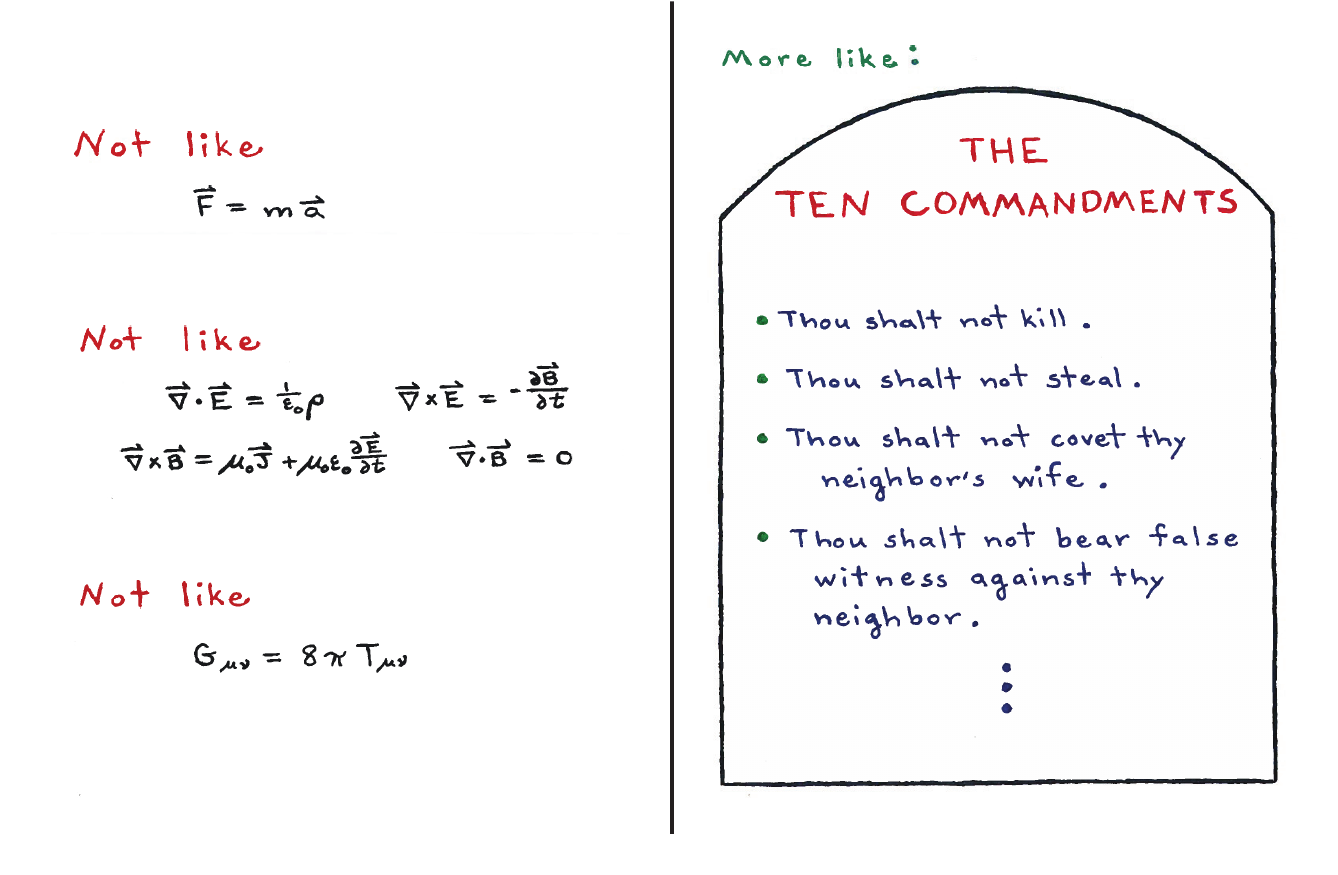}
\caption{ The Born rule is not like the other classic laws of physics.  Its normative nature means, if anything, it is more like the Biblical Ten Commandments.  The classic laws on the left give no choice in their statement: If a field is going to be an electromagnetic field at all, it must satisfy Maxwell's equations; it has no choice.  Similarly for the other classic laws.  Their statements are intended to be statements concerning nature {\it just exactly as it is}.  But think of the Ten Commandments.  ``Thou shalt not steal.''  People steal all the time.  The role of the Commandment is to say, ``You have the power to steal if you think you can get away with it, but it's probably not in your best interest to do so.  Something bad is likely to happen as a result.''  Similarly for ``Thou shalt not kill,''\ and all the rest.  It is the worshipper's choice to obey each or not, but if he does not, he ought to count on something potentially bad in return.  ``And I commend enjoyment,'' urges Ecclesiastes, ``for man has nothing better under the sun than to eat, to drink and to be merry.''  This is a guideline for behavior---but one conditioned on, and justified in terms of, the character of the natural world.  The Born rule guides, ``Gamble in such a way that all your probabilities mesh together through me.''  The agent is free to ignore the advice, but if he does so, he does so at his own peril.  Yet, as with the advice of Ecclesiastes, the specifics of the rule can tell us about the character of the world we inhabit.}
\end{center}
\end{figure*}

The only substantive {\it conceptual\/} issue left\footnote{Not to worry, there are still plenty of technical ones,  as well as plenty more conceptual ones waiting for after the vaccination.} before synthesizing a final vaccine is whether quantum mechanics is obligated to derive the notion of agent for whose aid the theory was built in the first place.  The answer comes from turning the tables:  Thinking of probability theory in the personalist Bayesian way, as an extension of formal logic, would one ever imagine that the notion of an agent, the user of the theory, could be derived out of its conceptual apparatus?  Clearly not.  How could you possibly get flesh and bones out of a calculus for making wise decisions?  The logician and the logic he uses are two different substances---they live in conceptual categories worlds apart.  One is in the stuff of the physical world, and one is somewhere nearer to Plato's heaven of ideal forms.  Look as one might in a probability textbook for the ingredients to reconstruct the reader herself, one will never find them.  So too, the QBist says of quantum theory.

What counts as a ``user'' of quantum theory? Must the user be
conscious? We find that an inopportune way of phrasing things, for it
takes the issue too far afield. Instead, we prefer to say it is
whatever it takes to be a user of probability theory. Dogs don't
collapse wave functions because dogs don't use wave functions. Upon
reading this argument, one correspondent immediately protested: ``But
ants already use probabilities, it has been shown. For the paths they
take, one can model the trajectories with an appropriate choice of
probabilities and utilities.''

To which the QBist involved replied, ``No, that's not what I mean. And that is
no proof whatsoever that ants use probability theory in the sense I mean
it. To use probability theory, I mean one must use it internally, and in a
{\it normative\/} sense.'' Probability assignments spring from
an attempt to organize one's previous experience for the purpose of future
actions. Ants are surely not using it normatively.  Modeling agents from the
outside (at least in the discussions we've seen so far) never takes into account
the normative struggle that is required for any but the most trivial
probability assignments.

With this we finally pin down the precise way in which quantum theory is ``different in character from any physical theory posed before.''
For the QBist, quantum theory is not something {\it outside\/} probability theory---it is not a picture of the world as it is, as say Einstein's program of a unified field theory hoped to be---but rather it is an {\it addition\/} to probability theory itself.  As probability theory is a {\it normative\/} theory, not saying what one {\it must\/} believe, but offering rules of consistency an agent should strive to satisfy within his overall mesh of beliefs, so it is the case with quantum theory.  If quantum theory is a user's manual, one cannot forget that the world is its author.  And from its writing style, one may still be able to tell something of the author herself.  The question is how to tease out the psychology of the style, frame it, and identify the underlying motif.

To take this idea into one's mindset is all the vaccination one needs against the threat that quantum theory carries something viral for theoretical physics as a whole.  A healthy body is made healthier still.  With this protection, we are for the first time in a position to ask, with eyes wide open to what the answer could not be, {\it just what after all is the world made of?}  Far from being the last word on quantum theory, QBism, we believe, is the start of a great adventure.  An adventure full of mystery and danger, with hopes of triumph \ldots\ and all the marks of life.


\section{Teleportation}

``Teleportation'' in the quantum information sense isn't so very much
like the Star Trek version as the press always wants to portray it.
It's not about getting things from here to there without going in
between, but about making your information stop being about {\it this\/} and
start being about {\it that\/} without being about {\it anything else\/} in between.
In slightly stuffier language: Quantum teleportation is the
transference of an agent's predictions about one object onto another
object that has never interacted with the first.

In the usual way the teleportation drama is staged, the cast of
characters includes an Alice and a Bob who share two systems in a
maximally entangled state, and implicitly, a Charlie who prepares a
third system in the state of his choice and then hands it off to
Alice.  Alice then performs a measurement on the two systems in her
possession and announces the result of the measurement to Bob.  The
teleportation process is completed with Bob performing an operation on
his system conditioned upon his newly acquired information.

In what sense is it completed?  Only in this:  If Charlie has the
promise that Alice and Bob went through all the actions described
above, then he can safely ascribe the same quantum state to Bob's
system that he had originally ascribed to the system he handed off to
Alice.

The way in which the story is typically told leads one to ask, ``How is so much information transmitted?''\  and ``Just how does the information get from Alice to Bob?'' The honest answer is that  {\it no\/} information is transmitted in the process of teleportation (excepting the two bits that tell Bob which action to perform).  The only nontrivial thing transferred in the process of teleportation is {\it reference}.  Charlie's information---that is, his compendium of Bayesian degrees of belief---stops being {\it about\/} the qubit he just handed off to Alice and starts being {\it about\/} Bob's.

Here's a corresponding classical example.  In place of entanglement,
let us equip Alice and Bob each with a coin (oriented heads or tails)
encased in a magical opaque box.  These magical opaque boxes have the
following properties:  1) though one cannot see how a coin is
oriented within it, one can nevertheless reach inside a box and turn
the coin over if one wishes, and 2) if one touches two of these boxes
together, they will glow green if the coins within them have the same
orientations, and they will glow red if they have opposite
orientations---the glowing reveals nothing about the actual
orientation of either coin, only about their relationship.  Finally
let us stipulate the following for Alice and Bob:  That their opaque
boxes contain identically oriented coins, but Alice and Bob (or
anyone else for that matter) know nothing more about the coins beyond
that.  In other words, Alice and Bob possess HH or TT, but they do
not know which.

Now, as in quantum teleportation let us introduce a third character,
Charlie.  Charlie has an opaque box of his own.  But let us give him
some partial certainty about the orientation of his coin.
Particularly, let us suppose he ascribes a probability $p$ for his
coin to be heads.  This is a real number between 0 and 1, and in
principle it might take an arbitrarily huge number of bits to specify.

Here's the protocol.  Charlie hands off his coin (encased in a magical
opaque box) to Alice.  Alice touches her newly acquired box to her old
box.  The two glow red or green, and she communicates the result to
Bob.  If the result was green, Bob leaves his coin alone.  If the
result was red, he reaches into his opaque box and turns the coin
over.  Meanwhile, Alice randomizes the coins in her possession, i.e.,
she shuffles them so that Charlie no longer knows which is which.
Thus from Charlie's perspective, he now knows nothing about the coin
in the original box he gave Alice, and he would write down a 50--50
distribution for heads versus tails.  Charlie's original state for his
original coin is, in this way, ``destroyed''.  At that point the
``teleportation'' process is completed.

Again we can ask, ``In what sense is it completed?''  Only in this:
If Charlie has the promise that Alice and Bob went through all the
actions described above, then he can safely ascribe the same
probability $p$ to the coin in Bob's box (i.e., $p$ that it will be
heads) that he had originally ascribed to the coin in his own box.  In
other words, Charlie has everything it takes to update his {\it
  epistemic state\/} about the orientation of the coin in Bob's box to
what he had originally thought of the coin in his own box.

Is this wildly exciting?  The stuff that would make headlines in
papers all around the world and be called ``teleportation''?  At the
material cost of transferring a single bit from Alice to Bob, has
Charlie instantaneously transferred an arbitrarily stupendously big
number of bits (in the form of the real number $p$) between the two
sites?  Not at all!  The only thing that was materially transported
from one site to the other was a single bit (that the boxes glowed red
or green).  The rest was just ``conditionalizing'' or ``updating''.
And there is no shocking headline in that.

\section{The Meaning of No-Cloning}

It is often underappreciated that taking a stand on the interpretation of quantum states carries with it a fairly distinctive force on one's research.  For, the stand one takes implicitly directs the analogies (and disanalogies) one will seek for comparing classical physics to quantum physics.  At least this is the case for an information theoretic conception of quantum states.

In 1995, one of the authors (CAF) was quite taken with a point that both Asher Peres and Michael Berry emphasized in their discussions of quantum chaos and more broadly~\cite{Peres95}.  In making a comparison between quantum mechanics and classical Hamiltonian mechanics, the proper correspondence is not between quantum states and points in phase space, but between quantum states and {\it Liouville distributions\/} on the phase space.  The key insight is that the points of phase space are meant to represent states of reality, whereas the Liouville distributions are rather explicitly meant to represent one's uncertainty about the true state of reality.  In an information theoretic conception of quantum states, a quantum state too should not be a state of reality, but uncertainty about something (maybe not uncertainty of the true state of reality, but nonetheless uncertainty about  {\it something\/}).

If the analogy worked once, then it should be tested further afield!  Furthermore, maybe one could even take the insight the other way around, from quantum to classical.  This prompted the thought that despite all the hoopla over the no-cloning theorem,  there was nothing particularly quantum mechanical about it.  Classical Liouville evolution preserves phase space volume---this any graduate student versed in Goldstein's classical-mechanics book \cite{Goldstein65} knows---but a less emphasized consequence of it is that Hamiltonian evolution must preserve the overlap between Liouville distributions as well (as Peres and Berry had been stressing in their quantum chaos work).  But then the same argument that drives the proof of the no-cloning theorem for quantum states would also drive it for classical Liouville distributions---for the no-cloning theorem is a nearly immediate consequence of the fact that unitary evolutions preserve Hilbert-space inner products.  So, nonorthogonal quantum states cannot be cloned, but neither can nonorthogonal classical Liouville distributions \cite{Caves96}.

A historical aside:  The issue of no-cloning boils down to an almost immediate consequence of unitarity---inner products cannot decrease.  In fact, Wigner's theorem on symmetries \cite{Wigner12} even shows that the group of time-continuous, inner-product preserving maps on Hilbert space is strictly equivalent to the unitary group.  Therefore, it is an intriguing historical fact that Wigner himself just missed the no-cloning theorem!  In a 1961 paper, Wigner took on the question, ``How probable is life?''~\cite{Wigner61}. He did this by identifying the issue of self-reproduction with the existence of the types of maps required for the cloning of quantum states.  He didn't tackle the question of cloning for a completely {\it unknown\/} quantum state head on, but instead analysed the ``fraction'' of unitary operators on a tensor-product Hilbert space that can lead to a cloning transformation for at least some states.  Nevertheless, he states quite clearly that an arbitrary linear superposition of clonable states ought also to be clonable. But this, of course, cannot be.

The fact that both teleportation and no-cloning arise in classical statistical theories has implications for the project of reconstructing quantum theory.  In our search for deep principles, should we try to rederive quantum physics from the postulate that ``quantum information can be teleported,'' or that ``quantum information cannot be cloned''?  These phenomena being not at all fundamentally quantum makes them feel like poor candidates for the seed from which quantum theory grows.  It would be far better to seek that essential DNA in the answer to the question, {\it Information about what?}

\section{The Essence of Bell's Theorem, QBism Style}

\label{BellTheorem}

It is easy enough to {\it say\/} that a quantum system (and hence each piece of the world) is a ``seat of possibility.''  In a spotty way, certain philosophers have been saying similar things for 150 years.  What is unique about quantum theory in the history of thought is the way in which its mathematical structure has pushed this upon us to our very surprise.  It wasn't that all these grand statements on the philosophical structure of the world were built into the formalism, but that the formalism reached out and shook its users until they opened their eyes.  Bell's theorem and all its descendants are examples of that.

So when the users opened their eyes, what did they see?  From the look of several recent prominent expositions \cite{Gisin09,Albert09,Norsen06}, the lesson was indisputably what Tim Maudlin put so forcefully in \cite{Maudlin14},
\begin{quote}
What Bell proved, and what theoretical physics
has not yet properly absorbed, is that the physical
world itself is nonlocal.
\end{quote}
The world really is full of spooky action at a distance---live with it and love it!  But conclusions drawn from even the most rigorous of theorems can only be {\it additions\/} to one's prior understanding and beliefs when the theorems do not contradict those beliefs flat out.  Such was the case with Bell's theorem.  It has just enough room in it to not contradict a misshapen notion of probability, and that is the hook and crook that the sci-fi fans have thrived on.  A QBist, however, with a different understanding of probability and a commitment to the idea that quantum measurement outcomes are personal, draws quite a different conclusion from the theorem.  In fact it is a conclusion from the far opposite end of the spectrum:  It tells of a world unknown to most monist and rationalist philosophies:  The universe, far from being one big nonlocal block, should be thought of as a thriving community of connubial, but otherwise autonomous entities.   That the world should violate Bell's theorem remains, even for QBism, the deepest statement ever drawn from quantum theory.  It says that quantum measurements are {\it moments of creation}.

This language has already been integral to our presentation, but seeing it come about in a formalism-driven way like Bell's makes the issue particularly vivid.  Here we devote some effort to showing that the language of creation is a consequence of three things:  1) the quantum formalism, 2) a personalist Bayesian interpretation of probability, and 3) the elementary notion of what it means to be two objects rather than one.  We do not do it however with Bell's theorem precisely, but with an argument that more directly implicates the EPR ``criterion of reality'' as the source of trouble with quantum theory.  The thrust of it is that it is the EPR criterion that should be jettisoned, not locality.

Our starting point is like our previous setup---an agent and a system---but this time we make it two systems:  One of them, the left-hand one, is ready. The other, the right-hand one, is waiting.  The agent will eventually measure each in turn.\footnote{It should be noted how we depart from the usual presentation here:  There is only the single agent and his two systems.  There is no Alice and Bob accompanying the two systems.}  Simple enough to say, but things get hung at the start with the issue of what is meant by ``two systems?''  A passage from a 1948 paper of Einstein \cite{Einstein48} captures the essential issue well:
\begin{quote}
\noindent If one asks what is characteristic of the realm of physical ideas independently of the quantum-theory, then above all the following attracts our attention: the concepts of physics refer to a real external world, i.e., ideas are posited of things that claim a ``real existence'' independent of the perceiving subject (bodies, fields, etc.), and these ideas are, on the one hand, brought into as secure a relationship as possible with sense impressions. Moreover, it is characteristic of these physical things that they are conceived of as being arranged in a space-time continuum. Further, it appears to be essential for this arrangement of the things introduced in physics that, at a specific time, these things claim an existence independent of one another, insofar as these things ``lie in different parts of space.'' Without such an assumption of the mutually independent existence (the ``being-thus'') of spatially distant things, an assumption which originates in everyday thought, physical thought in the sense familiar
to us would not be possible. Nor does one see how physical laws could be formulated and tested without such a clean separation. \ldots 

For the relative independence of spatially distant things (A and B), this idea is characteristic: an external influence on A has no {\it immediate\/} effect on B; this is known as the ``principle of local action,'' \ldots. 
The complete suspension of this basic principle would make impossible the idea of (quasi-) closed systems and, thereby, the establishment of empirically testable laws in the sense familiar to us.
\end{quote}
We hope it is clear to the reader by now that QBism concurs with every bit of this.  Quantum states may not be the stuff of the world, but QBists never shudder from positing quantum systems as ``real existences'' external to the agent.  And just as the agent has learned from long, hard experience that he cannot reach out and touch anything but his immediate surroundings, so he imagines of every quantum system, one to the other.  What is it that A and B are spatially distant things but that they are causally independent?

This notion, in Einstein's hands,\footnote{Beware!  This is {\it not\/} to say in the hands of EPR---Einstein, Podolsky, and Rosen.  The present argument is not their argument.  For a discussion of Einstein's dissatisfaction with the one appearing in the EPR paper itself, see \cite{Fine96}.} led to one of the nicest, most direct arguments that quantum states cannot be states of reality, but must be something {\it more like\/} states of information, knowledge, expectation, or belief \cite{Harrigan07}.  The argument is important---let us repeat the whole thing from Einstein's most thorough version of it \cite{Einstein49}.  It more than anything sets the stage for a QBist development of a Bell-style contradiction.
\begin{quote}
\noindent Physics is an attempt conceptually to grasp reality as it is thought independently of its being observed.  In this sense on speaks of ``physical reality.'' In pre-quantum physics there was no doubt as to how this was to be understood.  In Newton's theory reality was determined by a material point in space and time; in Maxwell's theory, by the field in space and time.  In quantum mechanics it is not so easily seen.  If one asks: does a $\psi$-function of the quantum theory represent a real factual situation in the same sense in which this is the case of a material system of points or of an electromagnetic field, one hesitates to reply with a simple ``yes'' or ``no''; why?  What the $\psi$-function (at a definite) time asserts, is this:  What is the probability for finding a definite physical magnitude $q$ (or $p$) in a definitely given interval, if I measure it at time $t$?  The probability is here to be viewed as an empirically determinable, therefore certainly as a ``real'' quantity which I may determine if I create the same $\psi$-function very often and perform a $q$-measurement each time.  But what about the single measured value of $q$?  Did the respective individual sys\-tem have this $q$-value even before this measurement?  To this question there is no definite answer within the framework of the theory, since the measurement is a pro\-cess which implies a finite disturbance of the sys\-tem from the outside; it would therefore be thinkable that the system obtains a definite numerical value for $q$ (or $p$) the measured numerical value, only through the measurement itself.  For the further discussion I shall assume two physicists $A$ and $B$, who represent a different conception with reference to the real situation as described by the $\psi$-function.
\begin{itemize}
\item[$A$.]  The individual system (before the measurement) has a definite value of $q$ (or $p$) for all variables of the system, and more specifically, {\it that\/} value which is determined by a measurement of this var\-i\-able.  Proceeding from this conception, he will state: The $\psi$-function is no exhaustive description of the real situation of the system but an incomplete description; it expresses only what we know on the basis of former measurements concerning the system.
\item[$B$.] The individual system (before the measurement) has no definite value of $q$ (or $p$).  The value of the measurement only arises in cooperation with the unique probability which is given to it in view of the $\psi$-function only through the act of measurement itself.  Proceeding from this conception, he will (or, at least, he may) state:  The $\psi$-function is an exhaustive description of the real situation of the system.
\end{itemize}
We now present to these two physicists the following instance:  There is to be a system which at the time $t$ of our observation consists of two partial systems $S_1$ and $S_2$, which at this time are spatially separated and (in the sense of classical physics) are without significant reciprocity.  The total system is to be completely described through a known $\psi$-function $\psi_{12}$ in the sense of quantum mechanics.  All quantum theoreticians now agree upon the following:  If I make a complete measurement of $S_1$, I get from the results of the measurement and from $\psi_{12}$ an entirely definite $\psi$-function $\psi_2$ of the system $S_2$.  The character of $\psi_2$ then depends upon {\it what kind\/} of measurement I undertake on $S_1$.

Now it appears to me that one may speak of the real factual situation of the partial system $S_2$. Of this real factual situation, we know to begin with, before the measurement of $S_1$, even less than we know of a system described by the $\psi$-function.
But on one supposition we should, in my opinion, absolutely hold fast: The real factual situation of the system $S_2$ is independent of what is done with the system $S_1$, which is spatially separated from the former.  According to the type of measurement which I make of $S_1$, I get, however, a very different $\psi_2$ for the second partial system. Now, however, the real situation of $S_2$ must be independent of what happens to $S_1$.  For the same real situation of $S_2$ it is possible therefore to find, according to one's choice, different types of $\psi$-function. \ldots

If now the physicists, $A$ and $B$, accept this consideration as valid, then $B$ will have to give up his position that the $\psi$-function constitutes a complete description of a real factual situation.  For in this case it would be impossible that two different types of $\psi$-functions could be coordinated with the identical factual situation of $S_2$.
\end{quote}
Aside from asserting a frequentistic conception of probability, the argument is nearly perfect.  It tells us one important reason why we should not be thinking of quantum states as the $\psi$-ontologists do.  Particularly, it is one we should continue to bear in mind as we move to a Bell-type setting:  Even there, there is no reason to waiver on its validity.  It may be true that Einstein implicitly equated ``incomplete description'' with ``there must exist a hidden-variable account'' (though we do not think he did), but the argument as stated neither stands nor falls on this issue.

There is, however, one thing that Einstein does miss in his argument, and this is where the structure of Bell's thinking steps in.  Einstein says, ``to this question there is no definite answer within the framework of the theory'' when speaking of whether quantum measurements are ``generative'' or simply ``revealing'' of their outcomes.  If we accept everything he said above, then with a little clever combinatorics and geometry one can indeed settle the question.

Let us suppose that the two spatially separated systems in front of the agent are two ququarts (i.e., each system is associated with a four-dimensional Hilbert space ${\mathcal H}_4$), and that the agent ascribes a maximally entangled state to the pair, i.e., a state $|\psi\rangle$ in ${\mathcal H}_4\otimes{\mathcal H}_4$ of the form,
\begin{equation}
|\psi\rangle=\frac{1}{2}\sum_{i=1}^4|i\rangle|i\rangle\;.
\end{equation}
Then we know that there exist pairs of measurements, one for each of the separate systems, such that if the outcome of one is known (whatever the outcome), one will thereafter make a probability-one statement concerning the outcome of the other.  For instance, if a nondegenerate Hermitian operator $H$ is measured on the left-hand system, then one will thereafter ascribe a probability-one assignment for the appropriate outcome of the transposed operator $H^{\rm T}$ on the right-hand system.  What this means for a Bayesian agent is that after performing the first measurement he will bet his life on the outcome of the second.

But how could that be if he has already recognized two systems with no instantaneous causal influence between each other?  Mustn't it be that the outcome on the right-hand side is ``already there'' simply awaiting confirmation or registration?  It would seem Einstein's physicist $B$ is already living in a state of contradiction.

Indeed it must be this kind of thinking that led Einstein's collaborators Podolsky and Rosen to their famous sufficient criterion for an ``element of [preexistent] reality'' \cite{Fine96}:
\begin{quote}
\noindent If, without in any way disturbing a system, we can predict with certainty (i.e., with probability equal to unity) the value of a physical quantity, then there exists an element of reality corresponding to that quantity.
\end{quote}
Without doubt, no personalist Bayesian would ever utter such a notion:  Just because he believes something with all his heart and soul and would gamble his life on it, it would not make it necessarily so by the powers of nature---even a probability-one assignment is a state of belief for the personalist Bayesian.  But he might still entertain something not unrelated to the EPR criterion of reality.  Namely, that believing a particular outcome will be found with certainty on a causally disconnected system entails that one {\it also\/} believes the outcome to be ``already there'' simply awaiting confirmation.

But it is not so, and the QBist has already built this into her story of measurement.  Let us show this presently\footnote{Overall this particular technique has its roots in Stairs \cite{Stairs83}, and seems to bear some resemblance to the gist of Conway and Kochen's ``Free Will Theorem'' \cite{Conway06,Conway09}. } by combining all the above with a beautifully simple Kochen--Specker style construction discovered by Cabello, Estebaranz, and Garc\'{\i}a-Alcaine (CEGA) \cite{Cabello97}.  Imagine some measurement $H$ on the left-hand system; we will denote its potential outcomes as a column of letters, like this:
\begin{equation}
\veec abcd
\end{equation}
Further, since there is a fixed transformation taking any $H$ on the left-hand system to a corresponding $H^{\rm T}$ on the right-hand one, there is no harm in identifying the notation for the outcomes of both measurements.  That is to say, if the agent gets outcome $b$ (to the exclusion of $a$, $c$, and $d$) for $H$ on the left-hand side, he will make a probability-one prediction for $b$ on the right-hand side, even though that measurement strictly speaking is a different one, namely $H^{\rm T}$.  If the agent further subscribes to (our Bayesian variant of) the EPR criterion of reality, he will say that he believes $b$ to be TRUE of the right-hand system as an element of reality.

Now let us consider two possible measurements, $H_1$ and $H_2$ for the left-hand side, with potential outcomes
\begin{equation}
\veec abcd \quad\qquad\mbox{and}\qquad\quad \veec efgh
\end{equation}
respectively.  Both measurements cannot be performed at once, but it might be the case that if the agent gets a specific outcome for $H_1$, say $c$ particularly, then not only will he make a probability-one assignment for $c$ in a measurement of $H_1^{\rm T}$ on the right-hand side, but also for $e$ in a measurement of $H_2^{\rm T}$ on it.  Similarly, if $H_2$ were measured on the left, getting an outcome $e$; then he will make a probability-one prediction for $c$ in a measurement of $H_1^{\rm T}$ on the right.  This would come about if $H_1$ and $H_2$ (and consequently $H_1^{\rm T}$ and $H_2^{\rm T}$) share a common eigenvector.  Supposing so and that $c$ was actually the outcome for $H_1$ on the left, what conclusion would the EPR criterion of reality draw?  It is that both $c$ and $e$ are elements of reality on the right, and none of $a$, $b$, $d$, $f$, $g$, or $h$ are.  Particularly, since the right-hand side could not have known whether $H_1$ or $H_2$ was measured on the left, whatever $c$ and $e$ stands for, it must be the same thing, the same property.  In such a case, we discard the extraneous distinction between $c$ and $e$ in our notation and write
\begin{equation}
\veec abcd \quad\qquad\mbox{and}\qquad\quad \veec cfgh
\end{equation}
for the two potential outcome sets for a measurement on the right.

We now have all the notational apparatus we need to have some fun.  The genius of CEGA was that they were able to find a set of nine ``interlocking'' Hermitian operators $H_1$, $H_2$, \ldots, $H_9$ for the left, whose set of potential outcomes for the corresponding operators on the right would look like this:
\begin{equation}
\veec abcd \qquad \veec aefg \qquad \veec hicj\qquad \veec hkgl\qquad \veec bemn\qquad \veec ikno\qquad \veec pqdj\qquad \veec prfl\qquad \veec qrmo \label{BurlIves}
\end{equation}
Take the second column as an example.  It means that if $H_2$ were measured on the left-hand system, only one of $a$, $e$, $f$, or $g$ would occur---the agent cannot predict which---but if $a$ occurred, he would be absolutely certain of it also occurring in a measurement of $H_1^{\rm T}$ on the right.  And if $e$ were to occur on the left, then he would be certain of getting $e$ as well in a measurement of $H_5^{\rm T}$ on the right.  And similarly with $f$ and $g$, with their implications for $H_8^{\rm T}$ and $H_4^{\rm T}$.

The wonderful thing to note about (\ref{BurlIves}) is that every letter $a$, $b$, $c$, \ldots, $r$ occurs exactly twice in the collection.  But the EPR criterion of reality (or our Bayesian variant of it) would require exactly one letter to have the truth value TRUE in each column, with the other three having the value FALSE.  In total, nine values of TRUE:  A clean contradiction!  For if every letter occurs exactly twice in the collection, whatever the total number of TRUE values is, it must be an even number.

To emphasize the point, let us sketch a similar argument for the case where the two halves are both {\it qutrits,} quantum systems of dimension 3.  Again, the argument is a little variation of the EPR thought experiment.  This time, we begin by ascribing to our bipartite system the maximally entangled state
\begin{equation}
|\psi\rangle=\frac{1}{2}\sum_{i=1}^3|i\rangle|i\rangle\;.
\end{equation}
The experimentalist---let us call her Alice---considers making a
measurement on the left-hand particle in some basis.  If Alice obtains
outcome number~2 on the left side, then she can predict that if she
were to make a particular measurement on the right side, she would get
outcome number~2.  (The bases for the left-hand and right-hand
measurements are related by a transpose operation.)  So, under the
assumption of Einsteinian locality and the EPR criterion, Alice would
say, ``Aha!  It must be the case that there is an element of reality
on the right side corresponding to outcome number~2 of that
measurement.  It's something inherent in that body.''  But we can play
this game with any basis: If Alice were to get outcome~$i$ for a
measurement on one particle, she would predict with certainty that she
would get outcome~$i$ for that measurement, transposed, on the other
particle.

We're talking about noncommuting variables here; by the EPR criterion
and locality, Alice would conclude it must be the case that there were
elements of reality associated with those noncommuting observables.
But it is no fun to consider only one basis, or only two.  Instead, we
proceed to analyze a whole set of them, corresponding to one of the
qutrit Kochen--Specker constructions, for instance the one that Asher
Peres found~\cite{Peres95}.  In order to present this construction in
a concise way, let us use an abbreviated notation in which a ``2''
stands for $\sqrt{2}$ and an overline means a minus sign.  Thus,
``$1\bar{1}2$'' stands for the ray $(1, -1, \sqrt{2})^{\rm T}$.  We
tabulate the following ten sets of three orthogonal rays apiece:
\begin{equation}
\begin{array}{ccc}
001 & 100 & 010 \\
101 & \bar{1}01 & 010 \\
011 & 0\bar{1}0 & 100 \\
1\bar{1}2 & \bar{1}12 & 110 \\
102 & \bar{2}01 & 010 \\
211 & 0\bar{1}1 & \bar{2}11 \\
201 & 010 & \bar{1}02 \\
112 & 1\bar{1}0 & \bar{1}\bar{1}2 \\
012 & 100 & 0\bar{2}1 \\
121 & \bar{1}01 & 1\bar{2}1
\end{array}
\end{equation}

The EPR criterion tells us that for each basis, there exists an
``element of physical reality'' that determines what the outcome of a
measurement in that basis will be.  This applies to each row above:
According to the EPR criterion, in any set of three orthogonal rays,
one must be marked TRUE and the other two marked FALSE.  But if we start
labeling our rays, marking each one orthogonal to a TRUE ray with FALSE,
eventually we hit a point where we can't do it consistently.  We end
up marking a set of three orthogonal rays all FALSE, which is against the
rules.

For each of these bases, Alice would have said, ``By making a
measurement here, I draw an inference about the element of reality
over there.''  She can do it for one, she can do it for another, and
another \ldots\ but she can't do it for all of them without running
into trouble!

\begin{figure}[ht]
\begin{center}
\includegraphics[width=8cm]{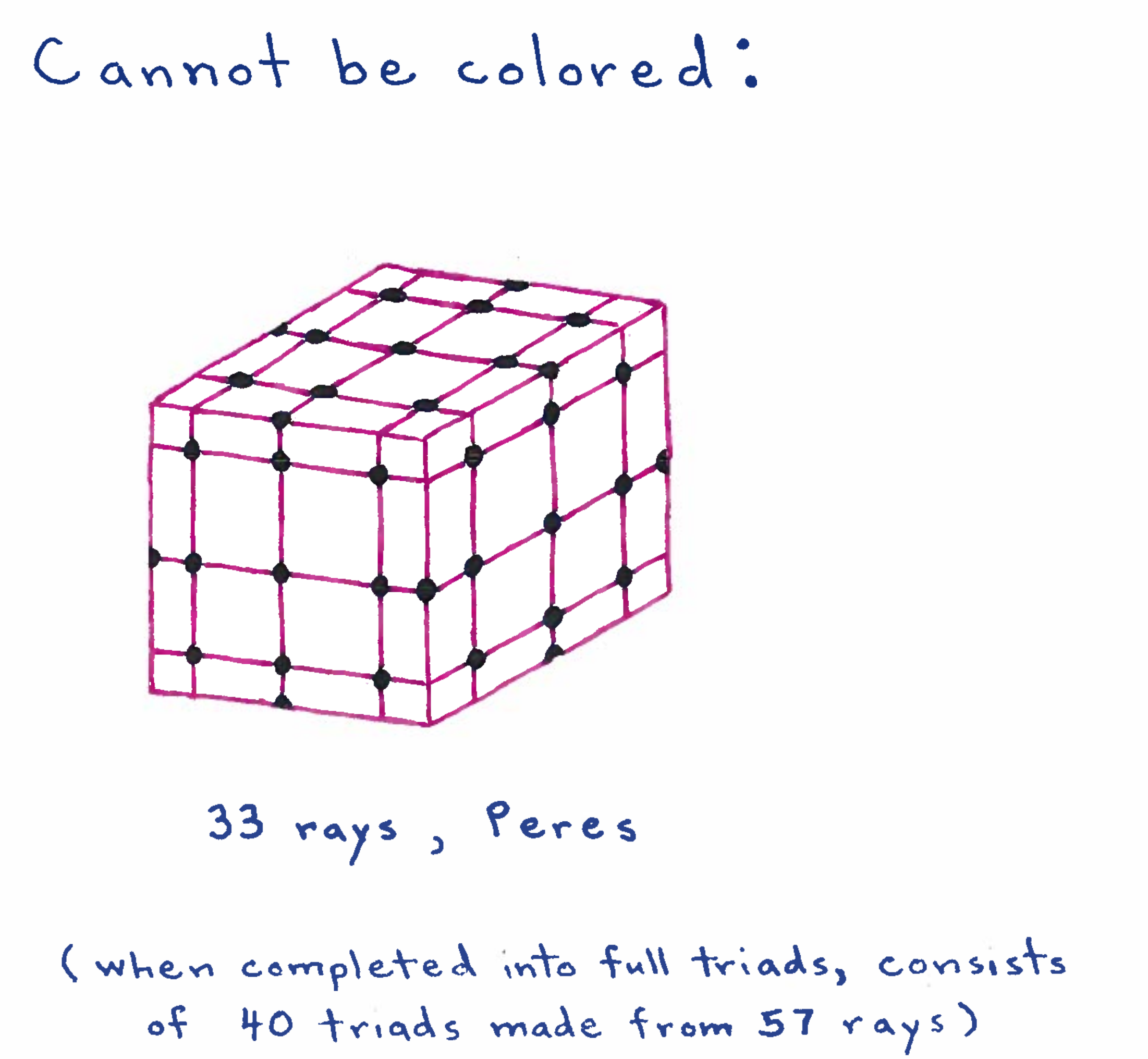}
\end{center}
\caption{\label{fig:peres} Each ray in Asher Peres's qutrit Kochen--Specker construction can be written as a ray in $\mathbb{R}^3$ and illustrated by where those rays intersect a cube.}
\end{figure}

Something must give.  The quick reaction of most of the quantum foundations community has been to question the causal independence of the two systems under consideration.  But if one gives up on the autonomy of one system from the other---after very explicitly assuming it---this surely amounts to saying that there were never two systems there after all; the very idea of separate systems is a broken concept.  This first raises a minor conundrum:  Why then would the quantum formalism engender us to formulate our description from beginning to end in terms of ${\mathcal H}_3\otimes{\mathcal H}_3$, rather than simply a raw nine-dimensional space ${\mathcal H}_{9}$?  Why is that separating symbol $\otimes$, apparently marking some kind of conceptual distinction, always hanging around?

Reaching much deeper however, if one is willing to throw away one's belief in systems' autonomy from each other, why would one ever believe in one's own autonomy?  All stringent reason for it gets lost, and indeed as Einstein warns, what now is the meaning of science?  As Hans Primas once emphasized \cite{Primas90},
\begin{quote}
\noindent It is a tacit assumption of all engineering sciences that nature can be {\it manipulated\/} and that the initial conditions required by experiments can be brought about by interventions of the world external to the object under investigation.  That is, {\it we assume that the experimenter has a certain freedom of action which is not accounted for by first principles of physics}.  Without this freedom of choice, experiments would be impossible.  Man's free will implies the ability to carry out actions, it constitutes his essence as an actor.  We act under the idea of freedom, but the topic under discussion is neither man's sense of personal freedom as a subjective experience, nor the question whether this idea could be an illusion or not, nor any questions of moral philosophy, but that {\it the framework of experimental science requires the freedom of action as a constitutive though tacit presupposition}.
\end{quote}
If the left-hand system can manipulate the right-hand system, {\it even when by assumption it cannot}, then who is to say that the right-hand system cannot manipulate the agent herself?  To put it still differently:  If one is never allowed to assume causal independence between separated systems because of a contradiction in the term, then one can never assume it of oneself either, even with respect to the components of the world that one thinks one is manipulating.  It would be a wackier world than even the one QBism entertains.

But QBism's world is not such a bad world, and some of us find its
openness to possibility immensely exciting.  What gives way in this
world is not the idea of reality, but simply the narrow-minded EPR
criterion for it.  We jettison  {\it both\/} the idea that a probability-one assignment implies there is a pre-existent outcome (property) ``over there'' waiting to be revealed and, barring that, that it must have been ``over here'' pre-existent, waiting to be transferred and then revealed.  The solution lies closer to one of John Wheeler's quips, ``No question? No answer.''  A probability-one assignment lays no necessary claim on what the world {\it is}, but what the agent using it believes with all her heart and soul.  In the case of our present example, what the agent believes is that if an outcome $b$ {\it came about\/} as a result of her action $H$ on the left-hand system, an outcome $b$ {\it would come about\/} if she were to perform the action $H^{\rm T}$ on the right-hand system.  But if she does not walk over to the right-hand system and take the action, there is no good sense in which the outcome (or property) $b$ is already there.  Measurement is not a passive process, but instead a fundamentally {\it participatory\/} one.

At the instigation of a quantum measurement, something new comes into the world that was not there before; and that is about as clear an instance of {\it creation\/} as one can imagine.  Sometimes one will have no strong beliefs for what will result from the creation (as with the measurement of $H$), and sometimes one will have very strong beliefs (as with the subsequent measurement of $H^{\rm T}\,$), but a free creation of nature it remains.

\vfill

\section{The Quantum de Finetti Theorem}

\begin{flushright}
\baselineskip=13pt
\parbox{2.8in}{\baselineskip=13pt\footnotesize

You know how men have always hankered after unlawful magic, and you know
what a great part in magic {\it words\/} have always played. If you
have his name, \ldots\ you can
control the spirit, genie, afrite, or whatever the power may be.
Solomon knew the names of all the spirits, and having their names, he
held them subject to his will.  So the universe has always appeared
to the natural mind as a kind of enigma, of which the key must be
sought in the shape of some illuminating or power-bringing word or
name.  That word names the universe's {\it principle}, and to possess
it is after a fashion to possess the universe itself.\medskip

But if you follow the pragmatic method, you cannot look on any such
word as closing your quest.  You must bring out of each word its
practical cash-value, set it at work within the stream of your
experience.  It appears less as a solution, then, than as a program
for more work, and more particularly as an indication of the ways in
which existing realities may be {\it changed}.\medskip

{\it Theories thus become instruments, not answers to enigmas, in
which we can rest.}  We don't lie back upon them, we move forward,
and, on occasion, make nature over again by their aid.}
\medskip\\
 --- William James
\end{flushright}

Since the beginning, those who brought Bayesian probability into quantum physics have been on the run proving technical theorems whenever required to close a gap in their logic or negate an awkwardness induced by their new way of speaking.  It was never enough to ``lie back upon'' the pronouncements:  They had to be shown to have substance, something that would drive physics itself forward.  A case in point is the {\it quantum de Finetti theorem} \cite{Fuchs04,Caves02b}.

The term ``unknown state'' is ubiquitous in quantum information:  Unknown quantum states are teleported, protected with quantum error correcting codes, used to check for quantum eavesdropping, and arise in innumerable other applications.  For a QBist, though, the phrase can only be an oxymoron:  If quantum states are compendia of beliefs, and not states of nature, then the state is known to someone, at the very least the agent who holds it.  But if so, then what are experimentalists doing when they say they are performing quantum-state tomography in the laboratory?  The very goal of the procedure is to characterize the unknown quantum state a piece of laboratory equipment is repetitively preparing. There is certainly no little agent sitting on the inside of the device devilishly sending out quantum systems representative of his beliefs, and smiling as the experimenter on the outside slowly homes in on those private thoughts through his experiments.

The quantum de Finetti theorem is a result that allows the story of quantum-state tomography to be told purely in terms of a single agent---namely, the experimentalist in the laboratory.  In a nutshell, the theorem is this.  Suppose the experimentalist walks into the laboratory with the very minimal belief that, of the systems her device is spitting out (no matter how many), she could interchange any two of them and it would not change the statistics she expects for any measurements she might perform.  Then the theorem says that ``coherence with this belief alone'' requires her to make a quantum-state assignment $\rho^{(n)}$ (for any $n$ of those systems) that can be represented in the form:
\begin{equation}
\rho^{(n)}=\int P(\rho)\, \rho^{\otimes n}\, d\rho\;,
\label{MushuPork}
\end{equation}
where $P(\rho)\, d\rho$ is some probability measure on the space of single-system density operators and $\rho^{\otimes n}=\rho\otimes\cdots\otimes\rho$ represents an $n$-fold tensor product of identical quantum states.

To put it in words, this theorem licenses the experimenter to act {\it as if\/} each individual system has some state $\rho$ unknown to her, with a probability density $P(\rho)$ representing her ignorance of which state is the true one.  But it is only {\it as if}---the only active quantum state in the picture is the one the experimenter actually possesses in her mind, namely $\rho^{(n)}$.  When the experimenter performs tomography, all she is doing is gathering data system-by-system and updating, via Bayes rule \cite{Schack01}, the state $\rho^{(n)}$ to some new state $\rho^{(k)}$ on a smaller number of remaining systems. Particularly, one can prove that this form of quantum-state assignment leads the agent to expect that with more data, she will make her $P(\rho)$ more and more narrow, and thus she will approach ever more closely a posterior state of the form $\rho^{(k)}\approx \rho^{\otimes k}$.  This is the real, underlying reason that excuses the habit of speaking of tomography as revealing ``the unknown quantum state.''

One important consequence of this theorem is the following.  Suppose
that Alice is collaborating with Bob.  From her perspective, Bob is a
physical system.  Alice mathematically models Bob as having
expectations about the sequence of systems they are studying,
expectations that matter to Alice because she can ask Bob questions
and get answers.  In Alice's mental model of Bob, she writes a de
Finetti representation $P_B(\rho)$, satisfying the same general
properties as her own expectations, which she encodes into the
function $P_A(\rho)$.  Alice imagines that she and Bob are receiving
the same data.  It follows from the quantum de Finetti theorem that if
$P_A(\rho)$ and $P_B(\rho)$ initially agree to at least a small
extent, then Alice should expect that their expectations will come
into greater and greater agreement.

Furthermore, just as there is a de Finetti theorem to make sense of
``unknown states,'' there is a de Finetti theorem to make sense of
``unknown measurements'' and ``unknown processes''~\cite{Fuchs04c}.

Despite the explicitly foundational motivation, the quantum de Finetti theorem has nonetheless fared like a stand-alone result for quantum information theory.  Among other things, it turned out to be useful for proving the security of some quantum key distribution schemes \cite{Lo05,Renner07,Renner08}, it became an important component in the analysis of entanglement detection \cite{Doherty05,Enk07}, and even served in an analysis of the quantum state of propagating laser light~\cite{vanEnk02,vanEnk02b}.

During the effort to prove the quantum de Finetti theorem, Carlton Caves brought attention to a special type of quantum measurement.  Up to that point, those of us who were bringing Bayesianism to quantum physics were in the habit of regarding a quantum state $\rho$ as a ``catalogue of probabilities.''  This way of thinking rested on the old way of modeling measurements, that is, the von Neumann tradition, where a measurement corresponds to an orthonormal basis.  The Born rule lets us get probabilities out of a density matrix $\rho$, yet no single von Neumann basis can yield sufficient information to reconstruct $\rho$; hence, the ``catalogue of probabilities'' language.  But quantum information theory brought a new manner of thinking, in which a measurement can be any collection of positive operators that sum to the identity---a {\it Positive Operator-Valued Measure,} or POVM.  This change of perspective opened the possibility of an {\it informationally complete\/} experiment, i.e., a {\it single\/} measurement whose statistics suffice to reconstitute an entire density matrix $\rho$.  Once we have constructed an ``IC'' POVM, we can then replace any $\rho$ with a probability distribution.  One matrix $\rho$, one vector $\vec{p}$.

At first, Caves believed that proving the quantum de Finetti theorem would require a specific {\it type\/} of IC POVM.  Luckily, this turned out to be wrong:  Finding a construction of an IC POVM in arbitrary finite Hilbert-space dimension was enough.  However, the particular kind of IC POVM to which Caves called attention soon took on a life of its own, and it is to that kind which we now turn.

\section{Seeking SICs -- The Born Rule as Fundamental}

\label{SeekingSICs}

If quantum theory is so closely allied with probability theory, then why is it not written in a language that starts with probability, rather than a language that ends with it?  Why does quantum theory invoke the mathematical apparatus of complex amplitudes, Hilbert spaces, and linear operators?  This question quickly brings us to the research frontier.

For, actually there are ways to pose quantum theory purely in terms of probabilities---indeed, there are many ways, each with a somewhat different look and feel \cite{Ferrie09}.  The work of W.~K. Wootters is an example, and as he emphasized long ago \cite{Wootters86},
\begin{quote}
\noindent It is obviously {\it possible\/} to devise a formulation of quantum mechanics
without probability amplitudes. One is never forced to use any quantities in
one's theory other than the raw results of measurements. However, there is
no reason to expect such a formulation to be anything other than
extremely ugly. After all, probability amplitudes were invented for a reason.
They are not as directly observable as probabilities, but they make the
theory simple. I hope to demonstrate here that one {\it can\/} construct a
reasonably pretty formulation using only probabilities. It may not be quite
as simple as the usual formulation, but it is not much more complicated.
\end{quote}
What has happened in the intervening years is that the mathematical structures of quantum information theory have grown significantly richer than the ones he had based his considerations on---so much so that we may now be able to optimally re-express the theory.  What was once ``not much more complicated,'' now has the promise of being downright insightful.

The key ingredient is a hypothetical structure called a ``symmetric informationally complete positive-operator-valued measure,'' or SIC (pronounced ``seek'') for short.  This is a set of $d^2$ rank-one projection operators $\Pi_i=|\psi_i\rangle\langle\psi_i|$ on a finite $d$-dimensional Hilbert space such that
\begin{equation}
\big|\langle\psi_i|\psi_j\rangle\big|^2=\frac{1}{d+1}\quad \mbox{whenever} \quad i\ne j\;.
\label{Mojo}
\end{equation}
Because of their extreme symmetry, it turns out that such sets of operators, when they exist, have three very fine-tuned properties: 1) the operators must be linearly independent and span the space of Hermitian operators, 2) there is a sense in which they come as close to an orthonormal basis for operator space as they can under the constraint that all the elements in a basis be positive semi-definite \cite{Appleby07}, and 3) after rescaling, they form a resolution of the identity operator, $I=\sum_i \frac{1}{d}\Pi_i$.

The symmetry, positive semi-definiteness, and properties 1 and 2 are significant because they imply that an arbitrary quantum state $\rho$---pure or mixed---can be expressed as a linear combination of the $\Pi_i$.  Furthermore, the expansion is likely to have some significant features not found in other, more arbitrary expansions.  The most significant of these becomes apparent when one takes property 3 into account.  Because the operators $H_i=\frac1d \Pi_i$ are positive semi-definite and form a resolution of the identity, they can be interpreted as labeling the outcomes of a quantum measurement device---not a standard-textbook, von Neumann measurement device whose outcomes correspond to the eigenvalues of some Hermitian operator, but to a measurement device of the most general variety allowed by quantum theory, the POVMs \cite{Nielsen00,Peres95}.  Particularly noteworthy is the smooth relation between the probabilities $P(H_i)={\rm tr}\big(\rho H_i\big)$ given by the Born rule for the outcomes of such a measurement\footnote{There is a slight ambiguity in notation here, as $H_i$ is dually used to denote an operator and an outcome of a measurement. For the sake of simplicity, we hope the reader will forgive this and similar abuses.}
and the expansion coefficients for $\rho$ in terms of the $\Pi_i$:
\begin{equation}
\rho = \sum_{i=1}^{d^2}\left( (d+1)\,P(H_i) - \frac1d \right)\Pi_i\;.
\label{Ralph}
\end{equation}
There are no other operator bases that give rise to such a simple formula connecting probabilities with density operators.

Before getting to that, however, we should reveal what is so consternating about the SICs: It is the question of when they exist.  Despite years of growing effort since the definition was first introduced \cite{Zauner99,Caves99,Renes04}, no one has been able to show that they exist in completely general dimension.  All that is known firmly is that they exist in dimensions 2 through 147 inclusive, as well as in some scattered cases beyond those: 168, 172, 195, 199, 228, 259 and
323.  Dimensions 2--21, 24, 28, 30, 31, 35, 37, 39, 43 and 48 are known through direct or computer-assisted analytic proof; the remaining solutions are known through numerical calculation, satisfying Eq.~(\ref{Mojo}) to high precision (in some cases, up to 16,000 digits accuracy).\footnote{Numerical and some analytical solutions up through dimension 67 are detailed in \cite{Scott09}.  Others were reported to us by Andrew Scott and Marcus Appleby. Numerical solutions in dimensions 122 through 142 and in dimensions 144, 145 and 146 are due to Michael C.\ Hoang, in collaboration with the authors, using the Chimera supercomputer.}  For the remainder of the article we will proceed as if they do always exist for finite $d$.    At least this is the conceit of our story.  We note in passing, however, that the SIC existence problem is not without wider context:  If they do exist, they solve at least four other (more practical, non-foundational) optimality problems in quantum information theory \cite{Fuchs03,Scott06,Appleby07,Wootters07,Zhu2016}.  In addition, the results they imply for Lie and Jordan algebras indicate that if SICs didn't always exist, linear algebra itself would have a drastically different character from one dimension to another~\cite{Appleby15}.  It would be a nasty trick if SICs failed to exist!

\begin{figure}
\begin{center}
\includegraphics[height=3.3in]{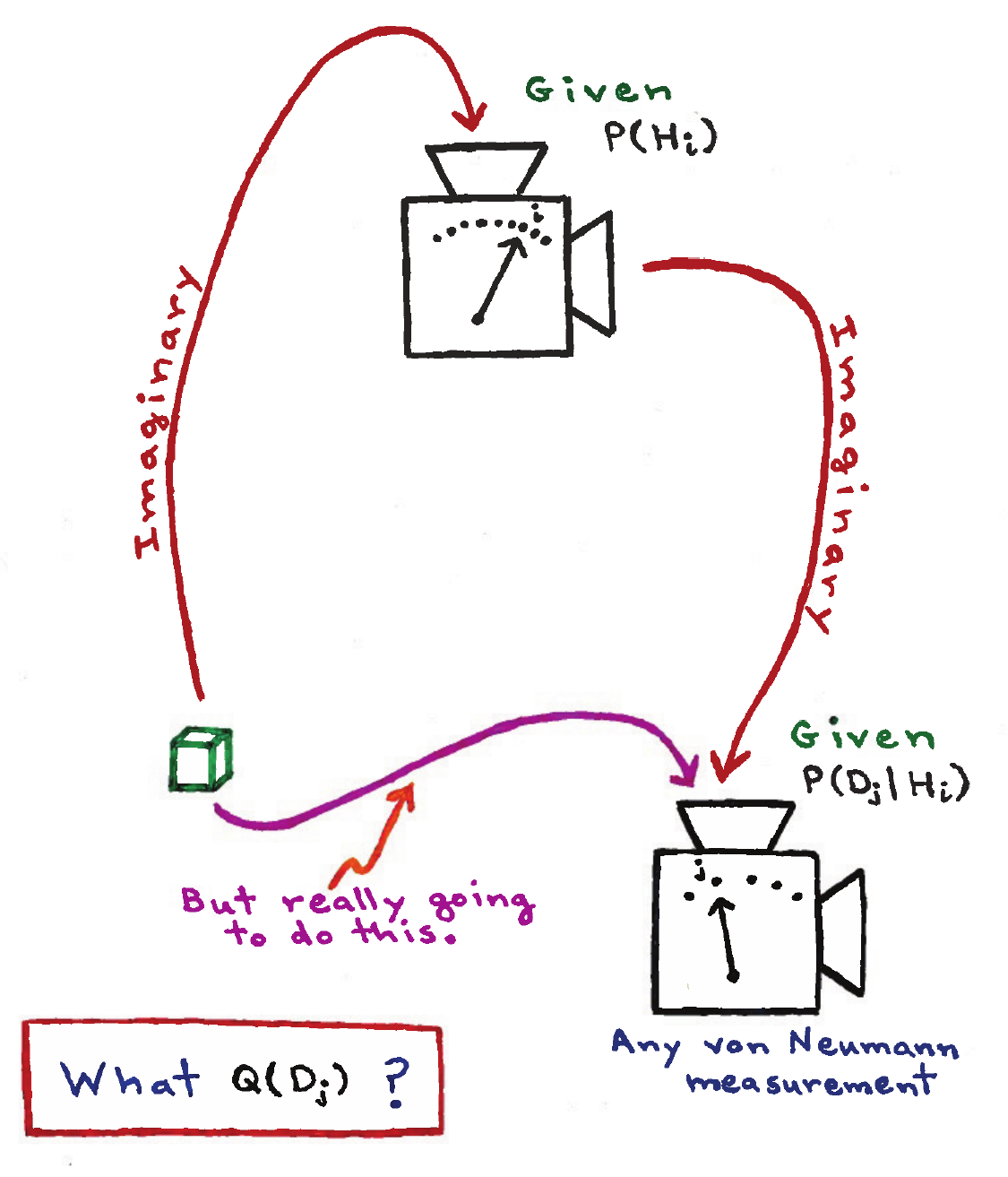}
\bigskip\caption{ Any quantum measurement can be conceptualized in two ways.
Suppose an arbitrary von Neumann measurement ``on the ground,'' with outcomes $D_j=1,\ldots,d$. Its probabilities $P(D_j)$ can be derived by cascading it with a fixed fiducial SIC measurement ``in the sky'' (of outcomes $H_i=1,\ldots,d^2$). Let $P(H_i)$ and $P(D_j|H_i)$ represent an agent's probabilities, assuming the measurement in the sky is actually performed. The probability $Q(D_j)$ represents instead the agent's probabilities under the assumption that the measurement in the sky is {\it not\/} performed. The Born rule, in this language, says that $P(D_j)$, $P(H_i)$, and $P(D_j|H_i)$ are related by the Bayesian-style Eq.~(\ref{ScoobyDoo}).}
\end{center}
\end{figure}

So suppose they do.  Thinking of a quantum state as {\it literally\/} an agent's probability assignment for the outcomes of a potential SIC measurement leads to a new way to express the Born rule for the probabilities associated with any {\it other\/} quantum measurement.  Consider the diagram in Figure 4.  It depicts a SIC measurement ``in the sky,'' with outcomes $H_i$, and any standard von Neumann measurement ``on the ground.''\footnote{Do not, however, let the designation ``SIC sitting in the sky'' make the device seem too exalted to be of any interest.  Already, announcements of experimental implementations have been made for qubits \cite{Ling06}, qutrits \cite{Medendorp10}, and still higher-dimensional systems \cite{BoydLeuchs}.}  For the sake of specificity, let us say the latter has outcomes $D_j=|j\rangle\langle j|$, the vectors $|j\rangle$ representing some orthonormal basis.  We conceive of two possibilities (or two ``paths'') for a given quantum system to get to the measurement on the ground:  ``Path 1'' is that it proceeds directly to the measurement on the ground.  ``Path 2'' is that it proceeds first to the measurement in the sky and only subsequently to the measurement on the ground---the two measurements are cascaded.

Suppose now, we are given the agent's personal probabilities $P(H_i)$ for the outcomes in the sky and his personal conditional probabilities $P(D_j|H_i)$ for the outcomes on the ground subsequent to the sky.  I.e., we are given the probabilities the agent would assign on the supposition that the quantum system follows Path 2.  Then ``coherence alone'' (in the Bayesian sense) is enough to tell what probabilities $P(D_j)$ the agent should assign for the outcomes of the measurement on the ground---it is given by the Law of Total Probability applied to these numbers:
\begin{equation}
P(D_j)=\sum_i P(H_i) P(D_j|H_i)\;.
\label{Magnus}
\end{equation}
That takes care of Path 2, but what of Path 1?  Is this enough information to recover the probability assignment $Q(D_j)$ the agent would assign for the outcomes on Path 1 via a normal application of the Born rule?  That is, that
\begin{equation}
Q(D_j)={\rm tr} (\rho D_j)
\label{EmmaPlayStarWars}
\end{equation}
for some quantum state $\rho$?  Maybe, but the answer will clearly not be $P(D_j)$.  One has
\begin{equation}
Q(D_j)\ne P(D_j)\;
\end{equation}
simply because Path 2 is {\it not\/} a coherent process (in the quantum sense!)\ with respect to Path 1---there is a measurement that takes place in Path 2 that does not take place in Path 1.

What is remarkable about the SIC representation is that it implies that, even though $Q(D_j)$ is not equal to $P(D_j)$, it is still a function of it.  Particularly,
\begin{eqnarray}
Q(D_j) &=& (d+1) P(D_j) - 1\nonumber
\\
&=&
(d+1)\sum_{i=1}^{d^2} P(H_i) P(D_j|H_i) - 1\;.
\label{ScoobyDoo}
\end{eqnarray}
The Born rule is nothing but a kind of Quantum Law of Total Probability!  No complex amplitudes, no operators---only probabilities in, and probabilities out.  Indeed, it is seemingly just a rescaling of the old law, Eq.~(\ref{Magnus}).  And in a way it is.

Earlier, we stressed the importance of considering quantum measurements in all their generality:  The notion of ``measurement'' should include all POVMs, not just the von Neumann ones.  So, what happens if the measurement on the ground is an arbitrary POVM, not necessarily given by an orthonormal basis?  Then,
\begin{equation}
Q(D_j) = \sum_{i=1}^{d^2} \left[(d+1) P(H_i) - \frac{1}{d}\right] P(D_j|H_i).
\label{ScoobyDoo2}
\end{equation}
This is only slightly more intricate than our previous expression, Eq.~(\ref{ScoobyDoo}), which is itself a special case of this more general formula.  And the general formula is still quite similar to the classical prescription for combining conditional probabilities:  We simply stretch the $P(H_i)$ and then shift them to preserve the overall normalization.

{\it But beware}:  One should not interpret Eq.~(\ref{ScoobyDoo2}) as invalidating probability theory itself in any way!  For the old Law of Total Probability has no jurisdiction in the setting of our diagram, which compares a two distinct, mutually exclusive hypothetical scenarios.  Path 1 is what Alice intends to do, but she recognizes that she could in principle follow Path 2 instead, and Eq.~(\ref{ScoobyDoo2}) sets the standard of consistency to which Alice should strive when meshing her probabilities together.\footnote{This is one place where we can point out a mild historical antecedent to QBism.  In the historical study \cite{Bacciagaluppi09}, it is pointed out that Born and Heisenberg, {\it already at the 1927 Solvay conference}, refer to the calculation $|c_n(t)|^2=\left|\sum_m S_{mn}(t)c_m(0)\right|^2$ and say, ``it should be noted that this `interference' does not represent a contradiction with the rules of the probability calculus, that is, with the assumption that the $|S_{nk}|^2$ are quite usual probabilities.'' Their reasons for saying this may have been different from our own, but at least they had come this far.}  Indeed as any Bayesian would emphasize, if there is a distinguishing mark in one's considerations---say, the fact of two distinct experiments, not one---then one ought to take that into account in one's probability assignments (at least initially so).  Thus there is a hidden, or at least suppressed, condition in our notation:  Really we should have been writing the more cumbersome, but honest, expressions $P(H_i|{\mathcal E}_2)$, $P(D_j|H_i,{\mathcal E}_2)$, $P(D_j|{\mathcal E}_2)$, and $Q(D_j|{\mathcal E}_1)$ all along.  With this explicit, it is no surprise that,
\begin{equation}
Q(D_j|{\mathcal E}_1)\ne\sum_iP(H_i|{\mathcal E}_2)P(D_j|H_i,{\mathcal E}_2)\;.
\end{equation}
The message is that quantum theory supplies some\-thing---a new form of ``Bayesian coherence,'' though empirically based (as quantum theory itself is)---that raw probability theory does not.  The Born rule in these lights is an addition to Bayesian probability, not in the sense of a supplier of some kind of more-objective probabilities, but in the sense of giving extra normative rules to guide the agent's behavior when she interacts with the physical world.

It is a normative rule for reasoning about the consequences of one's proposed actions in terms of the potential consequences of an alternative action.  It is like nothing else physical theory has contemplated before.  Seemingly at the heart of quantum mechanics from the QBist view is a statement about the impact of hypotheticals on our expectations for the actual.  The impact parameter is metered by a single, significant number associated with each physical system---its Hilbert-space dimension $d$.  The larger the $d$ associated with a system, the more $Q(D_j)$ must deviate from $P(D_j)$.  Of course this point must have been implicit in the usual form of the Born rule, Eq.~(\ref{EmmaPlayStarWars}).  What is important from the QBist perspective, however, is how the new form puts the significant parameter front and center, displaying it in a way that one ought to nearly trip over.

Understanding this as the goal helps pinpoint the role of SICs in our considerations.  The issue is not that quantum mechanics {\it must\/} be rewritten in terms of SICs, but that it {\it can\/} be.\footnote{If everything goes right, that is, and the damned things actually exist in all dimensions!}  Certainly no one is going to drop the usual operator formalism and all the standard methods learned in graduate school to do their workaday calculations in SIC language exclusively.  It is only that the SICs form an ideal coordinate system for a particular problem (an important one to be sure, but nonetheless a particular one)---the problem of {\it interpreting\/} quantum mechanics.  The point of all the various representations of quantum mechanics (like the various quasi-probability representations of \cite{Ferrie09}, the Heisenberg and Schr\"odinger pictures, and even the path-integral formulation) is that they give a means for isolating one or another aspect of the theory that might be called for by a problem at hand.  Sometimes it is really important to do so, even for deep conceptual issues and even if all the representations are logically equivalent.\footnote{Just think of the story of Eddington--Finkelstein coordinates in general relativity.  Once upon a time it was not known whether a Schwarzschild black hole might have, beside its central singularity, a singularity in the gravitational field at the event horizon.  Apparently it was a heated debate, yes or no. The issue was put to rest, however, with the development of the coordinate system.  It allowed one to write down a solution to the Einstein equations in a neighborhood of the horizon and check that everything was indeed all right.}  In our case, we want to bring into plain view the idea that quantum mechanics is an {\it addition\/} to Bayesian probability theory---not a generalization of it \cite{Bub07}, not something orthogonal to it altogether \cite{Jozsa04}, but an addition.  With this goal in mind, the SIC representation is a particularly powerful tool.  Through it, one sees the Born rule as a replacement for a usage of the Law of Total Probability that one would have made in another context (one mutually exclusive with the first).

Furthermore it is similarly so of unitary time evolution in a SIC picture.  To explain what this means, let us change considerations slightly and make the measurement on the ground a unitarily rotated version of the SIC in the sky.  In this setting, $D_j=\frac{1}{d}U\Pi_j U^\dagger$, which in turn implies a simplification of Eq.~(\ref{ScoobyDoo2}) to,
\begin{equation}
Q(D_j) = (d+1)\sum_{i=1}^{d^2} P(H_i) P(D_j|H_i) - \frac{1}{d}\;,
\label{Zizek}
\end{equation}
for the probabilities on the ground.  Note what this is saying!  As the Born rule is a replacement for the Law of Total Probability, unitary time evolution is a replacement for it as well.  For, if we thought in terms of the Schr\"odinger picture, $P(H_i)$ and $Q(D_j)$ would be the SIC representations for the initial and final quantum states under an evolution given by $U^\dagger$. The similarity is no accident.  This is because in both cases the conditional probabilities $P(D_j|H_i)$ completely encode the identity of a measurement on the ground.

Moreover, it makes abundantly clear another point of QBism that has not been addressed so much in the present paper.  Since a personalist Bayesian cannot turn his back on the clarification that {\it all\/} probabilities are personal judgments, placeholders in a calculus of consistency, he certainly cannot turn his back on the greater lesson Eqs.~(\ref{ScoobyDoo2}) and (\ref{Zizek}) are trying to scream out.   Just as quantum states $\rho$ are personal judgments $P(H_i)$, quantum measurement operators $D_j$ and unitary time evolutions $U$ are personal judgments too---in this case $P(D_j|H_i)$.  The only distinction is the technical one, that one expression is an unconditioned probability, while the other is a collection of conditionals.  Most importantly, it settles the age-old issue of why there should be two kinds of state evolution at all.  When Hartle wrote, ``A quantum-mechanical state being a summary of the observers' information about an individual physical system changes both by dynamical laws, and whenever the observer acquires new information about the system through the process of measurement,'' what is his dynamical law making reference to?  There are not two things that a quantum state can do, only one:  Strive to be consistent with all the agent's other probabilistic judgments on the consequences of his actions, across all hypothetical scenarios.  The SICs emphasize and make this point clear.

In fact, much of the most intense research of the UMass Boston QBism group is currently devoted to seeing how much of the essence of quantum theory is captured by Eq.~(\ref{ScoobyDoo2}).  We are frankly quite happy to have an extremely hard problem about the structure of quantum states spaces leading our thinking!  (And, as we'll see in the next section, it is a problem that prompts a traveler to question the received wisdom about the boundary between physics and pure mathematics.)  For instance, one way to approach this is to take Eq.~(\ref{ScoobyDoo2}) as a fundamental axiom and ask what further assumptions are required to recover all of quantum theory.  To give some hint of how a reconstruction of quantum theory might proceed along these lines, note Eq.~(\ref{Ralph}) again.  What it expresses is that any quantum state $\rho$ can be reconstructed from the probabilities $P(H_i)$ the state $\rho$ gives rise to.  This, however, does not imply that plugging just any probability distribution $P(H_i)$ into the equation will give rise to a valid quantum state.  A general probability distribution $P(H_i)$ in the formula will lead to a Hermitian operator of trace one, but it may not lead to an operator with nonnegative eigenvalues.  Indeed it takes further restrictions on the $P(H_i)$ to make this true.  That being the case, the QBist starts to wonder if these restrictions might arise from the requirement that Eq.~(\ref{ScoobyDoo2}) simply always make sense.  For note, if $P(D_j)$ is too small in the special case of Eq.~(\ref{ScoobyDoo}), $Q(D_j)$ will go negative; and if $P(D_j)$ is too large, $Q(D_j)$ will become larger than 1.  So, $P(D_j)$ must be restricted.  But that in turn forces the set of valid $P(H_i)$ to be restricted as well.  And so the argument goes.  We already know how to reconstruct many features of quantum theory in this fashion \cite{RMP,Appleby09a,Fuchs09b,Appleby09b}.  The question now is how to get the whole theory in the most economical way~\cite{qplex}.  Should we succeed, we will have a new development of quantum theory, one that puts its beguiling deviation from classicality, as encoded in Eq.~(\ref{ScoobyDoo2}), front and center.

Another exciting development comes from loosening the form of Eq.~(\ref{ScoobyDoo2}) to something more generic:
\begin{equation}
Q(D_j)=\sum_{i=1}^n \big[\alpha P(H_i) -\beta\big] P(D_j|H_i);,
\end{equation}
where there is {\it initially\/} no assumed relation between $\alpha$, $\beta$, and $n$ as there is in Eq.~(\ref{ScoobyDoo2}).  Then, under a few further conditions with only the faintest hint of quantum theory in them---for instance, that there should exist measurements on the ground for which, under appropriate conditions, one can have certainty for their outcomes---one immediately gets a significantly more restricted form for what becomes the analogue of Eq.~(\ref{ScoobyDoo}):
\begin{equation}
Q(D_j)=\left(\frac{1}{2} qd+1\right)\sum_{i=1}^{n} P(H_i) P(D_j|H_i) - \frac12 q\;.
\label{ExNihiloOmnia}
\end{equation}
Here, very interestingly, the parameters $q$ and $d$ can only take on integer values, $q=0,1,2,\ldots,\infty$ and $d=2,3,4,\ldots,\infty$, and $n=\frac12 qd(d-1) + d$.

The $q=2$ case can be identified with the quantum me\-chanical one we have seen before.  On the other hand, the $q=0$ case can be identified with the usual vision of the classical world:  A world where hypotheticals simply do not matter, for the world just ``is.''  In this case, an agent is well advised to take $Q(D_j)=P(D_j)$, meaning that there is no operational distinction between experiments ${\mathcal E}_1$ and ${\mathcal E}_2$ for her. It should not be forgotten however, that this rule, trivial though it looks, is still an addition to raw probability theory.  It is just one that meshes well with what had come to be expected by most classical physicists.  To put it yet another way, in the $q=0$ case, the agent says to herself that the fine details of her actions do not matter. This to some extent authorizes the view that observation is a passive process in principle---again the classical worldview.  Finally, the cases $q=1$ and $q=4$, though not classical, track still other structures that have been explored previously:  They correspond to what the Born rule would look like if alternate versions of quantum mechanics, those over real~\cite{Stueckelberg60} and quaternionic~\cite{AdlerBook} vector spaces, were expressed in the equivalent of SIC terms.\footnote{The equivalent of SICs (i.e., informationally complete sets of equiangular projection operators) certainly do {\it not\/} exist in general dimensions for the real-vector-space case---instead these structures only exist in a sparse set of dimensions, $d=2,3,7,23,\ldots\,$. With respect to the quaternionic theory, it appears from numerical work that they do not generally exist in that setting either \cite{Khatirinejad08}.  Complex quantum mechanics, like baby bear's possessions, appears to be just right.  This raises the possibility that {\it if\/} one had reason to think that the user's manual should be one of those three alternatives---real, complex and quaternionic---demanding the existence of SIC-type structures in all dimensions could narrow down the choice exactly to the complex case.}

Several years ago, R\"udiger Schack gave a talk on this material in Zurich, and Rob Spekkens asked, ``Why that particular choice for modification of the Law of Total Probability?''  Schack replied, ``If one is going to modify it in any way at all, this is the simplest modification one can imagine.''  The remarkable fact is that the simplest possible modification to the Law of Total Probability carries with it so much interlocking structure.

If all that you desire is a story that you can tell about the current quantum formalism, then all this business about SICs and probabilistic representations might be of little moment.  Of our fellow QBists, we know of one who likely doesn't care one way or the other about whether SICs exist.  Another would like to see a general proof come to pass, but is willing to believe that QBism can just as well be developed without them---i.e., they are not part of the essential philosophical ideas---and is always quick to make this point.  On the other hand, we two are inclined to believe that QBism will become stagnant in the way of {\it all other\/} quantum foundations programs without a deliberate effort to rebuild the formalism.  SICs might not be the only path to this goal~\cite{FQXi}, but they engage our attention for the following reason.

Formula (\ref{ExNihiloOmnia}) from the general setting indicates more strongly than ever that it is the role of {\it dimension\/} that is key to distilling the motif of our user's manual.  Quantum theory, seen as a normative addition to probability theory, is just one theory (the second rung above classical) along an infinite hierarchy.  What distinguishes the levels of this hierarchy is the strength $q$ with which dimension ``couples'' the two paths in our diagram of Figure 2.  It is the strength with which we are compelled to deviate from the Law of Total Probability when we transform our thoughts from the consequences of hypothetical actions upon a $d$'s worth of the world's stuff to the consequences of our actual ones.  Settling upon $q=2$ (i.e., settling upon quantum theory itself) sets the strength of the coupling, but the $d$ variable remains.  Different systems, different $d$, different deviations from a naive application of the Law of Total Probability.

In some way yet to be fully fleshed out, each quantum system seems to be a seat of active creativity and possibility, whose outward effect is as an ``agent of change'' for the parts of the world that come into contact with it.  Observer and system, ``agent and reagent,'' might be a way to put it.  Perhaps no metaphor is more pregnant for QBism's next move than this:  If a quantum system is comparable to a chemical reagent, then $d$ is comparable to a valence.  But valence for {\it what\/} more exactly?

\section{Mathematical Intermezzo: The Sporadic SICs}

Before we leap off into cosmological speculations, let us take a
moment to make a few things more concrete.  What does a SIC {\it look
  like,} anyway?  How do we write one out explicitly?  One of the
ongoing challenges of SIC research is that the solutions look so
complicated: As we go up in the dimension, the vectors soon take many
pages of computer printout.  Moreover, there are not obvious relations
between a SIC vector in one dimension and one in another, so finding
one solution doesn't help with finding the next.  It is only recently
that we have been able to tease out some interconnections happening beneath
the surface.  In this section, we'll explore the patterns that bind
sets of SIC solutions together~\cite{RCF-SIC, stacey-sporadic}.

A SIC is \emph{group covariant} if it can be constructed by starting
with a single vector (the \emph{fiducial}) and acting upon that vector
with the elements of some group.  All known SICs are group covariant,
although since group covariance simplifies the search process, this
could be a matter of the light being under the lamppost.  Furthermore,
in all known cases but one, that group is a \emph{Weyl--Heisenberg group.}
Working in dimension $d$, let $\omega_d = e^{2\pi i / d}$, and define
the shift and phase operators
\begin{equation}
X\ket{j} = \ket{j+1}\qquad \mbox{and}\qquad Z\ket{j} = \omega_d^j \ket{j}\;,
\end{equation}
where the shift is modulo $d$.  Products of powers of $X$ and $Z$,
together with dimension-dependent phase factors that we can neglect
for the present purposes, define the Weyl--Heisenberg group for dimension $d$.

A historical aside:  This group dates back to the earliest days of
quantum physics.  Note that the two operators $X$ and $Z$ just fail to
commute, doing so up to a phase factor:
\begin{equation}
ZX = \omega XZ.
\label{Weyl-E-Coyote}
\end{equation}
In July of 1925, Max Born had the idea that he could solve one of the equations in Heisenberg's seminal 1925 paper if he made an {\it ansatz\/} that the position observable $\hat{q}$ and momentum observable $\hat{p}$ satisfied the commutation relation \cite{Bernstein05,Fedak09}
\begin{equation}
\hat{q}\hat{p} - \hat{p}\hat{q}= i\hbar\hat{I}\;.
\label{Tombstone}
\end{equation}
Pascual Jordan later proved that Born's ansatz was the only one that could work, and the paper they wrote up together was received by {\sl Zeitschrift f\"ur Physik\/} on 27 September 1925.  It was titled ``On Quantum Mechanics.'' On the same day, Max Born received a letter from Hermann Weyl \cite{Scholz08} saying that a previous discussion they had earlier in the month inspired him to generalize Born's relation (\ref{Tombstone}) to Eq.~(\ref{Weyl-E-Coyote}).  Part of what pleased Weyl was that his generalization was not dependent upon infinite dimensional spaces---Weyl's relation would always have a solution in the complex matrices, finite and infinite dimensional.  And by that circumstance he declared to have a way of ``defining a general quantum system''---to each would be associated a ``phase space.''  In the discrete case, the points of the phase space would be associated with the operators
\begin{equation}
(m,n) \quad \longrightarrow \quad X^m Z^n\;.
\end{equation}
See Weyl's 1927 textbook \cite{Weyl27}, as well as Julian Schwinger's later
development and extension of the idea in \cite{Schwinger70}.  If a SIC
covariant under the Weyl--Heisenberg group always exists, then it
would mean that not only could a phase space be associated with a
quantum system abstractly, but that the points of the phase space very
directly correspond to the outcomes of a potential measurement.  It is
interesting to see how some of the earliest math in one of the
earliest formulations of the theory---Weyl's---came so close to what
we are working with today!

In $d = 2$, we can draw a SIC in the Bloch representation.  Any qubit
SIC forms a tetrahedron inscribed in the Bloch sphere~\cite{Renes04}.
One such tetrahedron is, in terms of the Pauli matrices,
\begin{equation}
\Pi_{s,r} = \frac{1}{2}\left(I + \frac{1}{\sqrt{3}}
(s\sigma_x + r\sigma_y + sr\sigma_z)\right),
\end{equation}
where the sign variables $s$ and $r$ take the values $\pm 1$.  The outcome probabilities
for a state $\ket{\psi} = \alpha\ket{0} + \beta\ket{1}$ are given by
the Born rule, $p(s,r) = \tr(\Pi_{s,r} \ket{\psi}\bra{\psi})/2$.
Explicitly,
\begin{equation}
p(s,r) = \frac{1}{4}
 + \frac{\sqrt{3}}{12}sr
   \left(\left|\alpha\right|^2 - \left|\beta\right|^2\right)
 + \frac{\sqrt{3}}{6}\hbox{Re}
   \left[\alpha\beta^*(s + ir)\right].
\end{equation}
This expression simplifies in terms of the Cartesian coordinates
$(x,y,z)$ of points on the Bloch sphere:
\begin{equation}
p(s,r) = \frac{1}{4}
 + \frac{\sqrt{3}}{12}
   \left(sx + ry + srz\right).
\end{equation}

Two SICs in higher dimensions will be important for our purposes.
First is the \emph{Hesse SIC} in $d = 3$, constructed by applying the
Weyl--Heisenberg group to the fiducial
\begin{equation}
\ket{\psi_0^{\rm (Hesse)}}
 = \frac{1}{\sqrt{2}} (0, 1, -1)^{\rm T}.
\label{eq:hesse-fiducial}
\end{equation}
Second is the \emph{Hoggar SIC} in $d = 8$.  We have multiple choices
of fiducial in this case, but they all yield structures that are
equivalent up to unitary or antiunitary transformations, so for
brevity we speak of ``the'' Hoggar SIC~\cite{zhu-thesis}.  One such
fiducial~\cite{Szymusiak2015, stacey-hoggar} is
\begin{equation}
\ket{\psi_0^{\rm (Hoggar)}} \propto (-1+2i, 1, 1, 1,
                      1, 1, 1, 1)^{\rm T}.
\label{eq:hoggar-fiducial}
\end{equation}
The Hoggar SIC is the \emph{only} known case where the group that
constructs the SIC from the fiducial is not the Weyl--Heisenberg group
for $d$ dimensions itself~\cite{zhu-thesis}.  It is, however, of a
related kind: It is the tensor product of three copies of the qubit
Weyl--Heisenberg group.  The more we poke at the Hoggar SIC, the more
odd and unusual things turn up about it~\cite{stacey-hoggar}.  We'll
take a look at one of them in this section.

The SICs in dimensions 2 and 3, as well as the Hoggar SIC in dimension
8, stand apart in some respects from the other known
solutions~\cite{RCF-SIC, stacey-qutrit}.  Recently, Appleby \emph{et
  al.}~\cite{RCF-SIC} found a link between SICs and \emph{algebraic
  number theory.}  Their results apply to Weyl--Heisenberg SICs in
dimensions 4 and larger.  The SICs in dimensions 2 and 3, as well as
the Hoggar SIC, fall outside of this category.  Either their
dimensions are too small, or (in the case of the Hoggar SIC) they have
the wrong symmetry group.  We can think of them as the \emph{sporadic
  SICs.}

\begin{figure*}
\begin{center}
\includegraphics[width=12cm]{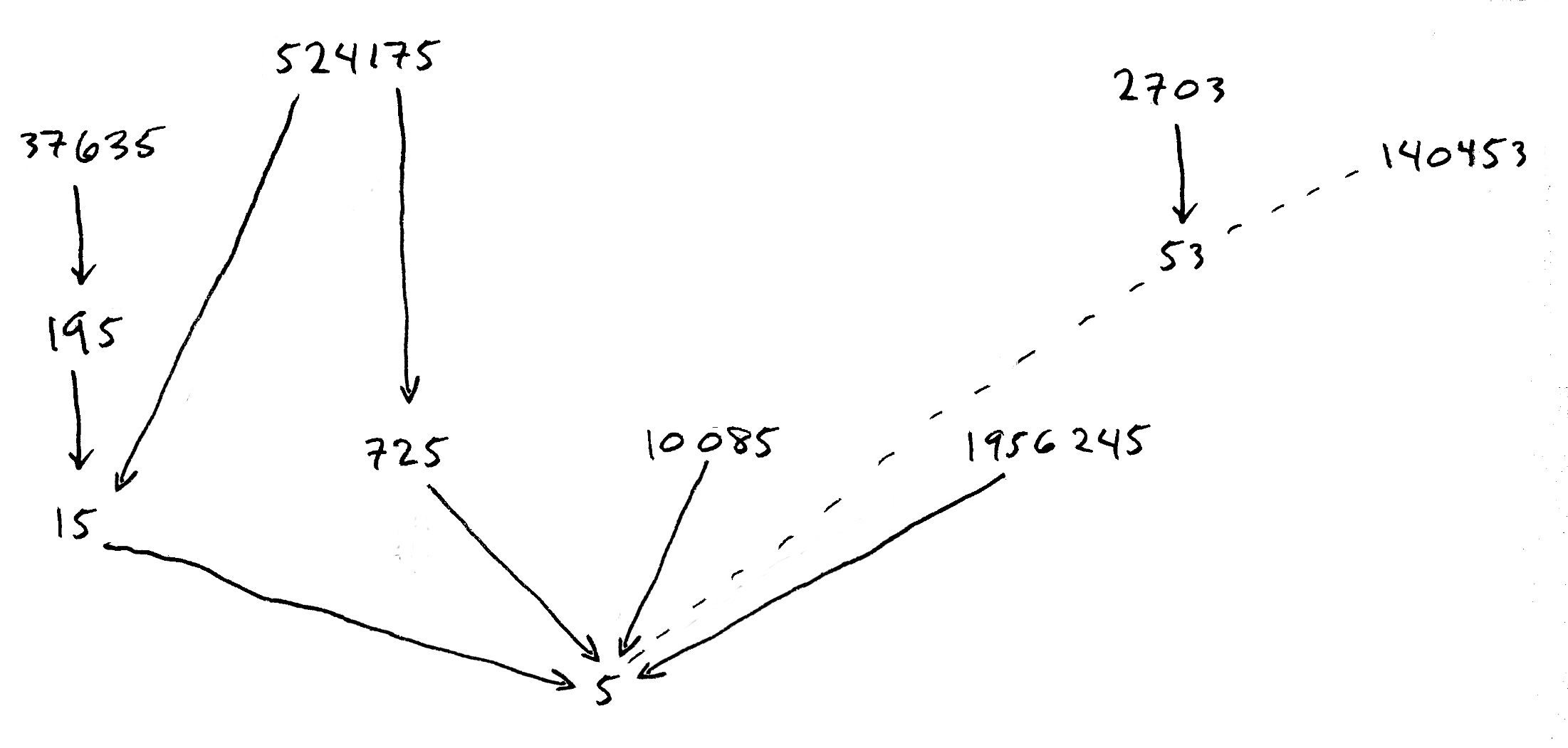}
\end{center}
\caption{\label{fig:dimension-tower} The lowest levels of a dimension
  tower.  Conjecturally, Weyl--Heisenberg SICs in these dimensions are
  related by way of algebraic number theory~\cite{RCF-SIC}.  The
  arithmetical meaning of the arrows is left as an exercise to the
  interested reader.}
\end{figure*}

First, let us sketch the picture for the SICs studied by Appleby
\emph{et al.}  The pattern, which is just beginning to come clear, is
a story about {\it number fields.}  To a physicist, a ``field'' means
something like the electric field, but to a number theorist, a field
is a set of numbers where addition and multiplication can both be
done, and where both additive and multiplicative inverses exist, and
everything plays together nicely.  The real numbers $\mathbb{R}$
constitute a field, as do the rational numbers $\mathbb{Q}$ within
them.  We can build up a field by starting with some base, like the
rationals, and augmenting it with a new element.  For example, let us
invent a number ``$\sqrt{3}$,'' about which all we know is that it is
a positive number that solves the equation $x^2 - 3 = 0$.  We then
consider all the numbers of the form $a + b\sqrt{3}$, where $a$ and
$b$ are rational.  This set is a new field, $\mathbb{Q}$ extended by
the new ingredient $\sqrt{3}$, which we write as
$\mathbb{Q}(\sqrt{3})$.

The connection between SICs and number fields happens when we take the
inner-product conditiont that defines a SIC,
\begin{equation}
\left|\braket{\psi_j}{\psi_k}\right|^2 = \frac{1}{d+1},
\end{equation}
and we leave off the magnitude-squared step:
\begin{equation}
\braket{\psi_j}{\psi_k} = \frac{e^{i\theta_{jk}}}{\sqrt{d+1}}.
\end{equation}
The phase factors $e^{i\theta_{jk}}$ turn out to live within very special
number fields, and they are particularly special numbers within those
fields.  They are {\it units of ray class fields or extensions
  thereof}---as Bengtsson quips, ``These words carry deep meaning for
algebraic number theorists''~\cite{Bengtsson16}.  They mean that we
are knocking on the door of {\it Hilbert's twelfth problem,} one of
the last remaining unsolved puzzles on history's most influential list
of mathematical challenges \cite{Schappacher98}.  And, remember, we got here because we
were trying to find a better way to talk about probability in quantum
mechanics!

Part of this still-emerging story~\cite{RCF-SIC} is that SICs in
different dimensions are related in a hidden way because their number
fields are related.  The SIC phase factors live in fields that are
extensions twice over of the rationals.  That is, for a
Weyl--Heisenberg SIC in dimension $d$, the phase factors
$e^{i\theta_{jk}}$ make their home in an extension of
$\mathbb{Q}\left(\sqrt{(d-3)(d+1)}\right)$.  Because we can factor perfect
squares out from under the radical sign, different values of $d$ can yield
the same extension of $\mathbb{Q}$.  This has led to the image of a {\it dimension
  tower,} an infinite sequence $\{d_1, d_2, d_3, \ldots\}$ where each
$d_j$ follows neatly from those before, and the number theory tells us
how to build a SIC in each $d_j$.  But this remains in the realm of
conjecture.

And what of the sporadic SICs?  Quite unexpectedly, the qubit SICs, the Hesse SIC and the Hoggar SIC
also connect to a subject in pure mathematics that has gone mostly un-utilized in physics.
Specifically, their symmetries are linked with a {\it lattice of integers\/} in
the set of numbers known as the {\it octonions}~\cite{stacey-sporadic}.
Physics students grow familiar with the complex numbers $\mathbb{C}$
by repeated exposure, internalizing the image of a number plane that
extends out on either side of the number line.  The quaternions
(usually written with an $\mathbb{H}$ for William Hamilton) and the octonions (denoted $\mathbb{O}$) arise when one tries to repeat
this dimension-doubling stunt, from two dimensions to four and then to
eight.  It so happens that familiar properties of arithmetic are lost
with each repetition: Multiplication of quaternions is not
commutative, but it is still associative.  And multiplication of
octonions is not even associative!  Nineteenth-century physics made much use
of quaternions to study 3D rotations, and they still find application
in geometry, for example in computer graphics.  Octonions are less
familiar still, and are perhaps best known for their relations to {\it
  exceptional structures\/} in mathematics.  John Baez
observed~\cite{baez-plus},
\begin{quote}
\noindent  Often you can classify some sort of gizmo, and you get
  a beautiful systematic list, but also some number of
  exceptions. Nine times out of 10 those exceptions are related to the
  octonions.
\end{quote}
It so transpires that this applies to SICs, too.  Moreover, the link with octonionic integers connects the sporadic SICs to the problem of {\it sphere packing,} that is, the question of how to fit Euclidean spheres of arbitrary dimension together in the most efficient way~\cite{viazovska2016}.

A few years ago, one of the authors (BCS) was attending an
interdisciplinary workshop and, over lunch, fell into conversation
with a mathematician who had that morning lectured on
higher-dimensional sphere packing.  After a little while, the
mathematician asked, ``And what are you working on?''

``Too many different things---but one problem has the same feel as
sphere packing, because solutions in one dimension don't seem to tell
you about solutions in others.  I guess the math people know it as the
problem of `complex equiangular lines'.''

``Ah! SICs!  You know, when I first heard about that conjecture, I
thought I could just sit down and solve it.  But that didn't quite
happen.''

As the mathematical properties of SIC dimensionalities grow more
intriguing, we are led back to the question of what the dimension of a
Hilbert space means physically.

\section{Hilbert-Space Dimension as a Universal Capacity}

\label{HSpaceDim}

\begin{flushright}
\baselineskip=13pt
\parbox{2.8in}{\baselineskip=13pt
\small It is entirely possible to conceive of a world composed of individual atoms, each as different from one another as one organism is from the next.}
\medskip\\
\small --- John Dupr\'e
\end{flushright}

\begin{figure}
\begin{center}
\includegraphics[height=2.0in]{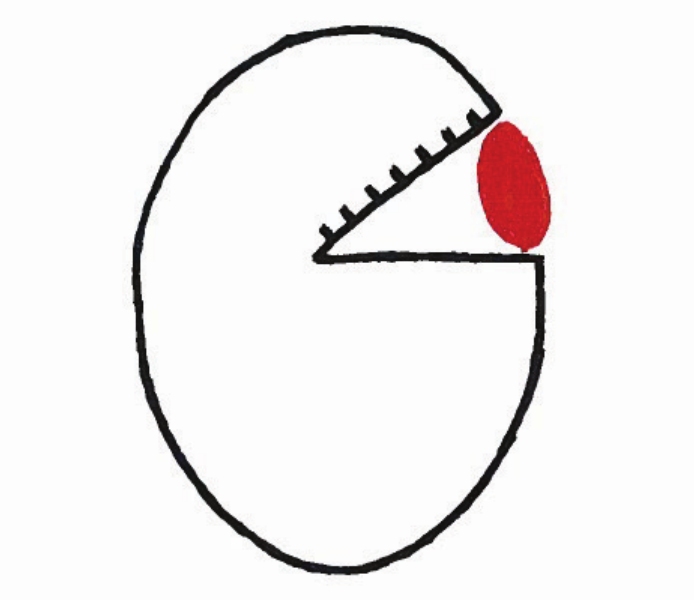}
\bigskip\caption{\protect {\bf Sarcasm.}~In a lecture bottlenecked by repeated accusations of QBism's solipsism, the authors sometimes use the following technique to move things along. Referring to the previous Figure 1, one asks the stubborn accuser, ``What about this diagram do you {\it not\/} get?  It shows an agent and a physical system external to him.  It says that a quantum state is a {\it state of belief\/} about what will come about as a consequence of his actions upon the system.   The quantum state is not a state of nature, but so what?  There is an agent with his belief; there is a system that is not part of him; and there is something that really, eventually comes about---it is called {\it the outcome}.  No agent, no outcome for sure, but that's not solipsism:  For, no system, no outcome either!  A quantum measurement without an external system participating would be like the sound of one hand clapping, a Zen koan.  If we were really expressing solipsism, wouldn't a diagram like the one above be more appropriate?  A big eyeball surveying nothing.  Now there's really no external system and nothing to act upon. {\it That's\/} solipsism.''}
\label{AllForNorsen}
\end{center}
\end{figure}

A common accusation heard by the QBist\footnote{For perhaps the loudest, see Ref.\ \cite{Norsen}.} is that the view leads straight away to solipsism, ``the belief that all reality is just one's  imagining of reality, and that one's self is the only thing that exists.''\footnote{This is the definition of {\sl The American Heritage New Dictionary of Cultural Literacy}, Third Edition (2005).  {\sl Encyclopedia Brittanica\/} (2008) expands, ``in philosophy \ldots\ the extreme form of subjective idealism that denies that the human mind has any valid ground for believing in the existence of anything but itself. The British idealist F.~H. Bradley, in {\sl Appearance and Reality\/} (1897), characterized the solipsistic view as follows: `I cannot transcend experience, and experience is my experience. From this it follows that nothing beyond myself exists; for what is experience is its (the self's) states.'$\,$''}  The accusation goes that, if a quantum state $|\psi\rangle$ only represents the degrees of belief held by some agent---say, the one portrayed in Figure 1---then the agent's beliefs must be the source of the universe.  The universe could not exist without him: This being such a ridiculous idea, QBism is dismissed out of hand, {\it reductio ad absurdum}.  It is so hard for the QBist to understand how anyone could think this (it being the antithesis of everything in his worldview) that a little of our own Latin comes to mind: {\it non sequitur}.  See Fig.~\ref{AllForNorsen}.

A fairer-minded assessment is that the accusation springs from our opponents ``hearing'' much of what we do say, but interpreting it in terms drawn from a particular conception of what physical theories {\it always ought to be}:  Attempts to directly represent (map, picture, copy, correspond to, correlate with) the {\it universe}---with ``universe'' here thought of in its totality as a pre-existing, static system; an unchanging, monistic something that just {\it is}.  From such a ``representationalist'' point of view, {\it if\/} {\bf a)} quantum theory is a proper physical theory, {\bf b)} its essential theoretical objects are quantum states, and {\bf c)} quantum states are states of belief, {\it then\/} the universe that ``just is'' corresponds to a state of belief.  This chain of deduction is logical and clear, but completely misguided.

QBism sidesteps the poisoned dart, as the previous sections have tried to convey, by asserting that quantum theory is just not a physical theory in the sense the accusers want it to be.  Rather it is an addition to personal, Bayesian, normative probability theory.  Its normative rules for connecting probabilities (personal judgments) were developed in light of the {\it character of the world}, but there is no sense in which the quantum state itself represents (pictures, copies, corresponds to, correlates with) a part or a whole of the external world, much less a world that {\it just is}.  In fact the very character of the theory seems to point to the inadequacy of the representationalist program when attempted on the particular world we live in.

There are no lofty philosophical arguments here that re\-presentationalism must be wrong always and in all possible worlds (perhaps because of some internal inconsistency\footnote{As, e.g., Rorty \cite{Rorty91} might try to argue.}).  Representationalism may well be true in this or that setting---we take no stand on the matter.  We only know that for nearly 90 years quantum theory has been actively resistant to representationalist efforts on {\it its\/} behalf.  This suggests that it might be worth exploring some philosophies upon which physics rarely sets foot.  Physics of course should never be constrained by any one philosophy (history shows it nearly always lethal), but it does not hurt to get ideas and insights from every source one can.  If one were to sweep the philosophical literature for schools of thought representative of what QBism actually is about, it is not solipsism one will find, but nonreductionism \cite{Dupre93,Cartwright99}, (radical) metaphysical pluralism \cite{James96a,Wahl25}, empiricism \cite{James40,James96b}, indeterminism and meliorism\footnote{Strictly speaking, meliorism is the doctrine ``that humans can, through their interference with processes that would otherwise be natural, produce an outcome which is an improvement over the aforementioned natural one.''  But we would be reluctant to take a stand on what ``improvement'' really means.  So said, all we mean in the present essay by meliorism is that the world before the agent is malleable to some extent---that his actions really can change it.  Adam said to God, ``I want the ability to write messages onto the world.''  God replied, ``You ask much of me.  If you want to write upon the world, it cannot be so rigid a thing as I had originally intended.  The world would have to have some malleability, with enough looseness for you to write upon its properties.  It will make your world more unpredictable than it would have been---I may not be able to warn you about impending dangers like droughts and hurricanes as effectively as I could have---but I can make it such if you want.'' And with that Adam brought all host of uncertainties to his life, but he gained a world where his deeds and actions mattered.} \cite{James1884}, and above all pragmatism \cite{Menand01,Thayer81}.

A form of nonreductionism can already be seen in play in our answer to whether the notion of agent should be derivable from the quantum formalism itself.  We say that it cannot be and it should not be, and to believe otherwise is to misunderstand the subject matter of quantum theory.  But nonreductionism also goes hand in hand with the idea that there is real particularity and ``interiority'' in the world.  Think again of the ``I-I-me-me mine'' feature that shields QBism from inconsistency in the ``Wigner's friend'' scenario.
When Wigner turns his back to his friend's interaction with the system, that piece of reality is hermetically sealed from him.  That phenomenon has an inside, a vitality that he takes no part in until he again interacts with one or both relevant pieces of it.  {\it With respect to Wigner}, it is a bit like a universe unto itself.

If one seeks the essence of indeterminism in quantum mechanics, there may be no example more directly illustrative of it than ``Wigner's friend.''  For it expresses to a tee William James's notion of indeterminism \cite{James1884}:
\begin{quote}
\noindent [Chance] is a purely negative and relative term, giving us no
information about that of which it is predicated, except that it
happens to be disconnected with something else---not controlled,
secured, or necessitated by other things in advance of its own actual
presence. \ldots\ What I say is that it tells us
nothing about what a thing may be in itself to call it ``chance.'' \ldots\
All you mean by calling it ``chance'' is that this is not guaranteed,
that it may also fall out otherwise. For the system of other things
has no positive hold on the chance-thing. Its origin is in a certain
fashion negative: it escapes, and says, Hands off!\ coming, when it
comes, as a free gift, or not at all.

This negativeness, however, and this opacity of the chance-thing when
thus considered {\it ab extra}, or from the point of view of previous
things or distant things, do not preclude its having any amount of
positiveness and luminosity from within, and at its own place and
moment. All that its chance-character asserts about it is that there
is something in it really of its own, something that is not the
unconditional property of the whole. If the whole wants this
property, the whole must wait till it can get it, if it be a matter
of chance. That the universe may actually be a sort of joint-stock
society of this sort, in which the sharers have both limited
liabilities and limited powers, is of course a simple and conceivable
notion.
\end{quote}
And once again \cite{JamesOSH},
\begin{quote}
\noindent Why may not the world be a sort of republican banquet of this sort, where all the qualities of being respect one another's personal sacredness, yet sit at the common table of space and time?

To me this view seems deeply probable.  Things cohere, but the act of cohesion itself implies but few conditions, and leaves the rest of their qualifications indeterminate.  As the first three notes of a tune comport many endings, all melodious, but the tune is not named till a particular ending has actually come,---so the parts actually known of the universe may comport many ideally possible complements. But as the facts are not the complements, so the knowledge of the one is not the knowledge of the other in anything but the few necessary elements of which all must partake in order to be together at all. Why, if one act of knowledge could from one point take in the total perspective, with all mere possibilities abolished, should there ever have been anything more than that act? Why duplicate it by the tedious unrolling, inch by inch, of the foredone reality? No answer seems possible. On the other hand, if we stipulate only a partial community of partially independent powers, we see perfectly why no one part controls the whole view, but each detail must come and be actually given, before, in any special sense, it can be said to be determined at all.  This is the moral view, the view that gives to other powers the same freedom it would have itself.
\end{quote}

The train of logic back to QBism is this.  If James and our analysis of ``Wigner's friend'' are right, the universe is not {\it one\/} in a very rigid sense, but rather more truly a pluriverse.\footnote{The term ``pluriverse'' is again a Jamesian one.  He used it interchangeably with the word ``multiverse,'' which he also invented \cite{JamesMultiverse}.  Unfortunately the latter has been coopted by the Everettian movement for their own---in the end monistic---purposes:  ``The world is one; it {\it is\/} the deterministically evolving universal quantum state, the `multiverse'.'' Too bad.  Multiverse is a tempting word, but we stick with pluriverse to avoid any confusion with the Everettian usage.}  To get some sense of what this can mean, it is useful to start by thinking about what it is not.  A good example can be found by taking a solution to the vacuum Maxwell equations in some extended region of spacetime.  Focus on a compact subregion and try to conceptually delete the solution within it, reconstructing it with some new set of values.  It can't be done.  The fields outside the region (including the boundary) uniquely determine the fields inside it.  The interior of the region has no identity but that dictated by the rest of the world---it has no ``interiority'' of its own.  The pluriverse conception says we'll have none of that.  And so, for any agent immersed in this world there will always be uncertainty for what will happen upon his encounters with it.  To wit, where there is uncertainty there should be Bayesian probabilities, and so on and so on until much of the story we have already told.

What all this hints is that for QBism the proper way to think of our world is as the empiricist or the radical metaphysical pluralist does.  Let us launch into making this clearer, for that process more than anything will explain how QBism hopes to interpret Hilbert-space dimension.

The metaphysics of empiricism can be put like this.  Every\-thing experienced, everything experienceable, has no less an ontological status than anything else.  A child awakens in the middle of the night frightened that there is a monster under her bed, one soon to reach up and steal her arm---that {\it we-would-call-imaginary\/} experience has no less a hold on onticity than a Higgs-boson detection event at the LHC, or the minuscule wobbles at LIGO that shook the scientific world.  They are of equal status from this point of view---they are equal elements in the filling out and making of reality.  There is indeed no doubt that we should call the child's experience {\it imaginary.}  That, however, is a statement about the experience's meaning and interpretation, not its existence.  The ex\-pe\-rience {\it as it is\/} exists, period. It is what it is.  Like the biblical burning bush, each ex\-pe\-rience declares, ``I am that I am.''  Most likely in the present example, the experience will be a little piece of the universe isolated, on its own, and of no great consequence.  But one never knows until all future plays out.  Some lucky dreams have built nations.
Maybe the same is true of some lucky Higgs-boson events.  Most though, surely, will be of the more minor fabric of existence.  All in all, the world of the empiricist is not a sparse world like the world of Democritus ({\it nothing but\/} atom and void) or Einstein ({\it nothing but\/} unchanging spacetime manifold equipped with this or that field), but a world overflowingly full of variety---a world whose details are beyond anything grammatical (rule-bound) expression can articulate.

Yet this is no statement that physics should give up, or that physics has no real role in coming to grips with the world.  It is only a statement that physics should better understand its function.  What is being aimed for here finds its crispest, clearest contrast in a statement Richard Feynman once made \cite{Feynman65}:
\begin{quote}
\noindent If, in some cataclysm, all of scientific knowledge were to be destroyed, and only one sentence passed on to the next generation of creatures, what statement would contain the most information in the fewest words?  I believe it is the atomic hypothesis (or the atomic fact) that all things are made of atoms---little particles that move around in perpetual motion, attracting each other when they are a little distance apart, but repelling upon being squeezed into one another. \ldots

Everything is made of atoms.  That is the key hypothesis.
\end{quote}
The issue for QBism hangs on the imagery that usually lies behind the phrase ``everything is made of.''  William James called it the great original sin of the rationalistic mind \cite{James97}:
\begin{quote}
\noindent Let me give the name of `vicious abstractionism' to a way of using concepts which may be thus described: We conceive a concrete situation by singling out some salient or important feature in it, and classing it under that; then, instead of adding to its previous characters all the positive consequences which the new way of conceiving it may bring, we proceed to use our concept privatively; reducing the originally rich phenomenon to the naked suggestions of that name abstractly taken, treating it as a case of `nothing but' that, concept, and acting as if all the other characters from out of which the concept is abstracted were expunged. Abstraction, functioning in this way, becomes a means of arrest far more than a means of advance in thought. It mutilates things; it creates difficulties and finds impossibilities; and more than half the trouble that metaphysicians and logicians give themselves over the paradoxes and dialectic puzzles of the universe may, I am convinced, be traced to this relatively simple source. {\it The viciously privative employment of abstract characters and class names\/} is, I am persuaded, one of the great original sins of the rationalistic
mind.
\end{quote}
What is being realized through QBism's peculiar way of looking at things is that physics {\it actually can be done\/} without any accompanying vicious abstractionism.  You do physics as you have always done it, but you throw away the idea ``everything is made of [Essence X]'' before even starting.

Physics---in the right mindset---is not about identifying the bricks with which nature is made, but about identifying what is {\it common to\/} the largest range of phenomena it can get its hands on.  The idea is not difficult once one gets used to thinking in these terms.  Carbon?  The old answer would go that it is {\it nothing but\/} a building block that combines with other elements according to the following rules, blah, blah, blah.  The new answer is that carbon is a {\it characteristic\/} common to diamonds, pencil leads, deoxyribonucleic acid, burnt pancakes, the space between stars, the emissions of Ford pick-up trucks, and so on---the list is as unending as the world is itself.  For, carbon is also a characteristic common to this diamond and this diamond and this diamond and this.  But a flawless diamond and a purified zirconium crystal, no matter how carefully crafted, have no such characteristic in common:  Carbon is not a {\it universal\/} characteristic of all phenomena.  The aim of physics is to find characteristics that apply to as much of the world in its varied fullness as possible.  However, those common characteristics are hardly what the world is made of---the world instead is made of this and this and this.  The world is constructed of every particular there is and every way of carving up every particular there is.

An unparalleled example of how physics operates in such a world can be found by looking to Newton's law of universal gravitation.  What did Newton really find?  Would he be considered a great physicist in this day when every news magazine presents the most cherished goal of physics to be a Theory of Everything?  For the law of universal gravitation is hardly that!  Instead, it {\it merely\/} says that every body in the universe tries to accelerate every other body toward itself at a rate proportional to its own mass and inversely proportional to the squared distance between them.  Beyond that, the law says nothing else particular of ob\-jects, and it would have been a rare thinker in Newton's time, if any at all, who would have imagined that all the complexities of the world could be derived from that limited law.  Yet there is no doubt that Newton was one of the greatest physicists of all time.  He did not give a theory of everything, but a Theory of One Aspect of Everything.  And only the tiniest fraction of physicists of any variety, much less the TOE-seeking variety, have ever worn a badge of that more modest kind.  It is as H.~C. von Baeyer wrote in one of his books \cite{Baeyer09},
\begin{quote}
Great revolutionaries don't stop at half measures if they can go all the way.  For Newton this meant an almost unimaginable widening of the scope of his new-found law.  Not only Earth, Sun, and planets attract objects in their vicinity, he conjectured, but all objects, no matter how large or small, attract all other objects, no matter how far distant.  It was a proposition of almost reckless boldness, and it changed the way we perceive the world.
\end{quote}
Finding a theory of ``merely'' one aspect of everything is hardly something to be ashamed of:  It is the loftiest achievement physics can have in a living, breathing non\-re\-duc\-tionist world.

Which leads us back to Hilbert space.  Quantum theory---that user's manual for decision-making agents immersed in a world of {\it some\/} yet to be fully identified character---makes a statement about the world to the extent that it identifies a quality common to all the world's pieces.  QBism says the quantum state is not one of those qualities.  But of Hilbert spaces themselves, particularly their distinguishing characteristic one from the other, {\it dimension},\footnote{Hardy \cite{Hardy01a,Hardy01b} and Daki\'c and Brukner \cite{Dakic09} are examples of foundational efforts that also emphasize this quantum analogue to what E\"otv\"os tested on platinum and copper \cite{Fuchs04b}.  Hardy put it this way in one of his axioms, ``There exist systems for which $N = 1, 2, \cdots$, and, furthermore, all systems of dimension $N$, or systems of higher dimension but where the state is constrained to an $N$ dimensional subspace, have the same properties.''} QBism car\-ries no such grudge.  Dimension is something one posits for a body or a piece of the world, much like one posits a mass for it in the Newtonian theory.  Dimension is something a body holds all by itself, regardless of what an agent thinks of it.

That this is so can be seen already from reasons internal to the
theory.  Just think of all the arguments rounded up for making the
case that quantum states should be interpreted as of the character of
Bayesian degrees of belief.  None of these work for Hilbert-space
dimension.  Take one example, an old favorite---Einstein's argument
about conditioning quantum states from afar.  In Section
\ref{BellTheorem} of this paper we repeated the argument verbatim, but
it is relevant to note that before Einstein could write down his
$\psi_{12}$, he would have had to associate some Hilbert spaces
${\mathcal H}_1$ and ${\mathcal H}_2$ with $S_1$ and $S_2$ and take
their tensor product ${\mathcal H}_1\otimes{\mathcal H}_2$.  Suppose
the dimensionalities of these spaces to be $d_1$ and $d_2$,
respectively.  The question is, is there anything similar to
Einstein's argument for changing the value of $d_2$ from a distance?
There isn't.  $\psi_2$ may be forced into this or that subspace by
choosing the appropriate measurement on $S_1$, but there is no
question of the whole Hilbert space ${\mathcal H}_2$ remaining intact.
When it is time to measure $S_2$ itself, one will still have the full
arsenal of quantum measurements appropriate to a Hilbert space of
dimension $d_2$ to choose from---none of those fall by the wayside.
In Einstein's terms, $d_2$ is part of the ``real factual situation''
of $S_2$.\footnote{Take a coin, and imagine flipping it. We generally
  write down a (subjective) probability distribution over two
  outcomes to capture our degrees of belief of which way the flip will
  go. But of course it is a judgement call that it can only go two
  ways. Steven van Enk would say it could always land on its side; so
  he would always write down a probability distribution over three
  outcomes.  If one takes $(p_0 , p_1 )$ as a subjective assignment,
  the number 2 is objective with respect to it: It is something we imagine
  or hypothesize about the coin. If one takes the $(p_0, p_1, p_2 )$
  as a subjective assignment, then the number 3 is objective with
  respect to it: It will fall one of three ways regardless of what we
  believe about which of the three ways it will fall. So
  objectivity/subjectivity comes in layers. We call something
  objective, and then make probability assignments in the subjective
  layer above it. But of course, the first ``calling something
  objective'' has a personal element in itself.  Recently, techniques
  have started to become available to ``test'' the supposition of a
  dimension against one's broader mesh of beliefs; see
  \cite{Brunner08,Wehner08,Wolf09}.}

The claim here is that quantum mechanics, when it came into existence, implicitly recognized a previously unnoticed capacity inherent in all matter---call it {\it quantum dimension}.  In one manifestation, it is the fuel upon which quantum computation runs \cite{Fuchs04b,BlumeKohout02}.  In another it is the raw irritability of a quantum system to being eavesdropped upon \cite{Fuchs03,Cerf02}.  In Eqs.~(\ref{ScoobyDoo}) and (\ref{ScoobyDoo2}) it was a measure of deviation from the Law of Total Probability induced by hypothetical thinking.  And in a farther-fetched scenario to which we will come back, its logarithm {\it might\/} just manifest itself as the squared gravitational mass of a Schwarzschild black hole \cite{Horowitz04,Gottesman04}.

When quantum mechanics was discovered, something was {\it added\/} to matter in our conception of it.  Think of the apple that inspired Newton to his law.  With its discovery the color, taste, and texture of the apple didn't disappear; the law of universal gravitation didn't reduce the apple privatively to {\it just\/} gravitational mass. Instead, the apple was at least everything it was before, but afterward even more---for instance, it became known to have something in common with the moon.  A modern-day Cavendish would be able to literally measure the fur\-ther attraction an apple imparts to a child already hungry to pick it from the tree.  So similarly with Hil\-bert-space dimension.  Those diamonds we have already used to illustrate the idea of nonreductionism, in very careful conditions, could be used as components in a quantum computer \cite{Prawer08}.  Diamonds have among their many properties something not envisioned before quantum mechanics---that they could be a source of relatively accessible Hilbert space dimension and as such have this much in common with any number of other proposed implementations of quantum computing.  Diamonds not only have something in common with the moon, but now with the ion-trap quantum-computer prototypes around the world.

Diamondness is not something to be derived from quantum mechanics.  It is that quantum mechanics is something we {\it add\/} to the repertoire of things we already say of diamonds, to the things we do with them and the ways we admire them.  This is a very powerful realization:  For diamonds already valuable, become ever more so as their qualities compound.  And saying more of them, not less of them as is the goal of all reductionism, has the power to suggest all kinds of variations on the theme.  For instance, thinking in quantum mechanical terms might suggest a technique for making ``purer diamonds''---though to an empiricist this phrase means not at all what it means to a reductionist.  It means that these similar things called diamonds can suggest exotic variations of the original objects with various pinpointed properties this way or that.  Purer diamond is not {\it more\/} of what it already was in nature.  It is a new species, with traits of its parents to be sure, but nonetheless stand-alone, like a new breed of dog.

To put it still differently, and now in the metaphor of music, a jazz musician might declare that a tune once heard thereafter plays its most crucial role as a substrate for something new. It is the fleeting solid ground upon which something new can be born.  The nine tracks titled {\sl Salt Peanuts\/} in CAF's mp3 player\footnote{Charlie Parker, Dizzy Gillespie, Charlie Parker, Charlie Parker, Charlie Parker, Joshua Redman, Miles Davis Quintet, Arturo Sandoval, ``The Quintet'' (Massey Hall, 1953).} are moments of novelty in the universe never to be recreated.  So of diamonds, and so of all this quantum world.  Or at least that is the path QBism seems to indicate.\footnote{A nice {\it logical\/} argument for this can be found in \cite{Ojima92}.}

To the reductionist, of course, this seems exactly backwards.  But then, it is the reductionist who must live with a seemingly infinite supply of conundrums arising from quantum mechanics.  It is the reductionist who must live in a state of arrest, rather than moving on to the next stage of physics.  Take a problem that has been a large theme of the quantum foundations meetings for the last 30 years.  To put it in a commonly heard question, ``Why does the world look classical if it actually operates according to quantum mechanics?''  The touted mystery is that we never ``see'' quantum superposition and entanglement in our everyday experience.  But have you ever seen a probability distribution sitting in front of you?  Probabilities in personalist Bayes\-ian\-ism are not the sorts of things that can be seen; they are the things that are thought.  It is {\it events\/} that are seen.

The real issue is this.  The expectation of the quantum-to-classical transitionists\footnote{See \cite{Schlosshauer07,Schlosshauer08} for particularly clear discussions of the subject.} is that quantum theory is at the bottom of things, and ``the classical world of our experience'' is something to be derived out of it.  QBism says ``No.  Experience is neither classical nor quan\-tum.  Experience is experience with a richness that classical physics of any variety could not remotely grasp.''  Quantum mechanics is something put on top of raw, unreflected experience.  It is additive to it, suggesting wholly new types of experience, while never invalidating the old.  To the question, ``Why has no one ever {\it seen\/} superposition or entanglement in diamond before?,'' the QBist replies:  It is simply because before recent technologies and very controlled conditions, as well as lots of refined analysis and thinking, no one had ever mustered a mesh of beliefs relevant to such a range of interactions (factual and hypothetical) with diamonds.  No one had ever been in a position to adopt the extra normative constraints required by the Born rule.  For QBism, it is not the emergence of classicality that needs to be explained, but the emergence of our new ways of manipulating, controlling, and interacting with matter that do.

In this sense, QBism declares the quantum-to-classical research program unnecessary (and actually obstructive\footnote{Without an ontic understanding of quantum states, quantum operations, and unitary time evolutions---all of which QBism rejects~\cite{Fuchs02,RMP,Leifer06}---how can the project even get off the ground?  As one can ask of the Big Bang, ``What banged?,'' the QBist must ask, ``In those days of the world before agents using quantum theory, what decohered?''}) in a way not so dissimilar to the way Bohr's 1913 model of the hydrogen atom declared another research program unnecessary (and actually obstructive).  Before Bohr, everyone thought that the only thing that could count as an explanation of the hydrogen atom's stable spectrum was a mechanical model.  Bohr's great genius in comparison to all the other physicists of his day was in being the first to say, ``Enough!  I shall not give a mech\-anistic explanation for these spectra we see.  Here is a way to think of them with no mechanism.''  Researchers had wasted years seeking an unfulfillable vision of the world, and that certainly was an obstruction to science.

All is not lost, however, for the scores of decoherentists this policy would unforgivingly unemploy.  For it only suggests that they redirect their work to the opposite task.  The thing that needs insight is not the quantum-to-classical transition, but the classical-to-quantum!  The burning question for the QBist is how to model in Hilbert-space terms the common sorts of measurements we perform just by opening our eyes, cupping our ears, and extending our fingers.%

Take a professional baseball player watching a ball fly toward him:  He puts his whole life into when and how he should swing his bat.  But what does this mean in terms of the immense Hilbert space a quantum theoretical description would associate with the ball?  Surely the player has an intuitive sense of both the instantaneous position and instantaneous momentum of the baseball before he lays his swing into it---that's what ``keeping his eye on the ball'' means.  Indeed it is from this intuition that Newton was able to lay down his laws of classical mechanics.   Yet, what can it mean to say this given quantum theory's prohibition of simultaneously measuring complementary observables?  It means that whatever the baseball player is measuring, it ain't that---it ain't position {\it and\/} momentum as usually written in operator terms.  Instead, a quantum model of what he is doing would be some interesting, far-from-extremal {\it single\/} POVM---perhaps even one that takes into account some information that does not properly live within the formal structure of quantum theory (the larger arena that Howard Barnum calls ``meaty quantum physics'' \cite{Barnum10}). For instance, that an eigenvector $|i\rangle$ of some Hermitian operator, though identically orthogonal to fellow eigenvectors $|j\rangle$ and $|k\rangle$ in the Hilbert-space sense, might be {\it closer\/} in meaning to $|k\rangle$ than to $|j\rangle$ for some issue at hand.

So the question becomes how to take a given common-day measurement procedure and {\it add to it\/} a consistent quantum description?  The original procedure was stand alone---it can live without a quantum description of it---but if one wants to move it to a new level or new direction, having added a consistent quantum description will be most helpful to those ends.  Work along these lines is nascent, but already some excellent examples exist~\cite{Kofler08}.  Of course, unconsciously it is what has been happening since the founding days of quantum mechanics.  Here, we find an affinity with a comment of John Stuart Bell, one buried in a letter to Rudolf Peierls and almost lost to physics history~\cite{Mermin14}:
\begin{quote}
 \noindent I have the impression as I write this, that a moment ago I heard the
bell of the tea trolley. But I am not sure because I was concentrating
on what I was writing. [\ldots\!] The ideal instantaneous measurements of the
textbooks are not precisely realized anywhere anytime, and more or less
realized, more or less all the time, more or less everywhere.
\end{quote}
QBism thinks of the textbook ``ideal instantaneous measurements'' as on the same continuum as listening for the tea trolley.  But what POVM elements should one write for the latter?  Only time, actively spent in new research, will tell.

The important question is how matter can be coaxed to do new things.  It is in the ways the world yields to our desires, and the ways it refuses to, that we learn the depths of its character.
\begin{verse}
I give you an object of this much gravitational mass.  What can you do with it?  What can you not?  And when you are not about, what does it cause?
\\
I give you an object of this much quantum dim\-ension.  What can you do with it?  What can you not?  And when you are not about, what does it cause?
\end{verse}

If taken seriously what do these questions imply by their very existence?  That they should have meaningful answers!  Here is one example.  A knee-jerk reaction in many physicists upon hearing these things is to declare that dimension as a capacity collapses to a triviality as soon as it is spoken.  ``All real-world systems possess infinite-dimensional Hilbert spaces.  And it doesn't take quantum field theory to be completely correct to make that true; a simple one-dimensional harmonic oscillator will do.  It has an infinite-dimensional Hilbert space.''  But maybe not.  Maybe no real-world quantum system has that much oomph.  Just as one can treat the Earth's inertial mass as infinite for many a freshman mechanics problem, or a heat bath as infinite for many a thermodynamical one, maybe this is all that has ever been going on with infinite-dimensional Hilbert spaces.  It is a useful artifice when a problem can be economically handled with a differential equation.  (Ask Schr\"odinger.)  It is worth noting that when the algebraists set about making a rigorous statement of what a quantum field theory ought to be, they seem only to be able to make progress by imposing a postulate that says, roughly, ``In a QFT, the states that are localized in space and bounded in energy form a finite-dimensional space''~\cite{Haag10}.

And with this, we come to nearly the farthest edge of QBism.  It is the beginning of a place where quantum mechanics must step past itself.  To make quantum dimension meaningful in ontic terms, as a quality common to all physical objects, is to say it should be finite---going up, going down from this object to the next, but always finite.  Every region of space where electromagnetism can propagate, finite.  Every region of space where there is a gravitational ``field,'' finite.

It means that despite its humble roots in nonrelativistic quantum mechanics, there is something already cosmological about QBism.  It tinkers with spacetime, saying that in every ``hole'' (every bounded region) there is an interiority not given by the rest of the universe and a common quality called dimension.  It says that there is probably something right about the ``holographic principles'' arising from other reaches of physics \cite{Bekenstein2008}.  Recognizing entropy as a personal concept (entropy is a function of probability), QBism would suspect that it is not an entropy bound that arises from these principles, but perhaps a dimension bound \cite{Fuchs04b,Fuchs10b}.

Invocations of a ``holographic principle'' in quantum gravity research
traditionally contain a statement along the lines of, ``The
information in a volume actually lives on its boundary.''  These
locutions grow more opaque the more closely they are studied.  To talk
of the ``degrees of freedom existing on the boundary'' is to trap
oneself within obsolete intuitions.  Fundamentally, probabilities do
not exist without a gambler, and likewise, ``information'' does not
exist without an agent concerned with communication and computation.
(The latter statement is the logarithm of the former.)  Two orthogonal
quantum states for a system are not two distinct physical
configurations, in the sense of classical physics.  Rather, they are
two maximally distinct hypotheses for its possible future behavior
consequent upon an agent's action.  Alice's quantum state for a
system---whether a benzene ring or a black hole---does not live on the system's boundary, nor in its bulk.  It lives in Alice's mesh of beliefs, along with all her other fears and aspirations.

A novel perspective requires new images and metaphors, which can in turn stimulate novel technical developments.  How do we distance ourselves from the language that Hilbert-space dimension quantifies ``the number of distinct states a system can be in''?  For this mode of thought is at the root of all the loose talk in trying to interpret those holographic principles.

Suppose that Alice has access to a localized physical phenomenon that she wishes to employ in a quantum communication scheme.  Her goal is to detect, as well as possible, whether her communiqu\'es are being eavesdropped upon.  A technical result from a few years ago indicates that the maximal achievable sensitivity to eavesdropping is a simple function of the Hilbert-space dimension~\cite{Fuchs03}.  (In fact, a SIC furnishes a set of states that saturates this bound.)  If we take the holographic principle to say that the maximal Hilbert-space dimension of a phenomenon grows with the area that bounds it, then we have a relation between optimal ``sensitivity to the touch'' and a boundary area.  Indeed, sensitivity being tied to boundary area is an appealing image: For the boundary is the only thing an agent can touch in the first place!

\section{Quantum Cosmology from the Inside}

\begin{flushright}
\baselineskip=13pt
\parbox{2.8in}{\baselineskip=13pt
\small Theodore Roosevelt's decision to build the Panama Canal shows that free will
moves mountains, which implies, by general relativity, that even the
curvature of space is not determined. The stage is still being built while
the show goes on.}
\medskip\\
\small --- John Conway and Simon Kochen~\cite{Conway06}
\end{flushright}

Let us, however, step back from that farthest edge for a moment and discuss cosmology as it is presently construed before taking a final leap!

\begin{figure}
\begin{center}
\includegraphics[height=3.6in]{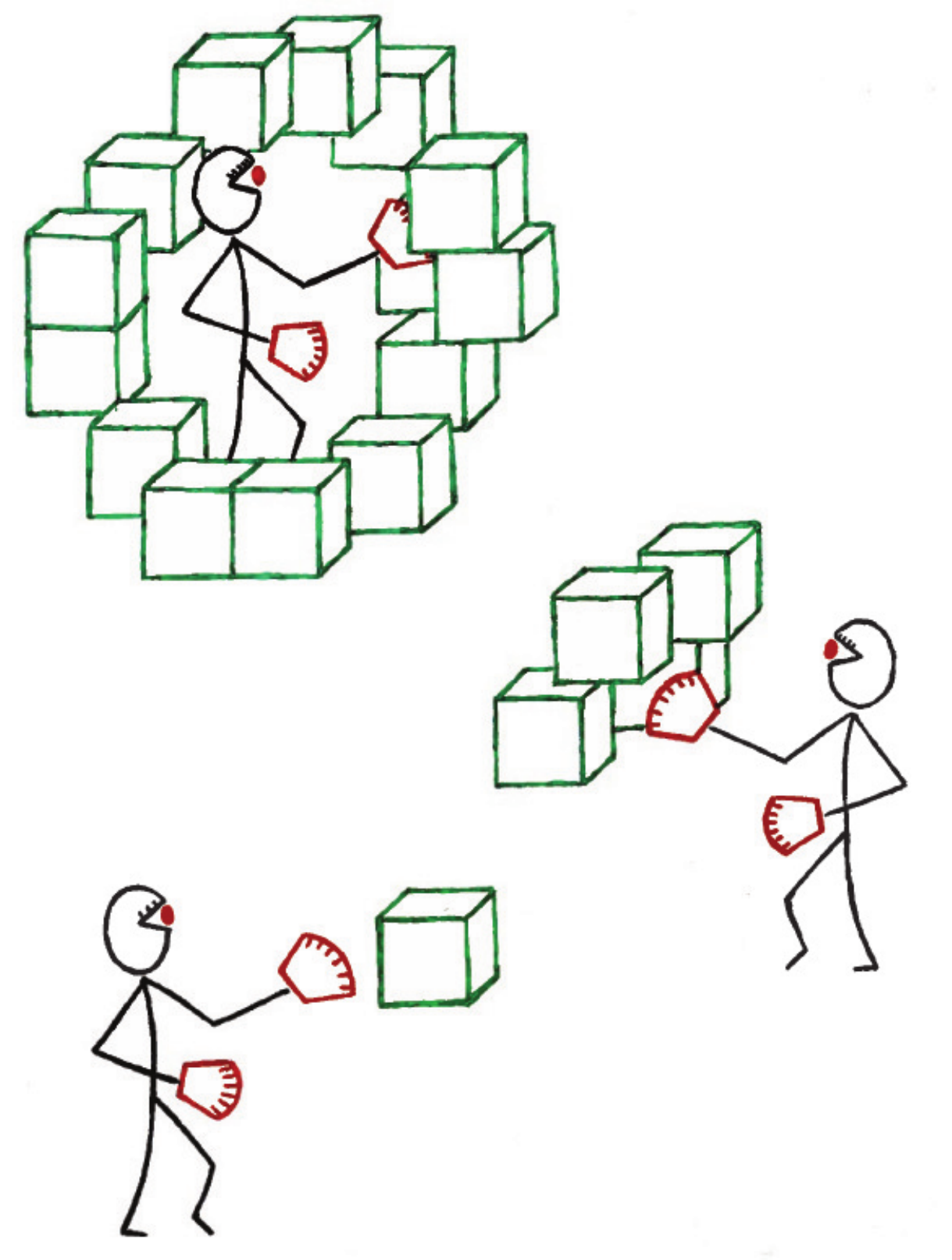}
\bigskip\caption{\protect {\bf Quantum Cosmology from the Inside.}~The agent in Figure 1 can consider measurements on ever larger systems. There is nothing in quantum mechanics to bar the systems considered from being larger and larger, to the point of eventually surrounding the agent.  Pushed far enough, this is quantum cosmology!  Why all this insistence on thinking that ``an agent must be outside the system he measures'' in the cosmological context should mean ``outside the physical universe itself''?  It means outside the system of interest, and that is the large-scale universe. Nor is there any issue of self-reference at hand.  One would be hard pressed to find a cosmologist who wants to include his beliefs about how the beats of his heart correlate with the sidereal cycles in his quantum-state assignment for the external universe.  The symbol $|\,\Psi_{\rm  universe}\,\rangle$ refers to the green boxes alone.}
\end{center}
\end{figure}

Sometimes it is claimed that a point of view about quantum theory like QBism's would make the enquiries of quantum cosmology impossible.  For instance, David Deutsch once wrote \cite{Deutsch86}:
\begin{quote}
 \noindent The best physical reason for adopting the Everett interpretation lies in quantum cosmology.  There one tries to apply quantum theory to the universe as a whole, considering the universe as a dynamical object starting with a big bang, evolving to form galaxies and so on. Then when one tries, for example by looking in a textbook, to ask what the symbols in the quantum theory mean, how does one use the wave function of the universe and the other mathematical objects that quantum theory employs to describe reality?  One reads there, `The meaning of these mathematical objects is as follows:  first consider an observer outside the quantum system under consideration \ldots.' And immediately one has to stop short.  Postulating an outside observer is all very well when we're talking about a laboratory:  we can imagine an observer sitting outside the experimental apparatus looking at it, but when the experimental apparatus---the object being described by quantum theory---is the entire universe, it's logically inconsistent to imagine an observer sitting outside it. Therefore the standard interpretation fails.  It fails completely to describe quantum cosmology.  Even if we knew how to write down the theory of quantum cosmology, which is quite hard incidentally, we literally wouldn't know what the symbols meant under any interpretation other than the Everett interpretation.
\end{quote}
But this is nonsense.  It is not hard to imagine how to measure the universe as a whole:  You simply live in it.

What are the typical observables and predictables of cosmology?  The Hubble constant, the cosmological constant, the degree of inhomogeneity of the cosmic microwave background radiation, total baryon number in this or that era of the universe, perhaps others.  To do quantum cosmology is to ask how an application of quantum mechanics can be made with regard to these quantities.  For the QBist quantum theory would be used {\it as it always is}:  As a normative calculus of consistency for all probability assignments concerned.  Quantum theory advises an agent to make all his probability assignments derivable from a single quantum state.  Write it like this if you wish---a big, fat wave function (it's for the whole universe after all):
\begin{equation}
\scalebox{1.5}{$|\,\Psi_{\rm universe}\,\rangle$}.
\label{Muggle}
\end{equation}
Why not?  We are swimming in this ocean called the universe, and we have to do physics from inside of it.  But then all the rest of the universe is outside each of us.  This wave function represents an agent's catalogue of beliefs for the relevant things outside.\footnote{There is one issue with assigning a state vector $|\,\Psi_{\rm universe}\,\rangle$. One doesn't even write down a pure quantum state for laser light when its phase is unknown; a mixed state is more appropriate \cite{vanEnk02}.  It is hard to imagine why one would write down a pure state for the large-scale universe.  Who would have beliefs that strict of it?  Be that as it may, a pure state is certainly allowed in principle.  Even people with the most unreasonable of initial beliefs (from an outsider's perspective) want to gamble consistently.}

The only point here is that QBism has every bit as much right to do cosmology as any other interpretation of quantum mechanics.  The only difference is that QBism does it from the inside.

More exciting is the possibility that once it does all that (its own version of what the other interpretations might have done), its power may not be exhausted.  For, noting how the Big Bang itself is a moment of creation with some resemblance to every individual quantum measurement, one starts to wonder whether even it ``might be on the inside.''  Certainly QBism has creation going on all the time and everywhere; quantum measurement is just about an agent hitching a ride and partaking in that ubiquitous process.

At the end of a long article it doesn't hurt to speculate.  We let William James and John Archibald Wheeler do the work for us.  First more sweepingly \cite{James22},
\begin{quote}
\noindent Our acts, our turning-places, where we seem to ourselves to
make ourselves and grow, are the parts of the world to which we are
closest, the parts of which our knowledge is the most intimate and
complete. Why should we not take them at their facevalue? Why may
they not be the actual turning-places and growing-places which they
seem to be, of the world---why not the workshop of being, where we
catch fact in the making, so that nowhere may the world grow in any
other kind of way than this?

Irrational!\ we are told. How can new being come in local spots and
patches which add themselves or stay away at random, independently of
the rest? There must be a reason for our acts, and where in the last
resort can any reason be looked for save in the material pressure or
the logical compulsion of the total nature of the world? There can be
but one real agent of growth, or seeming growth, anywhere, and that
agent is the integral world itself. It may grow all-over, if growth
there be, but that single parts should grow {\it per se\/} is
irrational.

But if one talks of rationality---and of reasons for things, and
insists that they can't just come in spots, what {\it kind\/} of a
reason can there ultimately be why anything should come at all?
\end{quote}
Then, more modernly \cite{Wheeler82c},
\begin{quote}
\noindent Each elementary quantum phenomenon is an elementary act of ``fact
creation.'' That is incontestable. But is that the only mechanism
needed to create all that is? Is what took place at the big bang the
consequence of billions upon billions of these elementary processes,
these elementary ``acts of observer-participancy,'' these quantum
phenomena? Have we had the mechanism of creation before our eyes all
this time without recognizing the truth? That is the larger question
implicit in your comment [``Is the big bang here?''].
\end{quote}
When cosmology hails from the inside, the world stands a chance of being anything it wants to be.

\section{The Future}

\begin{flushright}
\baselineskip=13pt
\parbox{2.8in}{\baselineskip=13pt
\small It is difficult to escape asking a challenging question. Is the
entirety of existence, rather than being built on particles or fields
of force or multidimensional geometry, built upon billions upon
billions of elementary quantum phenomena, those elementary acts of
``observer-participancy,'' those most ethereal of all the entities
that have been forced upon us by the progress of science?}
\medskip\\
\small  --- John Archibald Wheeler
\end{flushright}

Imagine our universe at a time when there were no agents about to use
the laws of probability theory as an aid in their gambles---i.e.,
before the cruel creativity of Darwinian selection had brought forth
any such agents.  Were there any quantum states in the universe then?
We QBists say {\it No.} It's not a matter of the quantum state of
the universe waiting until a qualified PhD student came along before
having its first collapse, as John Bell joked, but that there simply
weren't any quantum states.  Indeed, though we know little about
elsewhere under Heaven, here on Earth there weren't any quantum states
until 1926 when Erwin Schr\"odinger wrote the first one down. The
reason is simple: The universe is made of something else than
$\psi$-flavored gelatin. But then, what of the Born rule? To this, in
contrast, a QBist would say, ``Aha, now there's a sensible question.''
For the Born rule is among the set of relations an agent should strive
to attain in his larger mesh of probability assignments. That
normative rule indicates the character of the natural world, a
character that is present even when there are no agents to make use of
it. As Craig Callender once paraphrased it, in QBism, it is the normative rule which is {\it nature's whisper}, not
the specific terms within it.

Any of us can use quantum theory, but you can only use it for yourself. By way of analogy, consider the single-celled organisms called {\it Euglena.}  These are ``flagellate protists''---microbes with tails coming off of them. The tail arose from evolutionary pressures, so that a Euglena can move from environments where there are depleted nutrients to environments where there's an abundance of nutrients. It's a tool. Quantum mechanics is like the Euglena's tail. It's something we evolved in the 1920s, and since it's been shown to be such a good tool, we keep using it and we pass it on to our children. The tail of a Euglena is a single-user tail. But we can look at the tail and ask, ``What might we learn about the environment by studying the tail's structure?'' We might notice the tail is not completely circular, and that might tell us something about the viscosity of the medium the Euglena travels through. We might look at the ratio of the length of it to the width of it in various places, and that might tell us about features of the microbial environment. Likewise, quantum mechanics is a single-user theory, but by dissecting it, we can learn something about the world that all of us are immersed in.

In this way, QBism is carrying out what Einstein called the program of the real~\cite{Einstein49b}:
\begin{quote}
\noindent  A basic conceptual distinction, which is a necessary
prerequisite of scientific and pre-scientific thinking, is the
distinction between ``sense-impressions'' (and the recollection of
such) on the one hand and mere ideas on the other.  There is no such
thing as a conceptual definition of this distinction (aside from
circular definitions, i.e., of such as make a hidden use of the object
to be defined). Nor can it be maintained that at the base of this
distinction there is a type of evidence, such as underlies, for
example, the distinction between red and blue. Yet, one needs this
distinction in order to be able to overcome solipsism. Solution: we
shall make use of this distinction unconcerned with the reproach that,
in doing so, we are guilty of the metaphysical ``original sin.'' We
regard the distinction as a category which we use in order that we
might the better find our way in the world of immediate
sensations. The ``sense'' and the justification of this distinction
lies simply in this achievement. But this is only a first step. We
represent the sense-impressions as conditioned by an ``objective'' and
by a ``subjective'' factor. For this conceptual distinction there also
is no logical-philosophical justification. But if we reject it, we
cannot escape solipsism. It is also the presupposition of every kind
of physical thinking. Here too, the only justification lies in its
usefulness. We are here concerned with ``categories'' or schemes of
thought, the selection of which is, in principle, entirely open to us
and whose qualification can only be judged by the degree to which its
use contributes to making the totality of the contents of
consciousness ``intelligible.'' The above mentioned ``objective
factor'' is the totality of such concepts and conceptual relations as
are thought of as independent of experience, viz., of perceptions. So
long as we move within the thus programmatically fixed sphere of
thought we are thinking physically. Insofar as physical thinking
justifies itself, in the more than once indicated sense, by its
ability to grasp experiences intellectually, we regard it as
``knowledge of the real.''

After what has been said, the ``real'' in physics is to be taken as a type of program,
to which we are, however, not forced to cling a priori.
\end{quote}

There is so much still to do with the physics of QBism, and this article has just started scratching the surface.  Just one example:  The technical problems with SICs are manifest.  For instance, there must be a reason a proof of their existence has been so recalcitrant.  An optimist would say it is because they reach so deeply into the core of what the quantum is telling us!  In any case, we do suspect that when we get the structure of SICs down pat, Eq.~(\ref{ScoobyDoo2}), though already so essential to QBism's distillation of quantum theory's message, will seem like child's play in comparison to the vistas the further knowledge will open up.

But the technical also complements and motivates the conceptual.  So far we have only given the faintest hint of how QBism should be mounted onto a larger empiricism.  It will be noticed that QBism has been quite generous in treating agents as physical objects when needed.  ``I contemplate you as an agent when discussing your experience, but I contemplate you as a physical system before me when discussing my own.''  Our solution to ``Wigner's friend'' is the great example of this.  Precisely because of this, however, QBism knows that its story cannot end as a story of gambling agents---that is only where it starts.  Agency, for sure, is not a derivable concept as the reductionists and vicious abstractionists would have it, but QBism, like all of science, should strive for a Copernican principle whenever possible.  We have learned so far from quantum theory that before an agent the world is really malleable and ready through their intercourse to give birth.  Why would it not be so for every two parts of the world?  And this newly defined valence, quantum dimension, might it not be a measure of a system's potential for creation when it comes into relationship with those other parts?

In this article, we have focused on what QBism has to say about small quantum systems.  This is, in part, an example of QBism showing its ancestry in quantum information theory.  But, as the science journalist James Burke once said, {\it the revolutionary ideal admits no half-measures}:  The lessons of QBism must apply more broadly than qubits and qutrits.  What does taking the principles of QBism on board mean for how one thinks about special relativity~\cite{Mermin13}?  What about the practice of renormalization in statistical physics and field theory~\cite{DeBrota16}?  Or classical probability, information theory and machine learning~\cite{Fuchs11, stacey-thesis}?  The study of SICs has already changed the shape of the boundary between physics and pure mathematics~\cite{stacey-hoggar, RCF-SIC, Bengtsson16}.  How far must the changes go?  What, we might even ask, do Jamesian pragmatism, empiricism and radical metaphysical pluralism mean for the nature of mathematical truth?

It is a large research program whose outline is just taking shape.  It hints of a world, a pluriverse, that consists of an all-pervading ``pure experience,'' as William James called it.\footnote{Aside from James's originals \cite{James96a,James96b}, further reading on this concept and related subjects can be found in Refs.~\cite{Lamberth99,Taylor96,Wild69,Gieser05,Russell08,Banks03,Heidelberger04}.}  Expanding this notion, making it technical, and letting its insights tinker with spacetime itself is the better part of future work.  Quantum states, QBism declares, are not the stuff of the world, but quantum {\it measurement\/} might be.  Might a one-day future Shakespeare write with honesty,

\begin{flushright}
\baselineskip=13pt
\parbox{3.4in}{
\begin{verse}
Our revels are now ended.  These our actors, \\
As I foretold you, were all spirits and \\
Are melted into air, into thin air skip \ldots \\
We are such stuff as \\
\hspace*{0.5cm} quantum measurement is made on.
\end{verse}}
\end{flushright}

\vfill

\acknowledgments

This research was supported by the Foundational Questions Institute Fund on the Physics of the Observer (grant FQXi-RFP-1612), a donor advised fund at the Silicon Valley Community Foundation.  Work reported in Section \ref{SeekingSICs} was supported in part by the U.~S. Office of Naval Research (Grant No.\ N00014-09-1-0247).  

\bibliographystyle{utphys}

\bibliography{arxiv-varenna-qbism}

\providecommand{\href}[2]{#2}\begingroup\raggedright\begin{thebibliography}{100}

\bibitem{RMP}
C.~A. Fuchs and R.~Schack, ``Quantum-{B}ayesian coherence,'' {\em Rev. Mod.
  Phys.} {\bf 85} (2013)  1693, \href{http://arxiv.org/abs/0906.2187}{{\tt
  arXiv:0906.2187 [quant-ph]}}.

\bibitem{Fuchs10}
C.~A. Fuchs, {\em Coming of Age with Quantum Information}.
\newblock Cambridge University Press, 2010.

\bibitem{Mermin13}
N.~D. Mermin, ``{QBism} as {CBism}: Solving the problem of `the now',''
  \href{http://arxiv.org/abs/1312.7825}{{\tt arXiv:1312.7825 [quant-ph]}}.

\bibitem{Fuchs10b}
C.~A. Fuchs, ``My struggles with the block universe,''
  \href{http://arxiv.org/abs/1405.2390}{{\tt arXiv:1405.2390 [quant-ph]}}.

\bibitem{AJP}
C.~A. Fuchs, N.~D. Mermin, and R.~Schack, ``An introduction to {QBism} with an
  application to the locality of quantum mechanics,'' {\em Am.\ J.\ Phys.} {\bf
  82} (2014) no.~8, 749--54, \href{http://arxiv.org/abs/1311.5253}{{\tt
  arXiv:1311.5253 [quant-ph]}}.

\bibitem{Mermin14}
N.~D. Mermin, ``Why {QBism} is not the {Copenhagen} interpretation and what
  {John} {Bell} might have thought of it,'' in {\em Quantum [Un]Speakables II}.
\newblock Springer-Verlag, 2016.
\newblock \href{http://arxiv.org/abs/1409.2454}{{\tt arXiv:1409.2454
  [quant-ph]}}.

\bibitem{stacey-vonneumann}
B.~C. Stacey, ``Von {Neumann} was not a {Quantum} {Bayesian},'' {\em Phil.
  Trans. Roy. Soc. A} {\bf 374} (2016)  20150235,
  \href{http://arxiv.org/abs/1412.2409}{{\tt arXiv:1412.2409
  [physics.hist-ph]}}.

\bibitem{Baeyer16}
H.~C. von Baeyer, {\em QBism: The Future of Quantum Physics}.
\newblock Harvard University Press, 2016.

\bibitem{sep-quantum-bayesian}
R.~Healey, ``Quantum-{Bayesian} and pragmatist views of quantum theory,'' in
  {\em The Stanford Encyclopedia of Philosophy}, E.~N. Zalta, ed., p.~N/A.
\newblock Metaphysics Research Lab, Stanford University, {Winter} 2016~ed.,
  2016.
\newblock
  \url{https://plato.stanford.edu/archives/win2016/entries/quantum-bayesian/}.

\bibitem{Caves02}
C.~M. Caves, C.~A. Fuchs, and R.~Schack, ``Quantum probabilities as {Bayesian}
  probabilities,'' {\em Phys. Rev. A} {\bf 65} (2002)  022305,
  \href{http://arxiv.org/abs/quant-ph/0106133}{{\tt arXiv:quant-ph/0106133}}.

\bibitem{Fuchs02}
C.~A. Fuchs, ``Quantum mechanics as quantum information (and only a little
  more),'' in {\em Quantum Theory: Reconsideration of Foundations},
  A.~Khrennikov, ed., p.~463.
\newblock V\"axj\"o University Press, 2002.
\newblock \href{http://arxiv.org/abs/quant-ph/0205039}{{\tt
  arXiv:quant-ph/0205039}}.

\bibitem{Fuchs04}
C.~A. Fuchs and R.~Schack, ``Unknown quantum states and operations, a
  {Bayesian} view,'' in {\em Quantum Estimation Theory}, M.~G.~A. Paris and
  J.~\v{R}eh\'a\v{c}ek, eds., p.~151.
\newblock Springer-Verlag, 2004.
\newblock \href{http://arxiv.org/abs/quant-ph/0404156}{{\tt
  arXiv:quant-ph/0404156}}.

\bibitem{Caves07}
C.~M. Caves, C.~A. Fuchs, and R.~Schack, ``Subjective probability and quantum
  certainty,'' {\em Stud. Hist. Phil. Mod. Phys.} {\bf 38} (2007)  255,
  \href{http://arxiv.org/abs/quant-ph/0608190}{{\tt arXiv:quant-ph/0608190}}.

\bibitem{Feynman03}
R.~P. Feynman, F.~B. Morinigo, and W.~G. Wagner, {\em The Feynman Lectures on
  Gravitation}.
\newblock Westview Press, Boulder, CO, 2003.

\bibitem{Gisin15}
N.~Gisin, ``Why {Bohmian} mechanics? {One} and two-time position measurements,
  {Bell} inequalities, philosophy and physics,''
  \href{http://arxiv.org/abs/1509.00767}{{\tt arXiv:1509.00767 [quant-ph]}}.

\bibitem{Caves05}
C.~M. Caves and R.~Schack, ``Properties of the frequency operator do not imply
  the quantum probability postulate,'' {\em Ann. Phys.} {\bf 315} (2005) no.~1,
  123--46, \href{http://arxiv.org/abs/quant-ph/0409144}{{\tt
  arXiv:quant-ph/0409144}}.

\bibitem{Kent14}
A.~Kent, ``Does it make sense to speak of self-locating uncertainty in the
  universal wave function? {Remarks} on {Sebens} and {Carroll},'' {\em Found.
  Phys.} {\bf 45} (2014) no.~2, 211--17,
  \href{http://arxiv.org/abs/1408.1944}{{\tt arXiv:1408.1944}}.

\bibitem{Kastner14}
R.~E. Kastner, ```{Einselection}' of pointer observables: The new
  ${H}$-theorem?,'' {\em Stud. Hist. Phil. Mod. Phys.} {\bf 48} (2014)  56--58,
  \href{http://arxiv.org/abs/1406.4126}{{\tt arXiv:1406.4126 [quant-ph]}}.

\bibitem{Adlam14}
E.~Adlam, ``The problem of confirmation in the {Everett} interpretation,'' {\em
  Stud. Hist. Phil. Mod. Phys.} {\bf 47} (2014)  21--32.

\bibitem{Jansson16}
L.~Jansson, ``Everettian quantum mechanics and physical probability: Against
  the principle of `state supervenience','' {\em Stud. Hist. Phil. Mod. Phys.}
  {\bf 53} (2016)  45--53.

\bibitem{Saunders05}
S.~Saunders, ``What is probability?,'' in {\em Quo Vadis Quantum Mechanics?},
  p.~209.
\newblock Springer, 2005.
\newblock \href{http://arxiv.org/abs/quant-ph/0409144}{{\tt
  arXiv:quant-ph/0409144}}.

\bibitem{Wallace09}
D.~Wallace, ``A formal proof of the {Born} rule from decision theoretic
  assumptions,'' \href{http://arxiv.org/abs/0906.2718}{{\tt arXiv:0906.2718
  [quant-ph]}}.

\bibitem{Price08}
H.~Price, ``Decisions, decisions, decisions: Can {S}avage salvage {E}verettian
  probability?,'' in {\em Many Worlds? Everett, Quantum Theory and Reality},
  p.~369.
\newblock Oxford University Press, 2010.
\newblock \href{http://arxiv.org/abs/0802.1390}{{\tt arXiv:0802.1390
  [quant-ph]}}.

\bibitem{Kent09}
A.~Kent, ``{One World versus Many}:\ {The} inadequacy of {Everettian} accounts
  of evolution, probability, and scientific confirmation,'' in {\em Many
  Worlds? Everett, Quantum Theory and Reality}, p.~307.
\newblock Oxford University Press, 2010.
\newblock \href{http://arxiv.org/abs/0905.0624}{{\tt arXiv:0905.0624
  [quant-ph]}}.

\bibitem{Hartle68}
J.~B. Hartle, ``Quantum mechanics of individual systems,'' {\em Am. J. Phys.}
  {\bf 36} (1968)  704.

\bibitem{Spekkens07}
R.~W. Spekkens, ``Evidence for the epistemic view of quantum states:\ a toy
  theory,'' {\em Phys. Rev. A} {\bf 75} (2007)  032110.

\bibitem{LordVoldemort}
C.~A. Fuchs, ``{QBism}, the perimeter of {Quantum} {Bayesianism},''
  \href{http://arxiv.org/abs/1003.5209}{{\tt arXiv:1003.5209 [quant-ph]}}.

\bibitem{Good83}
I.~J. Good, ``46656 varieties of {Bayesians},'' in {\em Good Thinking: The
  Foundations of Probability and Its Applications}, p.~20.
\newblock University of Minnesota Press, 1983.

\bibitem{Menand01}
L.~Menand, {\em The Metaphysical Club: A Story of Ideas in America}.
\newblock Farrar, Straus and Giroux, 2001.

\bibitem{Bernardo94}
J.~M. Bernardo and A.~F.~M. Smith, {\em Bayesian Theory}.
\newblock Wiley, 1994.

\bibitem{Lindley06}
D.~V. Lindley, {\em Understanding Uncertainty}.
\newblock Wiley, 2006.

\bibitem{DeFinetti90}
B.~de~Finetti, {\em Theory of Probability}.
\newblock Wiley, 1990.

\bibitem{Mises22}
R.~von Mises, ``{\"Uber die gegenw\"artage Krise der Me\-chan\-ik},'' {\em Die
  Naturwissenschaften} {\bf 10} (1922) no.~25, n.p. Translated in
  M.~St\"oltzner, ``Vienna Indeterminism II: From Exner's Synthesis to Frank
  and von Mises,'' in {\sl Logical Empiricism. Historical and Contemporary
  Perspectives}, edited by P.\ Parrini, W.\ Salmon, and M.\ Salmon, (University
  of Pittsburgh Press, 2003), p.~194.

\bibitem{Wigner71}
E.~P. Wigner, ``Remarks on the mind-body question,'' in {\em Symmetries and
  Reflections: Scientific Essays of Eugene P.\ Wigner}, p.~171.
\newblock Ox Bow Press, 1979.

\bibitem{Albert94}
D.~Z. Albert, {\em Quantum Mechanics and Experience}.
\newblock Harvard University Press, 1994.

\bibitem{Nielsen00}
M.~A. Nielsen and I.~L. Chuang, {\em Quantum Computation and Quantum
  Information}.
\newblock Cambridge University Press, 2000.

\bibitem{Leifer06}
M.~S. Leifer, ``Conditional density operators and the subjectivity of quantum
  operations,'' in {\em Foundations of Probability and Physics -- 4},
  G.~Adenier {\em et al.}, eds., p.~438.
\newblock American Institute of Physics, Melville, NY, 2007.

\bibitem{Pauli94}
W.~Pauli, {\em Writings on Physics and Philosophy}.
\newblock Springer-Verlag, 1994.
\newblock Edited by C.\ P.\ Enz and K.\ von Meyenn.

\bibitem{Wheeler82c}
J.~A. Wheeler, ``Bohr, {Einstein}, and the strange lesson of the quantum,'' in
  {\em Mind in Nature:~Nobel Conference XVII, Gustavus Adolphus College,
  St.~Peter, Minnesota}, R.~Q. Elvee, ed., pp.~1--23.
\newblock Harper \& Row, San Francisco, CA, 1982.
\newblock With discussions pp.~23--30, 88--89, 112--113, and 148--149.

\bibitem{Jaynes03}
E.~T. Jaynes, {\em Probability Theory: The Logic of Science}.
\newblock Cambridge University Press, 2003.

\bibitem{Peres95}
A.~Peres, {\em Quantum Theory: Concepts and Methods}.
\newblock Kluwer, Dordrecht, 1995.

\bibitem{Goldstein65}
H.~Goldstein, {\em Classical Mechanics}.
\newblock Pearson Education India, 1965.

\bibitem{Caves96}
C.~M. Caves and C.~A. Fuchs, ``Quantum information: How much information in a
  state vector?,'' {\em Ann. Israel Phys. Soc.} {\bf 12} (1996)  226--57.

\bibitem{Wigner12}
E.~P. Wigner, {\em Group Theory and Its Application to the Quantum Mechanics of
  Atomic Spectra}.
\newblock Elsevier, 2012.

\bibitem{Wigner61}
E.~P. Wigner, ``The probability of the existence of a self-reproducing unit,''
  in {\em The Logic of Personal Knowledge: Essays Presented to Michael Polanyi
  on his Seventieth Birthday}, pp.~231--38.
\newblock Routledge \& Kegan Paul, London, 1961.

\bibitem{Gisin09}
N.~Gisin, ``Quantum nonlocality: How does nature do it?,'' {\em Science} {\bf
  326} (2009)  1357.

\bibitem{Albert09}
D.~Z. Albert and R.~Galchen, ``A quantum threat to special relativity,'' {\em
  Sci. Am.} {\bf 300} (2009) no.~3, 32.

\bibitem{Norsen06}
T.~Norsen, ``Bell locality and the nonlocal character of nature,'' {\em Found.
  Phys. Lett.} {\bf 19} (2006)  633.

\bibitem{Maudlin14}
T.~Maudlin, ``What {Bell} did,'' {\em J. Phys. A} {\bf 47} (2014)  424010.

\bibitem{Einstein48}
A.~Einstein, ``Quanten-{Mechanik} und {Wirklichkeit},'' {\em Dialectica} {\bf
  2} (1948)  320. Passage translated in D.\ Howard, ``Einstein on Locality and
  Separability,'' Stud.\ Hist.\ Phil.\ Sci.\ Pt.\ A, {\bf 16} (1985).

\bibitem{Fine96}
A.~Fine, {\em The Shaky Game: Einstein, Realism and the Quantum Theory}.
\newblock University of Chicago Press, second~ed., 1996.

\bibitem{Harrigan07}
N.~Harrigan and R.~W. Spekkens, ``Einstein, incompleteness, and the epistemic
  view of quantum states,'' {\em Found. Phys.} {\bf 40} (2010) no.~2, 125--57,
  \href{http://arxiv.org/abs/0706.2661}{{\tt arXiv:0706.2661 [quant-ph]}}.

\bibitem{Einstein49}
A.~Einstein, ``Autobiographical notes,'' in {\em Albert Einstein:
  Philosopher-Scientist}, P.~A. Schilpp, ed.
\newblock Tudor Publishing Co., New York, 1949.

\bibitem{Stairs83}
A.~Stairs, ``Quantum logic, realism, and value-definiteness,'' {\em Phil. Sci.}
  {\bf 50} (1983)  578.

\bibitem{Conway06}
J.~Conway and S.~Kochen, ``The free will theorem,'' {\em Found. Phys.} {\bf 36}
  (2006)  1441, \href{http://arxiv.org/abs/quant-ph/0604079}{{\tt
  arXiv:quant-ph/0604079}}.

\bibitem{Conway09}
J.~Conway and S.~Kochen, ``The strong free will theorem,'' {\em Not. AMS} {\bf
  56} (2009)  226, \href{http://arxiv.org/abs/0807.3286}{{\tt
  arXiv:0807.3286}}.

\bibitem{Cabello97}
A.~Cabello, J.~M. Estebaranz, and G.~Garc\'{\i}a-Alcaine,
  ``{Bell-Kochen-Specker} theorem: A proof with 18 vectors,'' {\em Phys. Lett.
  A} {\bf 212} (1996)  183.

\bibitem{Primas90}
H.~Primas, ``Beyond {Baconian} quantum physics,'' in {\em Kohti uutta
  todellisuusk\"asityst\"a. Juhlakirja professori Laurikaisen
  75-vuotisp\"aiv\"an\"a (Towards a New Conception of Reality. Anniversary
  Publication to Professor Laurikainen's 75th Birthday)}, U.~Ketvel, ed.,
  p.~100.
\newblock Yliopistopaino, Helsinki, 1990.

\bibitem{Caves02b}
C.~M. Caves, C.~A. Fuchs, and R.~Schack, ``Unknown quantum states:\ the quantum
  de {Finetti} representation,'' {\em J. Math. Phys.} {\bf 43} (2002)  4537,
  \href{http://arxiv.org/abs/quant-ph/0104088}{{\tt arXiv:quant-ph/0104088}}.

\bibitem{Schack01}
R.~Schack, T.~A. Brun, and C.~M. Caves, ``Quantum {Bayes} rule,'' {\em Phys.
  Rev. A} {\bf 64} (2001)  014305,
  \href{http://arxiv.org/abs/quant-ph/0008113}{{\tt arXiv:quant-ph/0008113}}.

\bibitem{Fuchs04c}
C.~A. Fuchs, R.~Schack, and P.~F. Scudo, ``De {Finetti} representation theorem
  for quantum-process tomography,'' {\em Phys. Rev. A} {\bf 69} (2004) no.~6,
  062305, \href{http://arxiv.org/abs/quant-ph/0307198}{{\tt
  arXiv:quant-ph/0307198}}.

\bibitem{Lo05}
H.~K. Lo, H.~F. Chau, and M.~Ardehali, ``Efficient quantum key distribution
  scheme and a proof of its unconditional security,'' {\em J. Crypto.} {\bf 18}
  (2005)  133--65.

\bibitem{Renner07}
R.~Renner, ``Symmetry of large physical systems implies independence of
  subsystems,'' {\em Nature Phys.} {\bf 3} (2007)  645--49.

\bibitem{Renner08}
R.~Renner, ``Security of quantum key distribution,'' {\em Int. J. Quant. Inf.}
  {\bf 6} (2008)  1--127.

\bibitem{Doherty05}
A.~C. Doherty, P.~A. Parrilo, and F.~M. Spedalieri, ``Detecting multipartite
  entanglement,'' {\em Phys. Rev. A} {\bf 71} (2005)  032333.

\bibitem{Enk07}
S.~J. van Enk, N.~L\"utkenhaus, and H.~J. Kimble, ``Experimental procedures for
  entanglement verification,'' {\em Phys. Rev. A} {\bf 75} (2007)  052318.

\bibitem{vanEnk02}
S.~J. van Enk and C.~A. Fuchs, ``Quantum state of an ideal propagating laser
  field,'' {\em Phys. Rev. Lett.} {\bf 88} (2002)  027902//1--4,
  \href{http://arxiv.org/abs/quant-ph/0104036}{{\tt arXiv:quant-ph/0104036}}.

\bibitem{vanEnk02b}
S.~J. van Enk and C.~A. Fuchs, ``Quantum state of a propagating laser field,''
  {\em Quant. Info. Comp.} {\bf 2} (2002)  151--65.

\bibitem{Ferrie09}
C.~Ferrie and J.~Emerson, ``Framed {Hilbert} space:\ hanging the
  quasi-probability pictures of quantum theory,'' {\em New J. Phys.} {\bf 11}
  (2009)  063040.

\bibitem{Wootters86}
W.~K. Wootters, ``Quantum mechanics without probability amplitudes,'' {\em
  Found. Phys.} {\bf 16} (1986)  391.

\bibitem{Appleby07}
D.~M. Appleby, H.~B. Dang, and C.~A. Fuchs, ``Symmetric
  informationally-complete quantum states as analogues to orthonormal bases and
  minimum uncertainty states,'' {\em Entropy} {\bf 16} (2007) no.~3, 1484--92,
  \href{http://arxiv.org/abs/0707.2071}{{\tt arXiv:0707.2071 [quant-ph]}}.

\bibitem{Zauner99}
G.~Zauner, {\em Quantum Designs -- Foundations of a Non-Commutative Theory of
  Designs}.
\newblock PhD thesis, University of Vienna, 1999.

\bibitem{Caves99}
C.~M. Caves, ``Symmetric informationally complete {POVMs}.'' Posted at
  \url{http://info.phys.unm.edu/~caves/reports/infopovm.pdf}, 1999.

\bibitem{Renes04}
J.~M. Renes, R.~Blume-Kohout, A.~J. Scott, and C.~M. Caves, ``Symmetric
  informationally complete quantum measurements,'' {\em J. Math. Phys.} {\bf
  45} (2004)  2171.

\bibitem{Scott09}
A.~J. Scott and M.~Grassl, ``{SIC-POVMs}: A new computer study,''
  \href{http://dx.doi.org/10.1063/1.3374022}{{\em Journal of Mathematical
  Physics} {\bf 51} (2009)  042203}, \href{http://arxiv.org/abs/0910.5784}{{\tt
  arXiv:0910.5784 [quant-ph]}}.

\bibitem{Fuchs03}
C.~A. Fuchs and M.~Sasaki, ``Squeezing quantum information through a classical
  channel:\ measuring the `quantumness' of a set of quantum states,'' {\em
  Quant. Info. Comp.} {\bf 3} (2003)  377,
  \href{http://arxiv.org/abs/quant-ph/0302092}{{\tt arXiv:quant-ph/0302092}}.

\bibitem{Scott06}
A.~J. Scott, ``Tight informationally complete quantum measurements,'' {\em J.
  Phys. A} {\bf 39} (2006)  13507.

\bibitem{Wootters07}
W.~K. Wootters and D.~M. Sussman, ``Discrete phase space and
  minimum-uncertainty states,'' \href{http://arxiv.org/abs/0704.1277}{{\tt
  arXiv:0704.1277 [quant-ph]}}.

\bibitem{Zhu2016}
H.~Zhu, ``Quasiprobability representations of quantum mechanics with minimal
  negativity,'' {\em Phys. Rev. Lett.} {\bf 117} (2016) no.~12, 120404,
  \href{http://arxiv.org/abs/1604.06974}{{\tt arXiv:1604.06974 [quant-ph]}}.

\bibitem{Appleby15}
D.~M. Appleby, C.~A. Fuchs, and H.~Zhu, ``Group theoretic, {Lie} algebraic and
  {Jordan} algebraic formulations of the {SIC} existence problem,'' {\em
  Quantum Information \& Computation} {\bf 15} (2015) no.~1--2, 61--94,
  \href{http://arxiv.org/abs/1312.0555}{{\tt arXiv:1312.0555 [quant-ph]}}.

\bibitem{Ling06}
A.~Ling, S.~K. Pang, A.~Lamas-Linares, and C.~Kurtsiefer, ``Experimental
  polarization state tomography using optimal polarimeters,'' {\em Phys. Rev.
  A} {\bf 74} (2006)  022309.

\bibitem{Medendorp10}
Z.~E.~D. Medendorp, F.~A. Torres-Ruiz, L.~K. Shalm, C.~A. Fuchs, and A.~M.
  Steinberg, ``Characterizing a qutrit directly with symmetric informationally
  complete ({SIC}) {POVM}s,'' in {\em Quantum Electronics and Laser Science
  Conference}.
\newblock Optical Society of America, 2010.

\bibitem{BoydLeuchs}
N.~Bent, H.~Qassim, A.~A. Tahir, D.~Sych, G.~Leuchs, L.~L. S\'anchez-Soto,
  E.~Karimi, and R.~W. Boyd, ``Experimental realization of quantum tomography
  of photonic qudits via symmetric informationally complete positive
  operator-valued measures,'' {\em Phys. Rev. X} {\bf 5} (2015) no.~4, 041006.

\bibitem{Bacciagaluppi09}
G.~Bacciagaluppi and E.~Crull, ``Heisenberg (and {Schr{\"o}dinger}, and
  {Pauli}) on hidden variables,'' {\em Stud. Hist. Phil. Mod. Phys.} {\bf 40}
  (2009)  374.

\bibitem{Bub07}
J.~Bub, ``Quantum probabilities as degrees of belief,'' {\em Stud. Hist. Phil.
  Mod. Phys.} {\bf 38} (2007)  232.

\bibitem{Jozsa04}
R.~Jozsa, ``Illustrating the concept of quantum information,'' {\em IBM J. Res.
  Dev.} {\bf 48} (2004)  79.

\bibitem{Appleby09a}
D.~M. Appleby, {\AA}.~Ericsson, and C.~A. Fuchs, ``Properties of {QBist} state
  spaces,'' \href{http://dx.doi.org/10.1007/s10701-010-9458-7}{{\em Found.
  Phys.} {\bf 41} (2011) no.~3, 564--79},
  \href{http://arxiv.org/abs/0910.2750}{{\tt arXiv:0910.2750 [quant-ph]}}.

\bibitem{Fuchs09b}
C.~A. Fuchs and R.~Schack, ``A {Quantum-Bayesian} route to quantum-state
  space,'' \href{http://dx.doi.org/10.1007/s10701-009-9404-8}{{\em Found.
  Phys.} {\bf 41} (2011) no.~3, 345--56},
  \href{http://arxiv.org/abs/0912.4252}{{\tt arXiv:0912.4252 [quant-ph]}}.

\bibitem{Appleby09b}
D.~M. Appleby, S.~T. Flammia, and C.~A. Fuchs, ``The {Lie} algebraic
  significance of symmetric informationally complete measurements,''
  \href{http://dx.doi.org/10.1063/1.3555805}{{\em J. Math. Phys.} {\bf 52}
  (2011)  022202}, \href{http://arxiv.org/abs/1001.0004}{{\tt arXiv:1001.0004
  [quant-ph]}}.

\bibitem{qplex}
M.~Appleby, C.~A. Fuchs, B.~C. Stacey, and H.~Zhu, ``Introducing the qplex: A
  novel arena for quantum theory,'' \href{http://arxiv.org/abs/1612.03234}{{\tt
  arXiv:1612.03234 [quant-ph]}}.

\bibitem{Stueckelberg60}
E.~C.~G. Stueckelberg, ``Quantum theory in real {Hilbert} space,'' {\em Helv.
  Phys. Acta} {\bf 33} (1960)  727.

\bibitem{AdlerBook}
S.~L. Adler, {\em Quaternionic Quantum Mechanics and Quantum Fields}.
\newblock Oxford University Press, 1995.

\bibitem{Khatirinejad08}
M.~Khatirinejad, {\em Regular Structures of Lines in Complex Space}.
\newblock PhD thesis, Simon Fraser University, 2008.

\bibitem{FQXi}
G.~Chiribella, A.~Cabello, and M.~Kleinmann, ``The observer observed: a
  {Bayesian} route to the reconstruction of quantum theory.''
  \url{http://fqxi.org/grants/large/awardees/view/__details/2016/chiribella},
  2016.

\bibitem{RCF-SIC}
D.~M. Appleby, S.~Flammia, G.~McConnell, and J.~Yard, ``Generating ray class
  fields of real quadratic fields via complex equiangular lines,''
  \href{http://arxiv.org/abs/1604.06098}{{\tt arXiv:1604.06098 [math.NT]}}.

\bibitem{stacey-sporadic}
B.~C. Stacey, ``Sporadic {SICs} and the normed division algebras,''
  \href{http://arxiv.org/abs/1605.01426}{{\tt arXiv:1605.01426 [quant-ph]}}.

\bibitem{Bernstein05}
J.~Bernstein, ``Max {Born} and the quantum theory,'' {\em Am. J. Phys.} {\bf
  73} (2005)  999--1008.

\bibitem{Fedak09}
W.~A. Fedak and J.~J. Prentis, ``The 1925 {Born} and {Jordan} paper `on quantum
  mechanics','' {\em Am. J. Phys.} {\bf 77} (2009)  128--39.

\bibitem{Scholz08}
E.~Scholz, ``Weyl entering the `new' quantum mechanics discourse,'' in {\em
  HQ-1: Conference on the History of Quantum Physics}, C.~Joas, C.~Lehner, and
  J.~Renn, eds., pp.~249--67.
\newblock Max Planck Institute for the History of Science, 2008.
\newblock \url{http://www.mpiwg-berlin.mpg.de/Preprints/P350.PDF\#page=262}.

\bibitem{Weyl27}
H.~Weyl, {\em The Theory of Groups and Quantum Mechanics}.
\newblock Dover, 1927.

\bibitem{Schwinger70}
J.~Schwinger, {\em Quantum Kinematics and Dynamics}.
\newblock 1970.

\bibitem{zhu-thesis}
H.~Zhu, {\em Quantum State Estimation and Symmetric Informationally Complete
  {POMs}}.
\newblock PhD thesis, National University of Singapore, 2012.
\newblock \url{https://scholarbank.nus.edu.sg/handle/10635/35247}.

\bibitem{Szymusiak2015}
A.~Szymusiak and W.~S\l{}omczy\'nski, ``Informational power of the {Hoggar}
  symmetric informationally complete positive operator-valued measure,'' {\em
  Phys. Rev. A} {\bf 94} (2016)  012122,
  \href{http://arxiv.org/abs/1512.01735}{{\tt arXiv:1512.01735 [quant-ph]}}.

\bibitem{stacey-hoggar}
B.~C. Stacey, ``Geometric and information-theoretic properties of the {Hoggar}
  lines,'' \href{http://arxiv.org/abs/1609.03075}{{\tt arXiv:1609.03075
  [quant-ph]}}.

\bibitem{stacey-qutrit}
B.~C. Stacey, ``{SIC-POVMs} and compatibility among quantum states,'' {\em
  Mathematics} {\bf 4} (2016) no.~2, 36,
  \href{http://arxiv.org/abs/1404.3774}{{\tt arXiv:1404.3774 [quant-ph]}}.

\bibitem{Bengtsson16}
I.~Bengtsson, ``The number behind the simplest {SIC-POVM},''
  \href{http://arxiv.org/abs/1611.09087}{{\tt arXiv:1611.09087 [quant-ph]}}.

\bibitem{Schappacher98}
N.~Schappacher, ``On the history of {Hilbert}'s twelfth problem: A comedy of
  errors,'' in {\em Mat{\'e}riaux pour l’histoire des math{\'e}matiques au
  XX$^{\rm e}$ si{\`e}cle}, p.~243.
\newblock Soc. Math. France, Paris,, 1998.

\bibitem{baez-plus}
J.~C. Baez and H.~Joyce, ``Ubiquitous octonions.'' {\it PLUS MAGAZINE,}
  \url{https://plus.maths.org/content/ubiquitous-octonions}, 2005.

\bibitem{viazovska2016}
M.~Viazovska, ``The sphere packing problem in dimension 8,''
  \href{http://arxiv.org/abs/1603.04246}{{\tt arXiv:1603.04246 [math.NT]}}.

\bibitem{Norsen}
T.~Norsen, ``Quantum solipsism and non-locality.''
  \url{http://www.ijqf.org/wps/wp-content/uploads/2014/12/Norsen-Bell-paper.pdf},
  2014.

\bibitem{Rorty91}
R.~Rorty, {\em Philosophical Papers, Volume 1: Objectivity, Relativism, and
  Truth}.
\newblock Cambridge University Press, 1991.

\bibitem{Dupre93}
J.~Dupr\'e, {\em The Disorder of Things: Metaphysical Foundations of the
  Disunity of Science}.
\newblock Harvard University Press, 1993.

\bibitem{Cartwright99}
N.~Cartwright, {\em The Dappled World: A Study of the Boundaries of Science}.
\newblock Cambridge University Press, 1999.

\bibitem{James96a}
W.~James, {\em A Pluralistic Universe}.
\newblock University of Nebraska Press, Lincoln, NB, 1996.

\bibitem{Wahl25}
J.~Wahl, {\em The Pluralist Philosophies of England and America}.
\newblock Open Court, London, 1925.
\newblock Translated by F.\ Rothwell.

\bibitem{James40}
W.~James, {\em Some Problems of Philosophy}.
\newblock Longmans, Green, and Co., London, 1940.

\bibitem{James96b}
W.~James, {\em Essays in Radical Empiricism}.
\newblock University of Nebraska Press, Lincoln, NB, 1996.

\bibitem{James1884}
W.~James, ``The dilemma of determinism,'' in {\em The Will to Believe and Other
  Essays in Popular Philosophy; Human Immortality---Both Books Bound as One},
  p.~145.
\newblock Dover, 1884.

\bibitem{Thayer81}
H.~S. Thayer, {\em Meaning and Action: A Critical History of Pragmatism}.
\newblock Hackett Publishing Co., Indianapolis, second~ed., 1981.

\bibitem{JamesOSH}
W.~James, ``On some {Hegelisms},'' in {\em The Will to Believe and Other Essays
  in Popular Philosophy; Human Immortality---Both Books Bound as One}, p.~263.
\newblock Dover, 1882.

\bibitem{JamesMultiverse}
W.~James, ``Is life worth living?,'' in {\em The Will to Believe and Other
  Essays in Popular Philosophy; Human Immortality---Both Books Bound as One},
  p.~32.
\newblock Dover, 1884.

\bibitem{Feynman65}
R.~P. Feynman, {\em The Character of Physical Law}.
\newblock MIT Press, 1965.

\bibitem{James97}
W.~James, ``Abstractionism and `relativismus','' in {\em The Meaning of Truth},
  p.~246.
\newblock Prometheus Books, Amherst, NY, 1997.

\bibitem{Baeyer09}
H.~C. von Baeyer, {\em Petite Le\c{c}ons de Physique dans les Jardins de
  Paris}.
\newblock Dunod, Paris, 2009.

\bibitem{Hardy01a}
L.~Hardy, ``Quantum theory from five reasonable axioms,''
  \href{http://arxiv.org/abs/quant-ph/0101012}{{\tt arXiv:quant-ph/0101012}}.

\bibitem{Hardy01b}
L.~Hardy, ``Why quantum theory?,'' in {\em Non-locality and Modality},
  T.~Placek and J.~Butterfield, eds., p.~61.
\newblock Kluwer, Dordrecht, 2002.

\bibitem{Dakic09}
B.~Daki\'c and \v{C}. Brukner, ``Quantum theory and beyond: Is entanglement
  special?,'' in {\em Deep Beauty: Understanding the Quantum World through
  Mathematical Innovation}, H.~Halvorson, ed., pp.~365--92.
\newblock Cambridge University Press, 2011.
\newblock \href{http://arxiv.org/abs/0911.0695}{{\tt arXiv:0911.0695
  [quant-ph]}}.

\bibitem{Fuchs04b}
C.~A. Fuchs, ``On the quantumness of a {Hilbert} space,'' {\em Quant. Info.
  Comp.} {\bf 4} (2004)  467, \href{http://arxiv.org/abs/quant-ph/0404122}{{\tt
  arXiv:quant-ph/0404122}}.

\bibitem{Brunner08}
N.~Brunner, S.~Pironio, A.~Acin, N.~Gisin, A.~A. M\'ethot, and V.~Scarani,
  ``Testing the {Hilbert} space dimension,'' {\em Phys. Rev. Lett.} {\bf 100}
  (2008)  210503.

\bibitem{Wehner08}
S.~Wehner, M.~Christandl, and A.~C. Doherty, ``Lower bound on the dimension of
  a quantum system given measured data,'' {\em Phys. Rev. A} {\bf 78} (2008)
  062112.

\bibitem{Wolf09}
M.~M. Wolf and D.~Perez-Garcia, ``Assessing dimensions from evolution,'' {\em
  Phys. Rev. Lett.} {\bf 102} (2009)  190504.

\bibitem{BlumeKohout02}
R.~Blume-Kohout, C.~M. Caves, and I.~H. Deutsch, ``Climbing mount scalable:\
  physical-resource requirements for a scalable quantum computer,'' {\em Found.
  Phys.} {\bf 32} (2002)  1641.

\bibitem{Cerf02}
N.~J. Cerf, M.~Bourennane, A.~Karlsson, and N.~Gisin, ``Security of quantum key
  distribution using $d$-level systems,'' {\em Phys. Rev. Lett.} {\bf 88}
  (2002)  127902.

\bibitem{Horowitz04}
G.~T. Horowitz and J.~Maldacena, ``The black hole final state,'' {\em J. High
  Energy Phys.} {\bf 2004} (2004) no.~2, 008.

\bibitem{Gottesman04}
D.~Gottesman and J.~Preskill, ``Comment on `the black hole final state','' {\em
  J. High Energy Phys.} {\bf 2004} (2004) no.~3, 026.

\bibitem{Prawer08}
S.~Prawer and A.~D. Greentree, ``Diamond for quantum computing,'' {\em Science}
  {\bf 320} (2008)  1601.

\bibitem{Ojima92}
I.~Ojima, ``Nature vs.\ science,'' {\em Acta Inst. Phil. Aesth.} {\bf 10}
  (1992)  55.

\bibitem{Schlosshauer07}
M.~Schlosshauer, {\em Decoherence and the Quantum-to-Classical Transition}.
\newblock Springer-Verlag, 2007.

\bibitem{Schlosshauer08}
M.~Schlosshauer and K.~Camilleri, ``Niels {Bohr} as philosopher of experiment:
  Does decoherence theory challenge {Bohr's} doctrine of classical concepts?,''
  {\em Stud. Hist. Phil. Mod. Phys.} {\bf 49} (2015)  73--83.

\bibitem{Barnum10}
H.~Barnum, ``Quantum knowledge, quantum belief, quantum reality. {Notes} of a
  {QBist} fellow traveler,'' \href{http://arxiv.org/abs/1003.4555}{{\tt
  arXiv:1003.4555 [quant-ph]}}.

\bibitem{Kofler08}
J.~Kofler, {\em Quantum Violation of Macroscopic Realism and the Transition to
  Classical Physics}.
\newblock PhD thesis, University of Vienna, 2008.
\newblock \href{http://arxiv.org/abs/0812.0238}{{\tt arXiv:0812.0238
  [quant-ph]}}.

\bibitem{Haag10}
R.~Haag, ``Some people and some problems met in half a century of commitment to
  mathematical physics,'' {\em Eur. Phys. J. H} {\bf 35} (2010)  263--307.

\bibitem{Bekenstein2008}
J.~Bekenstein, ``Bekenstein bound,'' {\em Scholarpedia} {\bf 3} (2008) no.~10,
  7374. \url{http://www.scholarpedia.org/article/Bekenstein_bound}.

\bibitem{Deutsch86}
P.~C.~W. Davies and J.~R. Brown, eds., {\em The Ghost in the Atom: A Discussion
  of the Mysteries of Quantum Physics}.
\newblock Cambridge University Press, 1986.

\bibitem{James22}
W.~James, {\em Pragmatism, a New Name for Some Old Ways of Thinking: Popular
  Lectures on Philosophy}.
\newblock Longmans, Green and Co., New York, 1922.

\bibitem{Einstein49b}
A.~Einstein, ``Remarks concerning the essays brought together in this
  co-operative volume,'' in {\em Albert Einstein: Philosopher-Scientist}, P.~A.
  Schilpp, ed., pp.~665--88.
\newblock Tudor Publishing Co., New York, 1949.

\bibitem{DeBrota16}
J.~B. DeBrota, ``A quantum information geometric approach to renormalization,''
  \href{http://arxiv.org/abs/1609.09440}{{\tt arXiv:1609.09440 [quant-ph]}}.

\bibitem{Fuchs11}
C.~A. Fuchs and R.~Schack, ``Bayesian conditioning, the reflection principle,
  and quantum decoherence,'' in {\em Probability in Physics}, pp.~233--47.
\newblock Springer, 2012.
\newblock \href{http://arxiv.org/abs/1103.5950}{{\tt arXiv:1103.5950
  [quant-ph]}}.

\bibitem{stacey-thesis}
B.~C. Stacey, {\em Multiscale Structure in Eco-Evolutionary Dynamics}.
\newblock PhD thesis, Brandeis University, 2016.
\newblock \href{http://arxiv.org/abs/1509.02958}{{\tt arXiv:1509.02958
  [q-bio.PE]}}.

\bibitem{Lamberth99}
D.~C. Lamberth, {\em William James and the Metaphysics of Experience}.
\newblock Cambridge University Press, 1999.

\bibitem{Taylor96}
E.~Taylor and R.~H. Wozniak, {\em Pure Experience: The Response to William
  James}.
\newblock Thoemmes Press, Bristol, UK, 1996.

\bibitem{Wild69}
J.~Wild, {\em The Radical Empiricism of William James}.
\newblock Doubleday, Garden City, NY, 1969.

\bibitem{Gieser05}
S.~Gieser, {\em The Innermost Kernel: Depth Psychology and Quantum
  Physics.~Wolfgang Pauli's Dialogue with C.~G. Jung}.
\newblock Springer, 2005.

\bibitem{Russell08}
B.~Russell, {\em The Analysis of Mind}.
\newblock Arc Manor, Rockville, MD, 2008.

\bibitem{Banks03}
E.~C. Banks, {\em Ernst Mach's World Elements: A Study in Natural Philosophy}.
\newblock Kluwer, Dordrecht, 2003.

\bibitem{Heidelberger04}
M.~Heidelberger, {\em Nature from Within: Gustav Theodor Fechner and His
  Psychophysical Worldview}.
\newblock University of Pittsburgh Press, 2004.

\end{thebibliography}\endgroup

\end{document}